\documentclass[fleqn,usenatbib]{mnras}

\usepackage{newtxtext,newtxmath}

\usepackage[T1]{fontenc}

\DeclareRobustCommand{\VAN}[3]{#2}
\let\VANthebibliography\thebibliography
\def\thebibliography{\DeclareRobustCommand{\VAN}[3]{##3}\VANthebibliography}

\usepackage[version=4]{mhchem}
\usepackage{chemmacros}
\chemsetup{
  modules = {reactions} ,
  formula = mhchem
}

\chemsetup[reactions]{
  before-tag = R ,
  tag-open = ( ,
  tag-close = )
}
\usepackage{graphicx}	
\usepackage[pdfpagelabels=false]{hyperref}	
\hypersetup{colorlinks=true,linkcolor=blue,citecolor=blue,filecolor=blue,urlcolor=blue,}

\title[Lightning-induced chemistry on tidally-locked Earth-like exoplanets]{Lightning-induced chemistry on tidally-locked Earth-like exoplanets}

\author[M. Braam et al.]{
Marrick Braam$^{1,2,3}$,\thanks{E-mail: \href{mailto:mbraam@ed.ac.uk}{mbraam@ed.ac.uk}}
Paul I. Palmer$^{1,2}$,
Leen Decin$^{3}$,
Robert J. Ridgway$^{4}$,
Maria Zamyatina$^{4}$,
\newauthor~Nathan J. Mayne$^{4}$,
Denis E. Sergeev$^{4}$,
and N. Luke Abraham$^{5,6}$\\
$^{1}$School of GeoSciences, University of Edinburgh, Edinburgh, EH9 3FF, UK\\
$^{2}$Centre for Exoplanet Science, University of Edinburgh, Edinburgh, EH9 3FD, UK\\
$^{3}$Institute of Astronomy, KU Leuven, 3001 Leuven, Belgium\\
$^{4}$Department of Physics and Astronomy, Faculty of Environment, Science and Economy, University of Exeter, Exeter, EX4 4QL, UK\\
$^{5}$Yusuf Hamied Department of Chemistry, University of Cambridge, Lensfield Road, Cambridge CB2 1EW, UK\\
$^{6}$National Centre for Atmospheric Science, UK
}

\date{Accepted XXX. Received YYY; in original form ZZZ}

\pubyear{2022}

\begin{document}
\label{firstpage}
\pagerange{\pageref{firstpage}--\pageref{lastpage}}
\maketitle

\begin{abstract}
Determining the habitability and interpreting atmospheric spectra of exoplanets requires understanding their atmospheric physics and chemistry. We use a 3-D Coupled Climate-Chemistry Model, the Met Office Unified Model with the UK Chemistry and Aerosols framework, to study the emergence of lightning and its chemical impact on tidally-locked Earth-like exoplanets. We simulate the atmosphere of Proxima Centauri b orbiting in the Habitable Zone of its M-dwarf star, but the results apply to similar M-dwarf orbiting planets. Our chemical network includes the Chapman ozone reactions and hydrogen oxide (HO$_{\rm{x}}$=H+OH+HO$_2$) and nitrogen oxide (NO$_{\rm x}$=NO+NO$_2$) catalytic cycles. 
We find that photochemistry driven by stellar radiation (177--850~nm) supports a global ozone layer between 20--50~km. We parameterise lightning flashes as a function of cloud-top height and the resulting production of nitric oxide (NO) from the thermal decomposition of N$_2$ and O$_2$. Rapid dayside convection over and around the substellar point results in lightning flash rates of up to 0.16 flashes $\mathrm{km}^{-2}\mathrm{yr}^{-1}$, enriching the dayside atmosphere below altitudes of 20~km in NO$_\mathrm{x}$.  Changes in dayside ozone are determined mainly by UV irradiance and the HO$_\mathrm{x}$ catalytic cycle. ${\sim}45$\% of the planetary dayside surface remains at habitable temperatures (T$_\mathrm{surf}{>}273.15$ K) and the ozone layer reduces surface UV radiation levels to 15\%.
Dayside-nightside thermal gradients result in strong winds that subsequently advect NO$_\mathrm{x}$ towards the nightside, where the absence of photochemistry allows NO$_\mathrm{x}$ chemistry to involve reservoir species. Our study also emphasizes the need for accurate UV stellar spectra to understand the atmospheric chemistry of exoplanets.
\end{abstract}




\section{Introduction}
We know that exoplanets are ubiquitous in the galaxy \citep[e.g.][]{kopparapu_habitable_2013, hsu_occurrence_2019}, but for those that support an atmosphere, we know little about the associated physical and chemical properties. This gap in our knowledge has implications for understanding whether these planets are potentially habitable and whether they could present false-positive biosignatures \citep[e.g.][]{scalo_m_2007, schwieterman_exoplanet_2018}. With the successful launch of the James Webb Space Telescope (JWST) in late 2021 and the construction of new ground-based facilities (such as the Extremely Large Telescope, ELT), we can expect the first insights into the atmospheres of some observationally favourable and potentially habitable exoplanets \citep[e.g.][]{lustig-yaeger_detectability_2019}. Simulating the physical and chemical properties of these atmospheres is essential to interpreting the observations. Here, we focus on understanding the atmospheric composition of an Earth-like exoplanet orbiting an M-dwarf star, which is likely to sustain habitable surface conditions for a range of atmospheric compositions \citep[][]{turbet_habitability_2016, boutle_exploring_2017, yates_ozone_2020}.

Earth-size planets are preferentially discovered in close-in orbits around M-dwarf stars \citep{dressing_occurrence_2015} and their potential habitability is an active area of research \citep[e.g.][]{shields_habitability_2016}. M stars are cooler and smaller compared to other types of stars. Consequently, the circumstellar Habitable Zone (HZ) \citep[][]{kasting_habitable_1993, kopparapu_habitable_2013} moves inward. A planet orbiting in this HZ is likely to be tidally locked \citep[e.g.][]{barnes_tidal_2017}, which results in large temperature differences between the dayside and nightside of a planet. Proxima Centauri b \citep[][]{anglada-escude_terrestrial_2016} is a nearby example of a planet orbiting in the HZ of an M star. Assuming the planet has an atmosphere, hemispheric mean temperatures from simulations can differ by $\sim$60~K for Proxima Centauri b \citep[][]{boutle_exploring_2017, sergeev_atmospheric_2020}, and the planet is a candidate for surface habitability \citep[see also][]{ribas_habitability_2016, turbet_habitability_2016, lewis_influence_2018}. The tidally-locked configuration has implications for winds and vertical transport that require an understanding of the full three-dimensional circulation including clouds. This motivates the adaptation of general circulation models (GCMs), used for weather and climate predictions for Earth, to exoplanets. The Met Office Unified Model (UM) has been adapted and applied to a wide range of exoplanets \citep[e.g.][]{mayne_using_2014, mayne_unified_2014, boutle_exploring_2017, drummond_effect_2018, drummond_implications_2020} and was also included in the recent THAI project \citep[][]{turbet_trappist-1_2022, sergeev_trappist-1_2022, fauchez_trappist-1_2022}, an intercomparison of GCM outputs for potentially habitable atmospheres on TRAPPIST-1 e. 

The main driver of atmospheric circulation on terrestrial exoplanets is the incoming stellar irradiation. The impact of different spectral energy distributions has been studied for fast-rotating exoplanets \citep[][]{shields_effect_2013, wolf_constraints_2017} and for tidally-locked planets \citep[][]{eager_implications_2020}. For tidally-locked planets, the day-night contrast in stellar radiation results, in many cases, in the development and maintenance of equatorial jets that redistribute heat to the nightside \citep[][]{showman_atmospheric_2002, showman_equatorial_2011, koll_temperature_2016}. This redistribution can help to prevent atmospheric collapse on the nightside \citep[][]{joshi_simulations_1997, turbet_modeling_2018}. Besides the jet, many 3-D GCM simulations show stationary gyres at mid-latitudes and moderate divergence at the substellar point \citep[e.g.][]{carone_connecting_2014, carone_connecting_2015, hammond_equatorial_2020, hammond_rotational_2021}. The circulation regime also varies with orbital period \citep[e.g.][]{merlis_atmospheric_2010, edson_atmospheric_2011, carone_connecting_2015, carone_stratosphere_2018} and time-dependent wave phenomena can further impact our ability to interpret observations \citep[][]{cohen_longitudinally_2022}. \citet{yang_stabilizing_2013} show that the dayside is covered by a thick cloud deck, resulting from vigorous convection centred at the substellar point. This results in a cloud albedo feedback that is also sensitive to the orbital period \citep[][]{yang_stabilizing_2013}. For Proxima Centauri b, nightside Rossby gyres develop on either side of the equatorial jet \citep[][]{turbet_habitability_2016, boutle_exploring_2017}. These gyres trap air, that experiences extensive radiative cooling \citep[][]{yang_low-order_2014}, as described by \citet{boutle_exploring_2017} and \citet{yates_ozone_2020}. As a consequence, the atmospheric pressure decreases locally and atmospheric constituents such as ozone are drawn downwards to lower altitudes thereby increasing local ozone column abundances \citep[][]{yates_ozone_2020}. Most GCM studies assume aquaplanets with a slab ocean, but ocean heat transport can also increase the habitable area of a (tidally-locked) planet \citep[][]{hu_role_2014, del_genio_habitable_2019}. The existence and distribution of landmasses further influence the planetary climate through changes in convection and water evaporation \citep[e.g.][]{abe_habitable_2011, lewis_influence_2018, rushby_effect_2020}. Finally, the chemical composition of the atmosphere also affects the radiative balance of the planet through scattering and absorption of incoming and outgoing radiation. Static chemical compositions have been investigated \citep[e.g.][]{pierrehumbert_palette_2010, turbet_habitability_2016, boutle_exploring_2017}, but there is a balance between the irradiation, the atmospheric physics and chemistry (in particular constituents that have significant opacities under the incident stellar radiation).

This balance motivates the need for coupled climate-chemistry models (CCMs) to study the relationships between radiatively active atmospheric constituents (gases and aerosols) and the atmospheric dynamics of the planet. There is a growing body of work investigating 3-D atmospheric photochemistry on exoplanets orbiting in the HZ, from understanding the impact of tidal locking on Earth's ozone distribution \citep[][]{proedrou_characterising_2016} to understanding how the stellar flux distributions of M-dwarfs influence both the magnitude and distribution of atmospheric biosignature gases \citep[][]{chen_biosignature_2018} and an ozone layer on Proxima Centauri b through changes in chemical production and loss rates \citep[][]{yates_ozone_2020}. Perturbations in the incident stellar radiation will influence the planetary atmospheric chemistry and physics. The magnitude of chemical perturbations due to stellar flares (through enhanced UV activity and proton events), for example, is determined by a combination of the planetary magnetic field, radiation environment and atmospheric circulation \citep[][]{chen_persistence_2021}. Perturbations are seen especially in distributions of NO, OH, and ozone for unmagnetized, tidally-locked planets around K and M stars. Planets that have protective magnetic fields are able to (partially) counteract the effect of flares \citep[][]{chen_persistence_2021}, emphasizing the potential role of stellar activity in determining the habitability of a planet. Furthermore, both inter-annual and seasonal variations in clouds and chemistry can impact the observability of spectral features on Earth-analogue exoplanets \citep[][]{cooke_variability_2022}.

On Earth, the presence of lightning discharges can lead to local perturbations in atmospheric chemistry. Lightning has sufficient energy to thermally decompose molecular nitrogen and oxygen to form an abiotic source of nitrogen oxides (NO$_\mathrm{x}$=NO+NO$_2$, \citealt{crutzen_influence_1970, schumann_global_2007}). Through atmospheric chemistry, NO$_\mathrm{x}$ can influence the distribution of atmospheric ozone. Ozone is the photochemical byproduct of molecular oxygen and thus depends on oxygen levels, as is shown by the 3-D CCM simulations of \citet{cooke_revised_2022} for various epochs in Earth's history. Furthermore, the nonlinear oxygen-ozone relationship depends on the host star's UV spectrum \citep[][]{kozakis_is_2022}. Since oxygen on Earth is largely produced by sources of biological origin \citep[][]{schwieterman_exoplanet_2018}, its photochemical product ozone can be seen as a potential biosignature. However, pathways to false positives for abiotic oxygen and ozone on planets orbiting M-dwarfs exist, including accumulation of O$_2$ following the photolysis of H$_2$O and subsequent H escape \citep[e.g.][]{tian_history_2015, wordsworth_redox_2018, lincowski_observing_2019} and CO$_2$ photolysis releasing oxygen atoms \citep[e.g.][]{kasting_evolution_2003, domagal-goldman_abiotic_2014, harman_abiotic_2015}. Understanding these abiotic influences is essential for the interpretation of spectral signatures \citep[][]{schwieterman_exoplanet_2018}. 

Studies using 1-D photochemical models have considered the impact of global thunderstorms on exoplanetary atmospheric chemistry \citep[][]{rimmer_chemical_2016, ardaseva_lightning_2017, harman_abiotic_2018}. It was found that the chemical effect of lightning would be hard to detect on Earth-like exoplanets \citep[][]{ardaseva_lightning_2017}. \citet{harman_abiotic_2018} showed that catalytic cycles following NO production from lightning enhance the reliability of O$_2$ as a biosignature, assuming Earth-like chemical composition. However, just like planetary atmospheres and climates, the emergence of lightning is a 3-D process, that depends on a combination of cloud formation, particle charging and charge separation \citep[][]{helling_exoplanet_2019}. In GCMs of Earth, lightning is usually parameterised in terms of convective parameters, such as cloud-top height \citep[][]{price_simple_1992, luhar_assessing_2021}, convective precipitation and mass flux \citep[][]{allen_evaluation_2002}, and upward cloud ice flux \citep[][]{finney_using_2014}. Lightning has already been observed on the giant planets in the Solar System \citep[e.g.][]{aplin_atmospheric_2006, hodosan_lightning_2016}, but is yet to be detected on exoplanets. However, GCMs predict a thick and convective cloud deck to cover the dayside of tidally-locked planets \citep[e.g.][]{yang_stabilizing_2013} and the first evidence for clouds on exoplanets is being obtained \citep[][]{pont_prevalence_2013, kreidberg_precise_2014, diamond-lowe_ground-based_2018}. Therefore, lightning on exoplanets is a reasonable expectation \citep[][]{helling_exoplanet_2019}, leading to potentially important disruptions of the atmospheric chemistry. 

Here, we investigate the impact of lightning-induced chemistry on a tidally-locked exoplanet in the HZ. We use the UM in the configuration of a planet orbiting an M-dwarf star, nominally Proxima Centauri b, building on previous studies \citep[][]{boutle_exploring_2017, yates_ozone_2020}. We also use the UK Chemistry and Aerosol framework (UKCA), coupled with the UM, to describe gas-phase chemistry, with lightning as the main source of nitric oxide (NO). In Section~\ref{sec:methods}, we describe the model setup, the process of making UKCA compatible with the M-dwarf setup and our choice of lightning parameterisation. In Section~\ref{sec:results} we present our results, briefly discussing the planetary climate before presenting the emergence of lightning and the resulting atmospheric chemistry. We discuss the importance of stellar fluxes, comparison to other results and potential observability in Section~\ref{sec:discussion}. Finally, we present the conclusions of our study in Section~\ref{sec:conclusion}. 

\section{Methods}\label{sec:methods}
In this Section, we describe the different components of the CCM, the lightning parameterisation, and, lastly, we specify the experimental setup of our simulations.
\subsection{Unified Model}
\label{sec:unmodel}
We use the UM, a 3-D GCM, in its Global Atmosphere 7.0 configuration \citep[][]{walters_met_2019}. The ENDGAME dynamical core solves the non-hydrostatic fully compressible deep-atmosphere equations of motion \citep[][]{wood_inherently_2014}. Parameterised sub-grid scale processes include convection that is described using a mass flux-based approach \citep[][]{gregory_mass_1990}, water cloud physics that is described using a prognostic condensate scheme \citep[][]{wilson_pc2_2008}, and turbulent mixing \citep[][]{lock_new_2000, brown_upgrades_2008}. The atmospheric radiative transfer is described by the Suite of Community Radiative Transfer codes based on Edwards and Slingo (SOCRATES) scheme, which uses the correlated-k method \citep{edwards_studies_1996, manners_socrates_2021}. The UM is typically used to study Earth's weather and climate but recently has been adapted to study different types of exoplanets \citep[e.g][]{mayne_unified_2014, mayne_results_2017, drummond_effect_2018, drummond_implications_2020, mayne_using_2014, boutle_exploring_2017, lewis_influence_2018, yates_ozone_2020, eager_implications_2020, sergeev_atmospheric_2020}. 

Here, we adapt the UM to investigate the climate dynamics and atmospheric chemistry of Proxima Centauri b \citep[][]{anglada-escude_terrestrial_2016}, in a circular, tidally-locked orbit around its host star following previous studies \citep[][]{boutle_exploring_2017, yates_ozone_2020}. The stellar, orbital and planetary parameters are listed in Table~\ref{tab:orbplan_params}. The horizontal resolution is $2$ by $2.5\degr$ in latitude and longitude, respectively. The atmosphere is divided into 60 vertical levels extending from the surface to 85~km, with quadratic stretching to enhance resolution near the surface, following \citet{yates_ozone_2020}. 
To describe stellar radiation, we use the composite spectrum at version 2.2 as presented by the MUSCLES spectral survey \citep{france_muscles_2016, youngblood_muscles_2016, loyd_muscles_2016}. This spectral energy distribution was created from archival data of XMM-Newton and the Hubble Space Telescope (HST) and covers wavelengths from 0.5~nm to 5.5~$\mu$m. Since the 6 `shortwave' bands of SOCRATES treat incoming radiation up to 10~$\mu$m, we extended our spectrum by using the spectrum presented by \citet{ribas_full_2017} for wavelengths between 5.5 and 10~$\mu$m. This final composite spectrum was used to recalculate correlated-k absorption coefficients. Previous UM studies of Proxima Centauri b used a synthetic BT-Settl spectrum, appropriate for the host star properties \citep[][]{boutle_exploring_2017, yates_ozone_2020}. We assume an aquaplanet covered by a 2.4~m slab ocean mixed layer with a total heat capacity of 10$^7$~J~K$^{-1}$~m$^{-2}$. Sea ice formation is not included in our simulations, but the associated ice-albedo feedback is weak for planets around M-dwarfs \citep[][]{joshi_suppression_2012, shields_effect_2013}. 

We assume a surface pressure of one bar for our simulations, building on previous work with the UM as well as other GCM studies \citep[e.g.][]{joshi_climate_2003, merlis_atmospheric_2010, yang_stabilizing_2013, carone_connecting_2014, kopparapu_inner_2016, boutle_exploring_2017, rushby_effect_2020, turbet_trappist-1_2022, sergeev_trappist-1_2022}. One bar of surface pressure is also a common assumption for many other photochemical models, ranging from 1-D \citep[e.g.][]{domagal-goldman_abiotic_2014, tian_high_2014, harman_abiotic_2015, harman_abiotic_2018} to global 3-D CCM studies \citep[][]{chen_biosignature_2018, chen_habitability_2019, chen_persistence_2021, yates_ozone_2020}. We acknowledge that stellar activity can potentially have detrimental effects on the atmospheric mass and thus surface pressure of Proxima Centauri b, as shown by \cite{garraffo_space_2016, garcia-sage_magnetic_2017, airapetian_how_2017, airapetian_impact_2020}. For this first investigation of lightning-induced chemistry, we opted for the case of a 1 bar surface pressure, since 1) we are currently unable to constrain any particular value of the surface pressure as the best value, so a sensible starting point for this first study is to use the atmospheric parameters that we understand in the greatest detail, and 2) the results apply more generally for tidally-locked planets that reside in the HZ of their M-dwarf host star. This may include planets with the potential to sustain a 1 bar N$_2$-dominated atmosphere, for example the outer planets of the TRAPPIST-1 system \citep[][]{turbet_review_2020}.

In this work, we focus on the impacts of lightning and use a time-averaged stellar spectrum. However, this work has been performed in close collaboration with a complementary study, using the UM but including the impact of stellar activity while omitting lightning. This study, Ridgway et al. (submitted), is also based on Proxima Centauri b, but uses SOCRATES to calculate the photolysis rates and a simplified idealised chemistry scheme to capture the ozone interactions \citep[][]{drummond_effects_2016, drummond_implications_2020}. The complementarity of these two studies has provided an excellent opportunity for mutual testing and development.

\begin{table}
	\centering
	\caption{Orbital and planetary parameters for the Proxima Centauri b setup, following \citet{boutle_exploring_2017}.}
	\label{tab:orbplan_params}
	\begin{tabular}{ll} 
		\hline
		Parameter & Value \\
		\hline
		Semi-major axis (AU) & 0.0485 \\
		Stellar Irradiance (W~m$^{-2}$) & 881.7 \\
		Orbital Period (days) & 11.186 \\
		Rotation rate (rad~s$^{-1}$) & $6.501\times10^{-6}$ \\
		Eccentricity & 0 \\
		Obliquity & 0 \\
		Radius (R$_\oplus$) & 1.1 \\
		Surface gravity (m~s$^{-2}$) & 10.9 \\
		\hline
	\end{tabular}
\end{table}

\subsection{UK Chemistry and Aerosol framework}\label{subsec:UKCA}
The UK Chemistry and Aerosol (UKCA) model \citep{morgenstern_evaluation_2009, oconnor_evaluation_2014, archibald_description_2020} is a framework that we use to describe the global atmospheric chemical composition of our simulated exoplanet. UKCA includes aerosol and gas-phase chemistry and is coupled to the UM dynamics. It uses the UM components for large scale advection, convective transport and boundary layer mixing of its aerosol and chemical tracers \citep{oconnor_evaluation_2014, archibald_description_2020}. UKCA contains a large number of gas-phase and heterogeneous chemical reactions,  some of which we have included in our chemical network. Furthermore, the chemistry schemes in UKCA describe wet and dry deposition \citep[][]{giannakopoulos_validation_1999}. In this study, we use the Stratospheric \citep[Strat,][]{morgenstern_evaluation_2009} and Stratospheric-Tropospheric \citep[StratTrop,][]{archibald_description_2020} chemistry schemes. Originally, StratTrop includes 75 chemical species that are connected by 283 reactions \citep{archibald_description_2020}. 
We used a reduced version of the UKCA chemistry schemes (Table~\ref{tab:chemsetups}), to quantify the impact of the different chemical mechanisms on the atmospheric chemistry of a tidally-locked exoplanet. First, we use a simple network that describes the Chapman mechanism of ozone formation \citep[][]{chapman_xxxv_1930}, following \citet{yates_ozone_2020}. Second, we add the reactive hydrogen (HO$_\mathrm{x}$) catalytic cycle, where HO$_\mathrm{x}$ denotes the ensemble of atomic hydrogen (H), the hydroxyl radical (OH) and the hydroperoxy radical (HO$_2$). We include this cycle to account for ozone chemistry following the oxidation and photolysis of water vapour. Lastly, we add the nitrogen oxide (NO$_\mathrm{x}$) catalytic cycle, including NO and nitrogen dioxide (NO$_2$), to the network. We also include other oxidised nitrogen species, such as nitrate (NO$_3$), nitrous oxide (N$_2$O), and the reservoirs nitric acid (HNO$_3$) and dinitrogen pentoxide (N$_2$O$_5$). Collectively, these nitrogen species belong to the NO$_\mathrm{y}$ family and can also influence ozone chemistry. In our simulations, lightning is the main source of NO that initiates further NO$_\mathrm{y}$ chemistry, as described in Section~\ref{subsec:lightning}. In the upper atmosphere, the slow termolecular reaction between N$_2$ and O($^1$D) provides another source of NO$_\mathrm{y}$, but this does not impact the lightning-induced chemistry that occurs at altitudes below 20~km.

\begin{table*}
 \caption{Specifications of the different chemistry schemes that were used in this study. The schemes presented are reduced versions of UKCA's Strat \citep{morgenstern_evaluation_2009} and StratTrop chemistry schemes \citep{archibald_description_2020}. Each row also includes the chemistry from the rows above, hence `+NO$_\mathrm{x}$' means NO$_\mathrm{x}$-chemistry added to HO$_\mathrm{x}$ and the Chapman mechanism. Tables~\ref{tab:initabundances} and \ref{tab:reactions} give a full overview of the chemical species and reactions included in each scheme.}
 \label{tab:chemsetups}
 \begin{tabular}{l|ccccccc}
  \hline
 Chemistry & Species & \multicolumn{1}{p{2cm}}{\centering Bimolecular \\ Reactions} & \multicolumn{1}{p{2cm}}{\centering Termolecular \\ Reactions} & \multicolumn{1}{p{2cm}}{\centering Photolysis \\ Reactions} & \multicolumn{1}{p{2cm}}{\centering Total \\ Reactions} & \multicolumn{1}{p{2cm}}{\centering Chemistry \\ Scheme} & Lightning-NOx \\
  \hline
Chapman & 6 & 6 & 1 & 4 & 11 & Strat & No \\
+HO$_\mathrm{x}$ & 12 & 23 & 4 & 6 & 33 & StratTrop & No \\
+NO$_\mathrm{x}$ & 21 & 42 & 14 & 15 & 71 & StratTrop & Yes \\
  \hline
 \end{tabular}
\end{table*}

The complete list of species, reactions, initial conditions and details about deposition states in each of the 3 reduced schemes can be found in Tables~\ref{tab:initabundances} and \ref{tab:reactions}, respectively. We also note which species are active in the SOCRATES radiation scheme. Following the Earth-like atmospheric setup from \citet{boutle_exploring_2017}, the initialisation of N$_2$, O$_2$ and CO$_2$ is based on pre-industrial Earth abundances, and these species are assumed to be well-mixed. Initial values for H$_2$O are based on evaporation from the slab ocean. To avoid the impact of initial conditions that are far from steady-state values, the remainder of the HO$_\mathrm{x}$ and the NO$_\mathrm{y}$ species are initialised at mass mixing ratios of $10^{-9}$ and $10^{-15}$, respectively. All the other species are initially set to zero and do not participate in the subsequent atmospheric chemistry. UKCA also includes surface emissions, but we set them to zero. Lightning discharges produce atmospheric emissions of NO, which we describe in Section~\ref{subsec:lightning}. Since lightning is the only source of NO in the lower 20~km of the atmosphere, this initiates the subsequent NO$_\mathrm{y}$ chemistry.

\subsection{Fast-JX photolysis code}\label{subsec:fastjx}
Besides participating in chemical reactions, atmospheric species can be photolysed by the interaction with ultraviolet (UV) and visible radiation \citep{bian_fast-j2_2002}. To describe atmospheric photolysis, UKCA uses the Fast-JX photolysis scheme \citep{wild_excitation_2000, bian_fast-j2_2002, neu_global_2007, telford_implementation_2013}. Fast-JX is an efficient photolysis scheme that takes into account the varying optical depths of Rayleigh scattering, absorbing gases, clouds and aerosols. In this way, Fast-JX provides an interactive treatment of photolysis in modelling (3-D) atmospheric compositions. The radiation is divided over 18 wavelength bins, with 11 bins covering 177--291~nm and seven bins covering 291--850~nm. These bins group regions of similar absorption as specified by \citet{bian_fast-j2_2002}. Fast-JX calculates how many photons of each wavelength are absorbed and/or scattered as light passes through the plane-parallel atmosphere \citep{telford_implementation_2013}. Photolysis rates are then calculated from the actinic flux, cross-sections and quantum yields in each bin. 

\begin{figure*}
	\includegraphics[width=2\columnwidth]{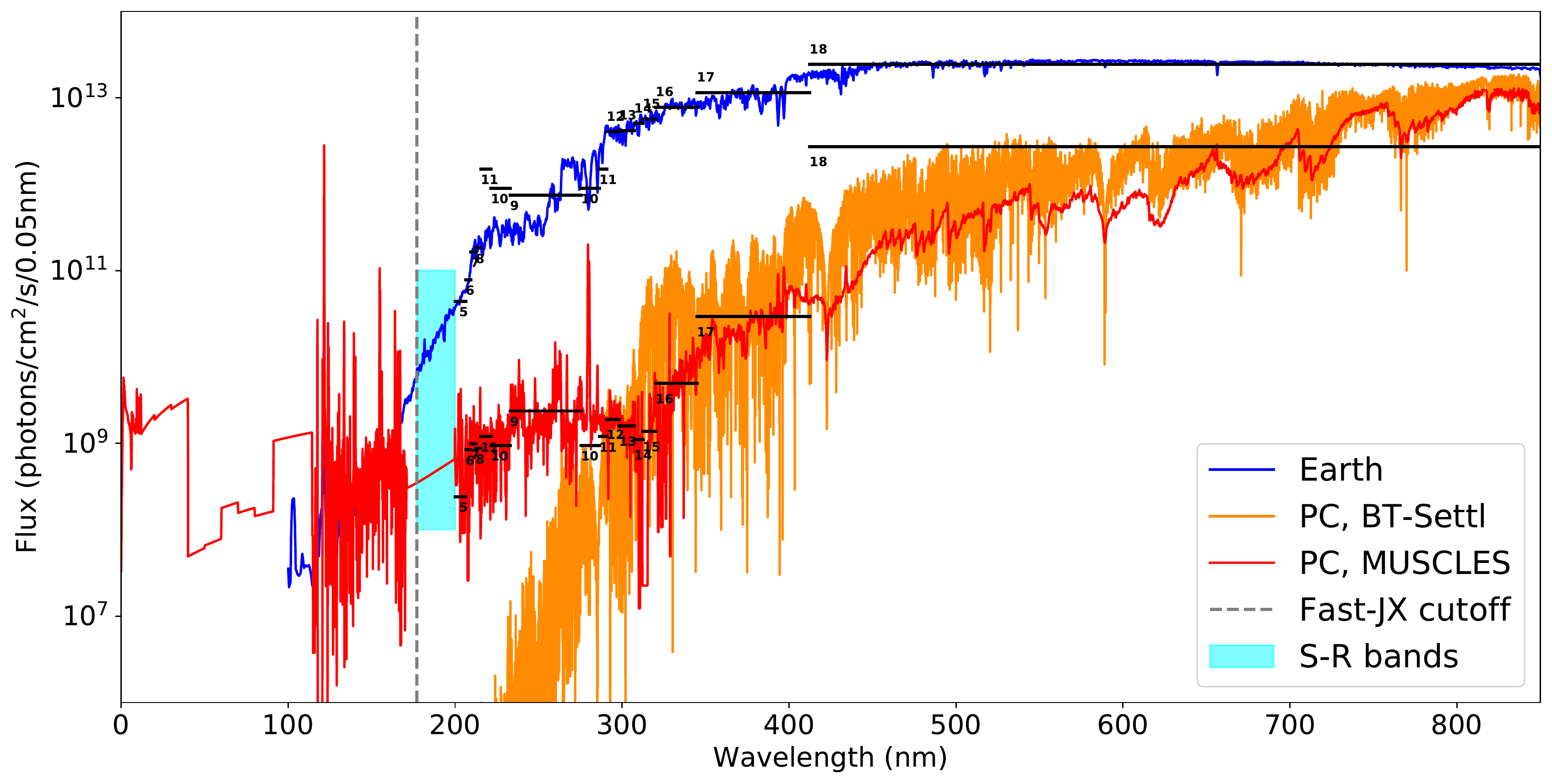}
    \caption{Top-of-the-atmosphere fluxes received on Earth and Proxima Centauri b. The red line shows the flux for the composite MUSCLES spectrum of Proxima Centauri, as used in our study; and the orange line denotes the stellar flux generated by the BT-Settl spectrum that was used by \citet{boutle_exploring_2017, yates_ozone_2020}. Fast-JX treats fluxes at wavelengths between 177 (the dashed vertical line) and 850~nm. The solid horizontal black lines denote the flux per bin for Fast-JX bins 5--18 and are numbered accordingly. Bins 1--4 and part of bin 5 contain a combination of the Schumann-Runge (S-R) bands \citep{bian_fast-j2_2002} and treat the fluxes falling in the shaded rectangular patch, or between 177 and 202.5 nm.}
    \label{fig:spectra_irrad}
\end{figure*}

For Fast-JX, we extract the radiation between 177 and 850~nm from the MUSCLES spectrum for Proxima Centauri. To describe non-Earth orbits in UKCA, we follow \citet{yates_ozone_2020} and scaled our M-dwarf fluxes to find the top-of-the-atmosphere (TOA) flux received by a planet at 1~AU. A synchronization to the orbital distance of Proxima Centauri b is added to UKCA to determine the TOA flux received by the planet. This is shown in Figure \ref{fig:spectra_irrad}, along with the TOA flux for Earth. 

We regroup the fluxes over the wavelength bins, following \citet{bian_fast-j2_2002}, to produce the TOA fluxes at 1~AU as shown in Table~\ref{tab:fluxpcb} (and as the horizontal black lines in Figure~\ref{fig:spectra_irrad}). Precursor work by \citet{yates_ozone_2020} using the BT-Settl spectrum did not include this regrouping and instead erroneously divided the fluxes over 18 bins in terms of increasing wavelength \citep[see Table 1 of][]{yates_ozone_2020}. Comparing our results with those from \citet{yates_ozone_2020} highlights significant differences between most of the bins, such as bin 18, for which the flux from \citet{yates_ozone_2020} is ${\sim}$2 orders of magnitude higher than either the BT-Settl or MUSCLES spectra. This is due to lumping together fluxes up to 2150~nm, beyond the 850~nm upper limit for Fast-JX. This affects, for example, ozone absorption in the Chappuis bands \citep{burrows_atmospheric_1999} to produce molecular oxygen and O($^3$P); correcting this error reduces photolysis rates for this reaction by a factor of 20 (see also Appendix~\ref{app:o3chem}). Hence, this removes the destruction of ozone due to unphysical reasons. As is also seen from Figure~\ref{fig:spectra_irrad}, the MUSCLES spectrum is stronger by up to 8 orders of magnitude at shorter wavelengths ($\lambda{<}$300~nm). This UV radiation plays an essential role in atmospheric chemistry by driving the photolysis of molecular oxygen ($\lambda{<}$240~nm and ozone ($\lambda{<}$320~nm).

\begin{table}
\caption[TOA flux PC]{Top-of-the-atmosphere flux for a planet orbiting at 1 AU around Proxima Centauri.}
\label{tab:fluxpcb}
\begin{tabular}{ll} \hline
\multicolumn{1}{p{0.8cm}}{Bin \# } & \multicolumn{1}{p{2.2cm}}{\centering TOA flux at 1 AU     \\(photons s$^{-1}$cm$^{-2}$)} \\ \hline
1  & $1.091\times10^8$  \\
2  & $1.011\times10^8$ \\
3  & $9.240\times10^7$ \\
4  & $4.550\times10^7$ \\
5  & $3.301\times10^8$ \\
6  & $1.189\times10^8$ \\
7  & $1.392\times10^8$ \\
8  & $1.371\times10^8$ \\
9  & $4.739\times10^9$ \\
10 & $5.773\times10^9$ \\
11 & $5.922\times10^8$ \\
12 & $6.467\times10^8$ \\
13 & $6.881\times10^8$ \\
14 & $2.596\times10^8$ \\
15 & $5.086\times10^8$ \\
16 & $5.761\times10^{9}$ \\
17 & $9.321\times10^{10}$ \\
18 & $5.625\times10^{13}$ \\
\hline
\end{tabular}
\end{table}

\subsection{Emissions of NO from lightning}\label{subsec:lightning}
We use a lightning parameterisation based on simple scaling relations between the size of a thundercloud and the electrical power output \citep[][]{vonnegut_facts_1963}. The scaling relations are derived from the laws of electricity, assuming a thunderstorm as an electric dipole separated by a distance characterised by the cloud dimension. Since the number of lightning flashes depends on the electrical power, the lightning flash rates (LFR) can be described in terms of the cloud-top height $H$ \citep{williams_large-scale_1985, price_simple_1992, boccippio_lightning_2002, luhar_assessing_2021}.  Parameterisations are derived for continental and oceanic LFRs. Since we assume an aquaplanet, we only use the oceanic parameterisation:
\begin{equation}\label{eq:lfr_o_l20}
    LFR_O=2.0 \times10^{-5}H^{4.38}.
\end{equation}
The flash rates are calculated at locations where the convective cloud depth exceeds 5~km, and the cloud depth is based on the base and top convective cloud heights from the convection scheme. The threshold of 5~km follows from the range of data used to develop the parameterisation \citep[][]{price_modeling_1994}. The flash rates are subsequently apportioned into cloud-to-ground (CG) and intracloud (IC) flashes, based on an empirical ratio between the two \citep[][]{price_what_1993}. Extending the parameterisation to extraterrestrial environments, justified by our assumption of an Earth-like atmosphere and thus a similar process of charging and charge separation on Proxima Centauri b, we can provide a first assessment of the spatial and temporal distributions of lightning flashes in tidally-locked environments and study the relation to the planet's convective activity. 

Atmospheric electric discharges (including lightning) provide high-temperature channels of up to 30,000~K. In these channels, new trace molecules can be produced from the ambient atmospheric constituents \citep{rakov_lightning_2003}. In Earth's atmosphere, the production of NO affects ozone photochemistry \citep[][]{crutzen_influence_1970} and the lifetimes of a range of other gases, e.g. CO and nitrous oxide (N$_2$O) \citep{rakov_lightning_2003, brune_extreme_2021, mao_global_2021}. As the air cools rapidly, the abundances from the high-temperature reactions are `frozen in' via the so-called Zel'dovich mechanism \citep{zeldovich_oxidation_1947}. The exact consequences for the atmospheric composition depend on the lightning flash rates and the amount of NO that is produced per flash, terrestrial constraints on the amount of NO per flash are presented by \citet{schumann_global_2007}.

We use the UKCA emission formulation to describe emissions of NO from individual flashes \citep{luhar_assessing_2021}. The production of NO per flash is a key uncertainty, ranging from ${\sim}$33--700 moles NO per flash with averages of 250 \citep[][]{schumann_global_2007} and 180 moles NO per flash \citep[][]{bucsela_midlatitude_2019}. To match the global lightning-NO$_\mathrm{x}$ production on Earth, NO production rates are scaled to $216$ moles NO per flash \citep[][]{archibald_description_2020, luhar_assessing_2021}. Furthermore, we do not distinguish between CG and IC flashes in the NO production rate. The emitted NO is redistributed vertically, between 500~hPa (or ${\sim}$4.4~km) and the cloud top for IC flashes and between the surface and 500~hPa for CG flashes, and added to the NO concentration in the chemistry scheme. With a global mean surface pressure equal to 1000~hPa, the midpoint in terms of atmospheric mass is at ${\sim}500$~hPa.

\subsection{Experimental setup}
To ensure model stability and to avoid violating the Courant-Friedrichs-Lewy conditions for strong high-altitude winds, we ran our simulations at timesteps of four minutes for atmospheric dynamics. The chemical timestep was left at the default of one hour, as in \citet{yates_ozone_2020}. From the initial conditions, as described in Sections~\ref{sec:unmodel} and \ref{subsec:UKCA}, we run our simulations to a steady state as determined by the balance of incoming and outgoing TOA radiation fluxes and the stabilisation of the surface temperature. We find that the steady state for Proxima Centauri b is generally reached by ${\sim}1000$ Earth days. We also check that chemical steady state is reached by examining the stabilisation of the total ozone column and volume mixing ratio of ozone, since ozone is a long-lived species. For Proxima Centauri b, this is typically reached within 15 Earth years, depending on the complexity of the chemical network being used and subject to stochastic changes in atmospheric dynamics. Following \citet{yates_ozone_2020}, we run for 5 more years to a total of 20 years of spin-up time and report our results as the mean of the following 120 days (or ${\sim}$10 orbits of Proxima Centauri b).

\section{Results}\label{sec:results}
This section starts with a description of the planetary climate, followed by the distribution of lightning flashes. Then we discuss the resulting atmospheric chemistry, first focusing on ozone chemistry and, finally, on lightning-induced chemistry.

\subsection{Background climate}\label{sec:plan_clim}
The overall climate resulting from our simulations is largely similar to that presented in \citet{boutle_exploring_2017}, which is unsurprising given the similarity in the model configurations. Here we include a description of the main climate elements relevant to this study and refer the reader to \citet{boutle_exploring_2017} for a more complete description.

We can see a clear dayside-nightside contrast in surface temperature, with time-mean temperatures ranging from a maximum of 291~K on the dayside to as low as 157~K over two nightside Rossby gyres on either side of the equator, due to persistent radiative cooling. In terms of sustaining liquid water on the planetary surface, ${\sim}$45\% of the planet's dayside remains at habitable surface temperatures (${>}273.15$ K). Generally, we find much higher atmospheric specific humidity on the dayside, due to evaporation (higher overall temperatures) and convection, than on the nightside \citep[][]{boutle_exploring_2017, yates_ozone_2020}. Clouds and radiatively active gases such as ozone impede the penetration of UV radiation to the surface so that levels are lower than those found on Earth \citep[e.g.][]{omalley-james_uv_2017}. The dayside hemispheric mean level of UV surface radiation ($\lambda{<}$320~nm) is reduced to 15\% of the TOA value. This is about 0.2\% of the level on Earth's surface \citep[][]{segura_ozone_2003}, comparable to the findings of \citet{segura_biosignatures_2005} for planets orbiting M-dwarfs.

\begin{figure}
	\includegraphics[width=\columnwidth]{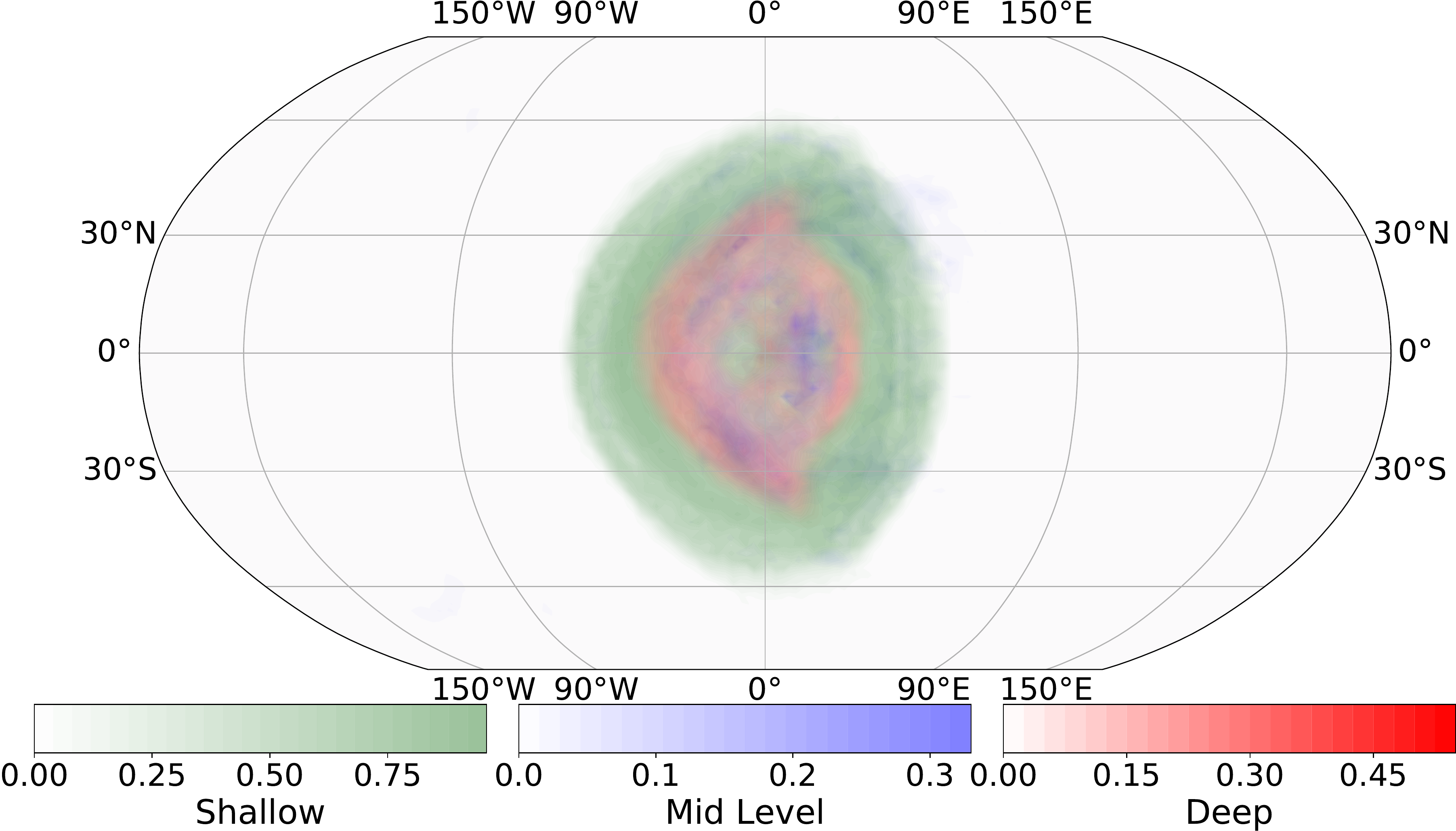}
    \caption{Time-mean (120 days) indicator of shallow, mid-level and deep convection over the planetary surface. The indicator is equal to 1 if a convection type is diagnosed or 0 if not, and the diagnosis of convection is based on undilute parcel ascent from the near surface, for grid boxes where the surface buoyancy flux is positive \citep[][]{walters_met_2019}.}
    \label{fig:pcb_conv}
\end{figure}

The distribution of clouds is influenced by the zonally asymmetric stellar heating of the planet due to the assumed tidally-locked configuration in our study. Intense heating at the substellar point drives deep convection and consequently water and ice cloud formation, resulting in a thick cloud deck that covers a large fraction of the dayside hemisphere centred on the substellar point \citep[e.g.][]{yang_stabilizing_2013, boutle_exploring_2017, sergeev_atmospheric_2020}. The equatorial jet advects the high clouds downstream, creating an asymmetry in the high-cloud cover that impacts the vertical extent of convective clouds \citep{boutle_exploring_2017}. From Figure~\ref{fig:pcb_conv} we can see that deep convection occurs around the substellar point and falls off radially. Furthermore, deep convective mixing tends to occur more intensely westward of the substellar point, coincident with a source of gravity waves \citep[][]{cohen_longitudinally_2022}. Due to a decreasing depth of convection  as a function of radial distance from the substellar point, lower-altitude clouds become more frequent \citep[][]{boutle_exploring_2017, sergeev_atmospheric_2020}. 

\begin{figure}
	\includegraphics[width=\columnwidth]{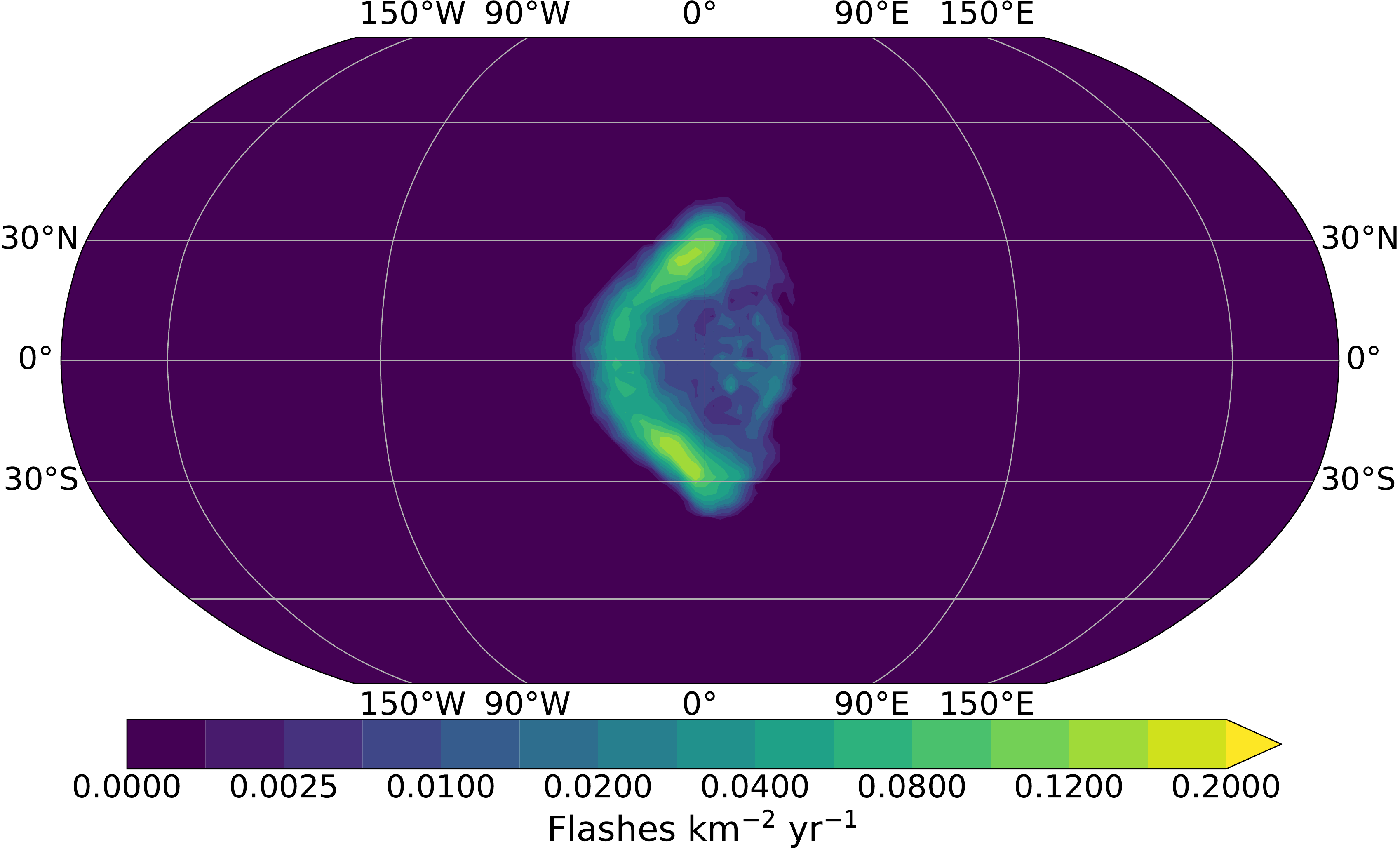}
    \caption{Annual lightning flash rates on Proxima Centauri b, following parameterisations shown in Equation~\ref{eq:lfr_o_l20} \citep{price_simple_1992, luhar_assessing_2021}. The mean of 120 days was taken from the high-frequency flash rate output and subsequently scaled to annual rates.}
    \label{fig:pcb_lfr}
\end{figure}

\subsection{Lightning flash rate estimates}\label{subsec:lightning_results}
From Earth, we expect lightning flashes to occur mainly in regions of vigorous convective activity and hence where high clouds develop, as is also evident from Equation~\ref{eq:lfr_o_l20}. Figure~\ref{fig:pcb_lfr} shows that, using the parameterisation in terms of cloud-top height, lightning flashes are indeed concentrated around the region of deep convective activity, associated with surface heating over the substellar point. The asymmetry in the vertical extent of convective cloud cover results in a crescent-like shape with rates of up to 0.16 flashes km$^{-2}$yr$^{-1}$. This LFR is generally lower than observed values over the Earth's oceans, where LFRs reach values of up to ${\sim}$0.01~flashes~km$^{-2}$day$^{-1}$ (or ${\sim}$3.65~flashes~km$^{-2}$yr$^{-1}$; \citealt{han_cloud_2021}). Accounting for the impacts of coastal regions and islands on oceanic flash rates on Earth \citep[e.g.][]{williams_islands_2004, liu_relationships_2012}, we retrieved the original LIS/OTD datasets on observed LFRs on Earth \citep{cecil_gridded_2014}. This was used to determine LFRs over parts of the oceans that are island-free, and resulted in LFRs between 0.11--0.57~flashes~km$^{-2}$yr$^{-1}$. Hence, our findings for Proxima Centauri b fall into the lower end of this range. 

\begin{figure}
	\includegraphics[width=\columnwidth]{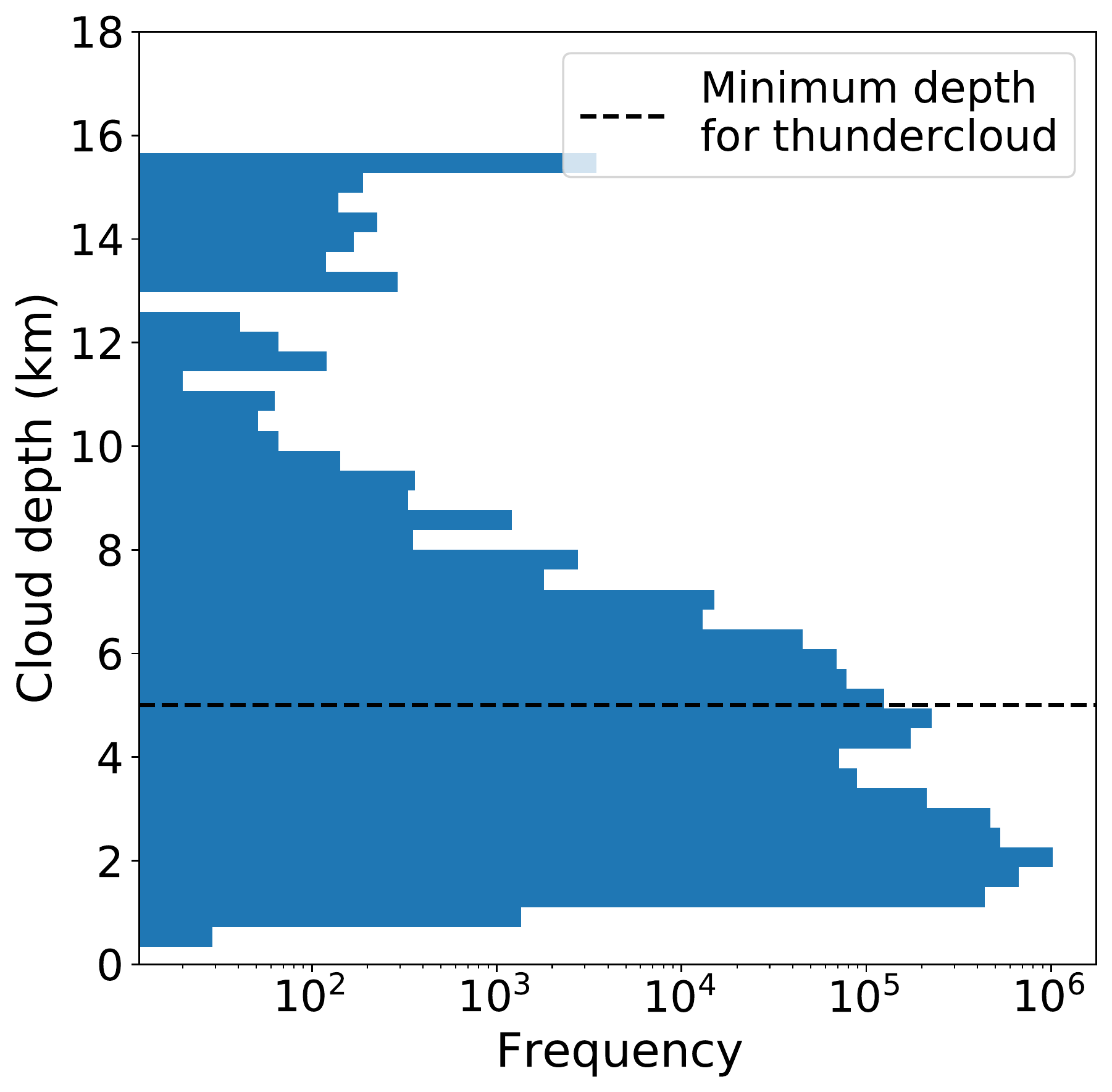}
    \caption{Convective cloud depths for Proxima Centauri b from 10 days of simulation at a high temporal resolution of 4~minutes. Depths were calculated as the difference between the convective cloud top and convective cloud base. Also indicated is the threshold for the classification of a thundercloud in the lightning scheme, as described in Section~\ref{subsec:lightning}.}
    \label{fig:pcb_clouddepth}
\end{figure}

To understand why LFRs for Proxima Centauri b are at the lower end of oceanic flash rates on Earth, we plot the convective cloud depths for Proxima Centauri b in Figure~\ref{fig:pcb_clouddepth}. The convective cloud depths are based on the analysis of output at a high temporal resolution (4 minutes) for 10 days of the simulation, giving us over $4{\times}10^{6}$ data points over the entire planetary surface. Cloud depths extend up to 15.7~km in altitude for Proxima Centauri b. For the oceanic regions in the tropics on Earth, clouds extend up to 17~km \citep[][]{dessler_tropical_2006, bacmeister_spatial_2011}. This suggests that the convection driving the formation of these clouds is weaker on Proxima Centauri b, resulting in a lower LFR. Besides that, Proxima Centauri b (T$_{\mathrm{surf}}{\sim}$291~K) is generally cooler than Earth, even without the continents (T$_{\mathrm{surf}}{\sim}$300~K). Atmospheric temperatures are generally also lower, resulting in a drier atmosphere and thereby limiting cloud formation and the initiation of lightning. As discussed below, the flash rates and the extended structures we see around the substellar point have a significant effect on the atmospheric chemistry of Proxima Centauri b, and more generally on similar tidally-locked planets. 

\subsection{Ozone chemistry}\label{subsec:o3chem}
\begin{figure*}
	\includegraphics[width=2\columnwidth]{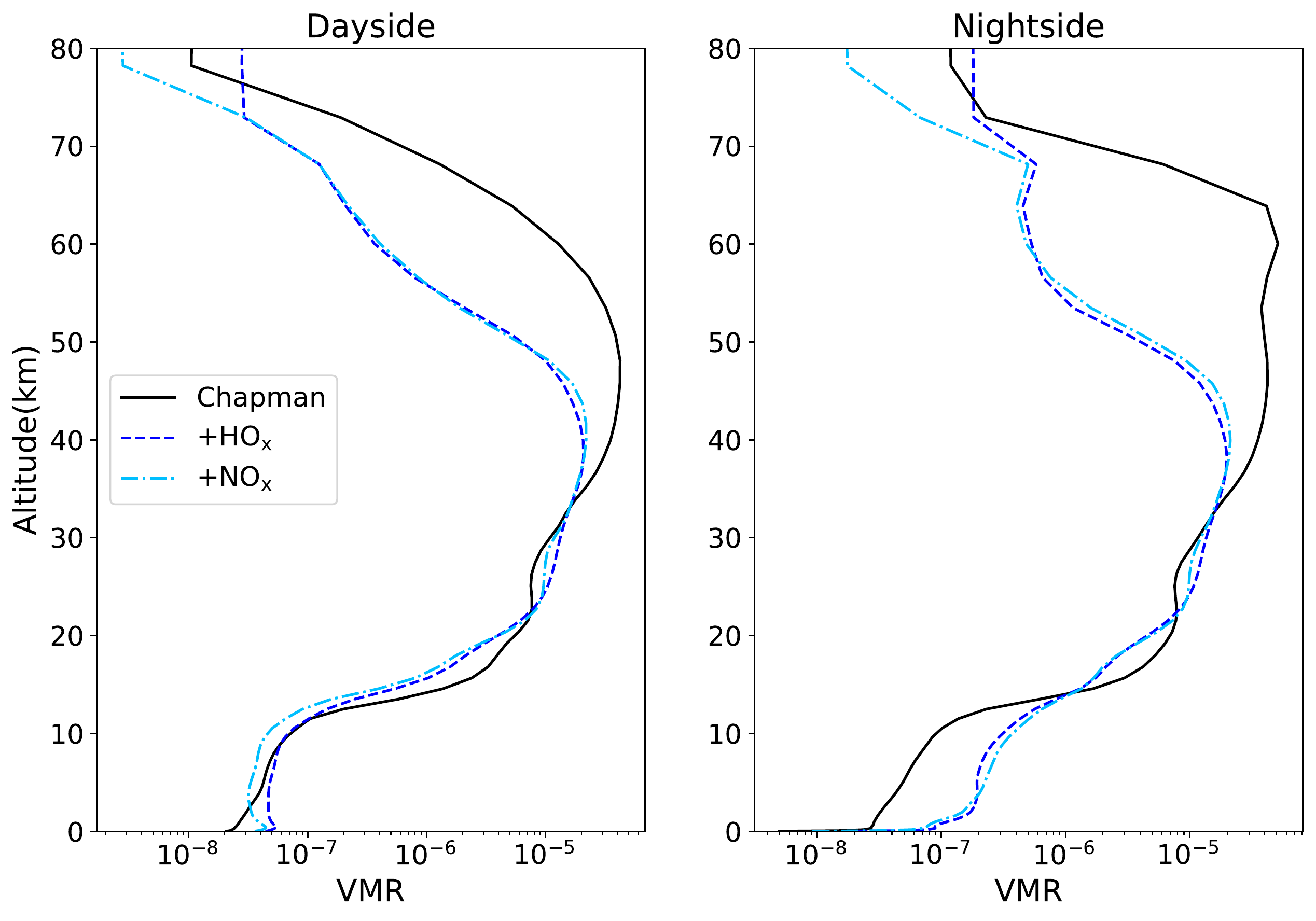}
    \caption{Hemispheric mean vertical ozone VMR profiles (mole~mole$^{-1}$) for different chemical schemes (Table~\ref{tab:chemsetups}). The left and right panels show the day- and nightside hemisphere respectively.}
    \label{fig:pcb_o3vmrz}
\end{figure*}

First, we reproduced the results of \citet{yates_ozone_2020} on atmospheric chemistry of Proxima Centauri b (Figure~\ref{fig:rflux_repy}). Subsequently, we use the refined spectral flux distribution as shown in Table~\ref{tab:fluxpcb}, and report our results as 120-day means after 20 years of spin-up from initial conditions. Figure~\ref{fig:pcb_o3vmrz} shows the hemispheric mean volume mixing ratios (VMR) of ozone as a function of altitude, corresponding to the three different chemistry schemes described in Table~\ref{tab:chemsetups}. In all three cases, incoming stellar radiation is sufficient to establish and maintain an ozone layer, initiated by the Chapman mechanism \citep{chapman_xxxv_1930}. This mechanism can be summarized by five chemical reactions:\\
\\
\ce{O$_2$ + h$\nu$ -> O($^3$P) + O($^3$P)}, \hfill (R1) \\
\ce{O($^3$P) + O$_2$ + M -> O$_3$ + M}, \hfill (R2) \\
\ce{O$_3$ + h$\nu$ -> O$_2$ + O($^3$P)}, \hfill (R3) \\
\ce{O$_3$ + h$\nu$ -> O$_2$ + O($^1$D)}, \hfill (R4) \\
\ce{O$_3$ + O($^3$P) -> O$_2$ + O$_2$}.  \hfill (R5) \\

Reaction R1 initiates the production of ozone and R5 represents the termination step for the Chapman mechanism. Reactions R2--4 describe the rapid interchange between O($^1$D), O($^3$P), O$_2$ and O$_3$. Furthermore, O($^1$D) is de-excited following the interaction with N$_2$, O$_2$ and CO$_2$.

Figure~\ref{fig:pcb_o3vmrz} shows that the Chapman mechanism results in the thickest ozone layer, peaking at 40--60~km with VMRs of 43 and 47~ppm on the dayside and nightside, respectively. The dayside ozone profile peaks at a lower VMR and a lower altitude (${\sim}$48~km) compared to the nightside peak at ${\sim}$60~km due to active photochemistry on the dayside. O($^3$P) is still transported to the nightside with the prevailing high-altitude horizontal winds. Combined, there is less photolysis of ozone through reactions R3 and R4 and the shifted balance between R2 and R5 determine the higher nightside VMR at altitudes above ${\sim}$50~km. 
Compared to \citet{yates_ozone_2020}, our refined bin distribution of the stellar spectra, described above, generally results in an increase in radiation with wavelengths ${<}$240~nm and consequently a twenty-fold increase in O$_2$ photolysis rates and increased ozone VMRs. The relatively high ozone VMR enhances transport of heat to the nightside, which increases the nightside surface temperature especially at the location of the cold traps, consistent with \citet{yang_low-order_2014} and the simulations from \citet{boutle_exploring_2017} for an Earth-like atmospheric composition. The increased nightside temperatures result in similar dayside and nightside ozone vertical profiles.

On Earth, several catalytic cycles destroy stratospheric ozone \citep[e.g.][]{grenfell_chemical_2006}. In this study, we consider two cycles that involve HO$_\mathrm{x}$ and one that involves NO$_\mathrm{x}$ driven by lightning. The two HO$_\mathrm{x}$ cycles are: \\
\\
HO$_\mathrm{x}$ cycle 1: \\
\ce{H$_2$O + O($^1$D) -> 2OH}, \hfill (R10)\\
\ce{OH + O$_3$ -> HO$_2$ + O$_2$}, \hfill (R11)\\
\ce{HO$_2$ + O$_3$ -> OH + 2O$_2$}, \hfill (R12)\\
\ce{OH + HO$_2$ -> H$_2$O + O$_2$}, \hfill (R13)\\
and\\
HO$_\mathrm{x}$ cycle 2: \\
\ce{H$_2$O + h$\nu$  -> OH + H}, \hfill (R14)\\
\ce{O($^3$P) + OH -> O$_2$ + H}, \hfill (R15)\\
\ce{H + O$_2$ + M -> HO$_2$ + M}, \hfill (R16)\\
\ce{HO$_2$ + O($^3$P) -> OH + O$_2$} \hfill (R17)\\
\ce{H + O$_3$ -> OH + O$_2$}, \hfill (R18)\\
\ce{OH + HO$_2$ -> H$_2$O + O$_2$}. \hfill (R13)\\
\\
These catalytic cycles are a consequence of water vapour in the atmosphere. Following O($^1$D) production from the photolysis of ozone in reaction R4, the oxidation of water vapour in reaction R10 initiates HO$_\mathrm{x}$ cycle 1 describing OH-HO$_2$ cycling, as previously studied by \citet{yates_ozone_2020}. The second HO$_\mathrm{x}$ cycle is initiated by the photolysis of water vapour and describes H-OH-HO$_2$ cycling.

\begin{figure*}
	\includegraphics[width=2\columnwidth]{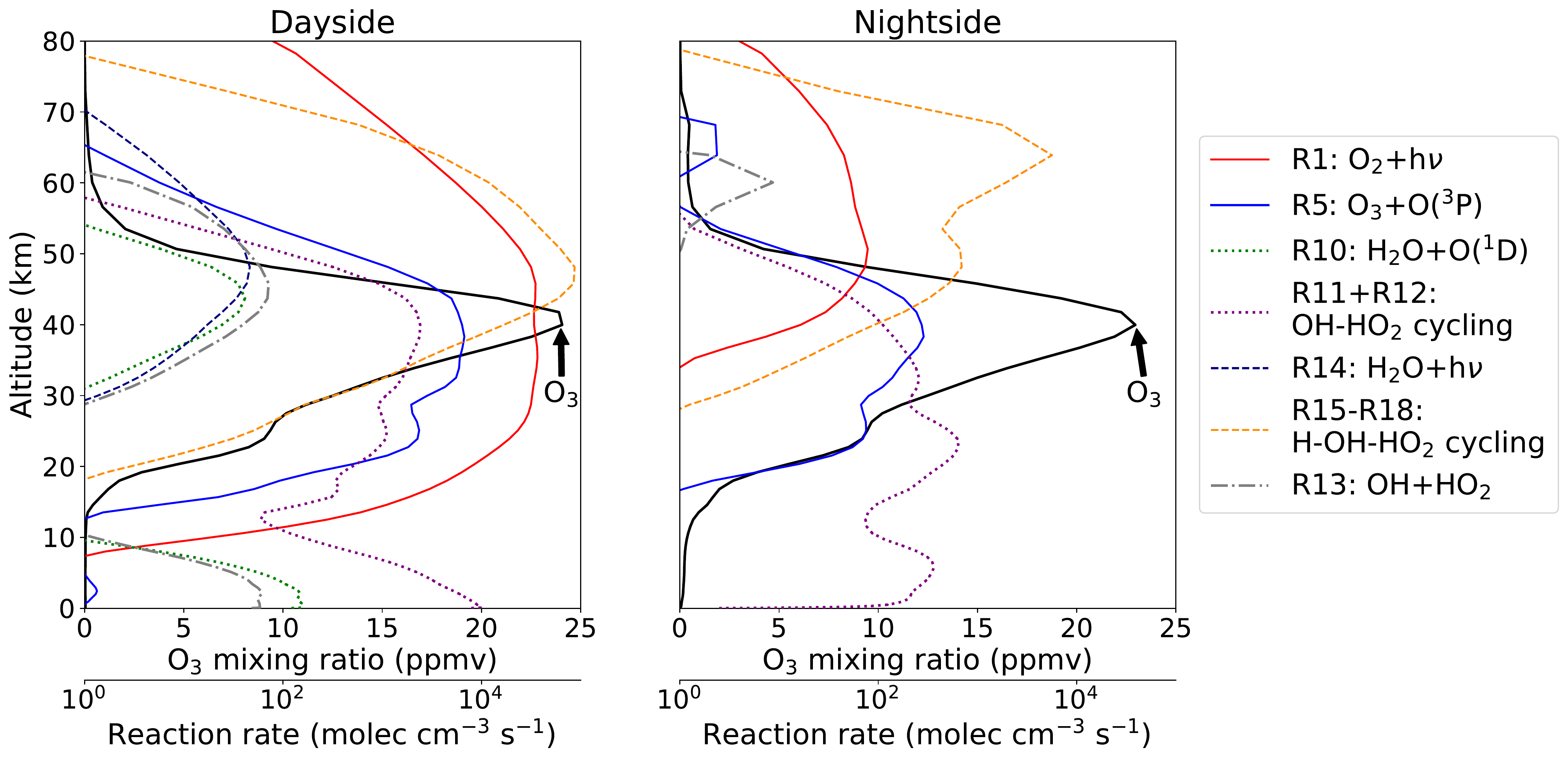}
    \caption{Dayside and nightside reaction rates (molecules~cm$^{-3}$~s$^{-1}$) and the corresponding ozone profile (VMR) in black using the improved spectral flux distribution. Results are from our third experiment, including the Chapman mechanism and HO$_\mathrm{x}$ and NO$_\mathrm{x}$ chemistry. Solid, dotted, and dashed lines denote reaction rates relevant to the Chapman mechanism, HO$_\mathrm{x}$ cycle 1, and HO$_\mathrm{x}$ cycle 2, respectively. The common termination reaction (R13) for both HO$_\mathrm{x}$ cycles is plotted as a dash-dotted line.}
    \label{fig:rflux_b_addhox}
\end{figure*}

Figure~\ref{fig:rflux_b_addhox} shows hemispheric mean reaction rates (molecules~cm$^{-3}$~s$^{-1}$) as a function of altitude on the dayside and nightside of the planet, for the Chapman+HO$_\mathrm{x}$+NO$_\mathrm{x}$ chemistry scheme. Using these rates, we can identify which reactions are most important throughout the atmosphere. The initiation (R1) and termination (R5) reactions of the Chapman mechanism are shown as solid lines. The reaction rate associated with oxygen photolysis (R1) mostly occurs on the dayside (nightside rates reflect those determined by scattered daytime radiation at the terminators) and falls off rapidly with decreasing altitude after the maximum in the ozone layer at 40~km. The peak rate of R5 between ${\sim}$19~km and ${\sim}$50~km reflects the coincident altitude of our ozone layer (Figure~\ref{fig:pcb_o3vmrz}) and a peak in O($^3$P) production from ozone and oxygen photolysis. On the dayside between ${\sim}$19 and ${\sim}$38.5~km, we find that ozone loss is dominated by reaction R5. The lower nightside rate of R5 follows from the dependence on photolysis reactions producing O($^3$P). Generally, nightside mean reaction rates are weaker than dayside mean reaction rates (Figure~\ref{fig:rflux_b_addhox}) due to the absence of stellar radiation and due to the slower progression of chemical reactions in the lower temperatures of the nightside hemisphere. 

Including the HO$_\mathrm{x}$ catalytic cycles decreases the dayside and nightside peaks in the ozone layer to 21 and 20 ppm, respectively (Figure~\ref{fig:pcb_o3vmrz}), and moves them to lower altitudes. Both HO$_\mathrm{x}$ cycles lead to the consumption of odd oxygen molecules (O($^3$P) and/or O$_3$), resulting in a significant depletion of the ozone layer above ${\sim}$38.5~km compared to the Chapman-only scheme (Figure~\ref{fig:pcb_o3vmrz}). The reaction rate of R1 increases between ${\sim}$20--30~km when adding in the HO$_\mathrm{x}$ chemistry, due to lower overhead ozone abundances and thus less UV shielding, resulting in a higher ozone VMR at these altitudes.
The dotted lines in Figure~\ref{fig:rflux_b_addhox} correspond to HO$_\mathrm{x}$ cycle 1, for which R10 is the initiation reaction. Reactions R11 and R12 denote the propagation steps responsible for the catalytic destruction of ozone, and reaction R13 denotes the termination step. The net effect of this OH-HO$_2$ cycling is the consumption of two ozone molecules. We find that this catalytic cycle dominates ozone loss in the lower atmosphere of our tidally-locked M-dwarf planet, in agreement with previous studies \citep[][]{yates_ozone_2020}. From the purple dotted line in Figure~\ref{fig:rflux_b_addhox}, we find that the propagation reactions R11 and R12 dominate below ${\sim}$19 and ${\sim}$30~km on the dayside and nightside, respectively. As described above, the termination step from the Chapman mechanism dominates ozone loss between ${\sim}$19 and ${\sim}$38.5~km on the dayside.

Above ${\sim}$38.5~km, HO$_\mathrm{x}$ cycle 2 begins to dominate, as can be seen from the dashed lines in Figure~\ref{fig:rflux_b_addhox}. In terms of increasing altitude, HO$_\mathrm{x}$ production is initially a combination of the oxidation (R10) and photolysis (R14) of water vapour. Above ${\sim}$44~km, H$_2$O photolysis dominates. In this cycle, both reactions R15 and R16 have to happen to convert OH into HO$_2$, and the cycle will then be completed by reaction R17 converting HO$_2$ back into OH. Another pathway is R15 for the conversion of OH to H followed immediately by R18 to return OH, completing the H-OH-HO$_2$ cycling. For the sake of readability, we combine these steps by plotting the sum of reactions R15, R16, R17 and R18 in Figure~\ref{fig:rflux_b_addhox}. We find that the propagation steps for H-OH-HO$_2$ cycling dominate above ${\sim}$38.5~km on the dayside. The common termination step for the two HO$_\mathrm{x}$ cycles is R13 (Figure~\ref{fig:rflux_b_addhox}). The reaction rates associated with the propagation reactions from both cycles can be up to 1000 times higher than the termination or the initiation steps (e.g. at 55~km). This means that one OH molecule can participate in these ozone-depleting catalytic cycles as many as 1000 times, before being removed by the termination step. 

On the nightside, the weak O$_2$ (see R1 in Figure~\ref{fig:rflux_b_addhox}) and O$_3$ photolysis resulting from scattered radiation at the terminator limit the formation of atomic oxygen. Consequently, advection from the dayside determines the abundance of O($^3$P) and OH, leading to significantly smaller nightside abundances. As a result, reaction rates can be decreased by a factor of 100 (e.g. the HO$_\mathrm{x}$ propagation steps, R11, R12, R15-R18) to 100,000 (R10). Below ${\sim}$20~km, only small amounts of ozone (tens of ppbs) are found, resulting from relatively weak oxygen photolysis, losses due to deposition, and the effectiveness of HO$_\mathrm{x}$ cycle 1 in the troposphere (0--20~km). 

\begin{figure*}
	\includegraphics[width=2\columnwidth]{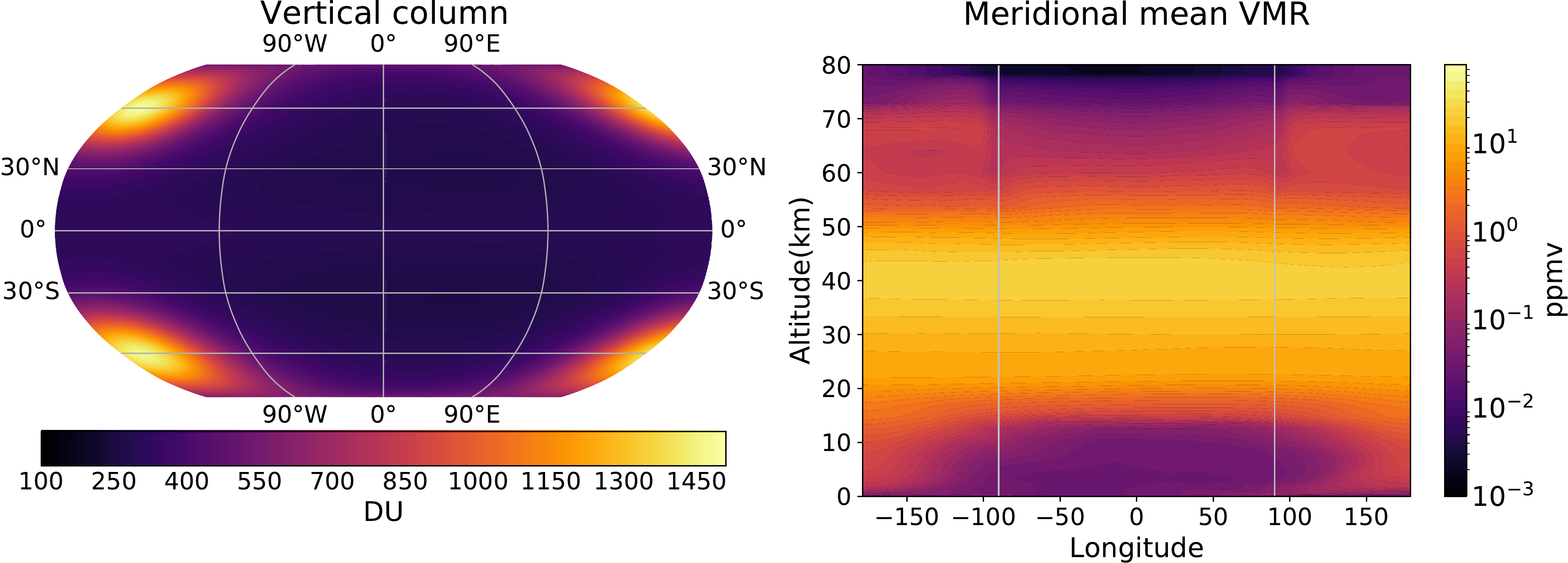}
    \caption{Spatial distribution of O$_3$ around the planet, for our third experiment in Table~\ref{tab:chemsetups}. The panel on the left shows the vertical column densities in DU. Right-hand side panels show the meridional mean VMR (mean over latitude) as a function of longitude and altitude. Relative depletion on the dayside and accumulation of O$_3$ in the nightside cold traps (centred at ${\sim}$150$^\circ$W) is clearly seen.}
    \label{fig:o3distrib_nox}
\end{figure*}

The net result of the Chapman mechanism and HO$_\mathrm{x}$ cycles is a 3-D ozone distribution similar to previous work \citep[][]{yates_ozone_2020}, shown in Figure~\ref{fig:o3distrib_nox}. However, the corrected spectral distribution of stellar radiation leads to larger ozone production across the planet. Ozone column densities are as thin as 269~Dobson Units (DU, where 1~DU${=}2.687\times10^{20}$~molecules~m$^{-2}$) on the dayside and peak at $1490$ DU over the nightside Rossby gyres located at midlatitudes and centred at ${\sim}$150$^\circ$W (Figure~\ref{fig:o3distrib_nox}). This increased thickness is caused by the Rossby gyres trapping air, which is subsequently exposed to extensive radiative cooling (so-called cold traps). Due to this cooling, the atmosphere reduces in thickness locally, transporting the ozone in the column down to the troposphere thereby increasing the column abundance \citep{yates_ozone_2020}. The mean reaction rates associated with the HO$_\mathrm{x}$ cycle are up to 100 times smaller on the nightside below ${\sim}$20~km. Combined, these two effects lead to higher nightside ozone abundances at these altitudes (Figure~\ref{fig:pcb_o3vmrz}). 

\subsection{Lightning-induced NO$_\mathrm{y}$ chemistry}\label{subsec:noxchem}
\begin{figure}
	\includegraphics[width=\columnwidth]{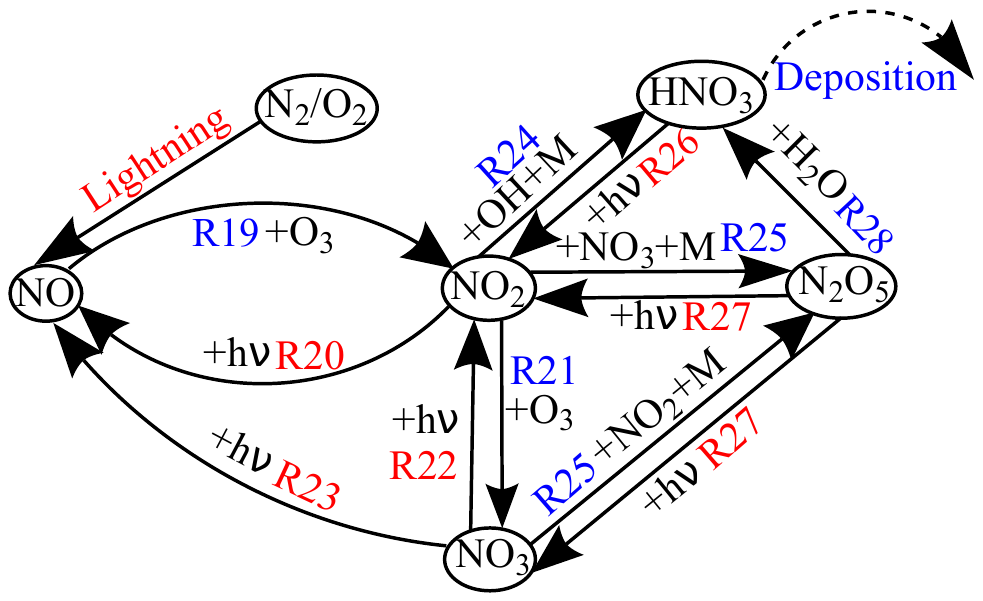}
    \caption{Overview of the main chemical reactions that are initiated by lightning-induced NO. Arrows run from the reactant to the product. The chemical reaction is numbered as R{XX}, corresponding to the main text. Reactions in red only occur on the dayside of the planet, reactions in blue can occur on both hemispheres. HONO, HO$_2$NO$_2$, and N$_2$O are omitted in this diagram but also included in the network.}
    \label{fig:nox_full_mech}
\end{figure}

Our third experiment focuses on NO$_\mathrm{y}$ chemistry in the lower atmosphere (${<}$20~km), where lightning is the only source of NO at 216~moles NO per flash, and is denoted `NO$_\mathrm{x}$' in Table~\ref{tab:chemsetups} and Figure~\ref{fig:pcb_o3vmrz}. The magnitude and distribution of these flash rates are described in Section~\ref{subsec:lightning_results} and shown in Figure~\ref{fig:pcb_lfr}. The resulting distribution of lightning-produced NO can initiate catalytic cycles that result in the loss of ozone and consist of the following reactions:
\\
\ce{NO + O$_3$ -> NO$_2$ + O$_2$} \hfill (R19)\\
\ce{NO$_2$ + h$\nu$ -> NO + O($^3$P)} \hfill (R20)\\
\ce{NO$_2$ + O$_3$ -> NO$_3$ + O$_2$} \hfill (R21)\\
\ce{NO$_3$ + h$\nu$ -> NO$_2$ + O($^3$P)} \hfill (R22)\\
\ce{NO$_3$ + h$\nu$ -> NO + O$_2$}. \hfill (R23)\\
\\
NO quickly reacts with O$_3$ to form NO$_2$ (R19). On the dayside, NO$_2$ can be photolysed back into NO again (R20). Together, they form a first catalytic cycle leading to the destruction of one ozone molecule. With a relatively small abundance of O($^3$P) below 20~km, NO$_2$ is likely to react with O$_3$ again through reaction R21 to produce NO$_3$ that is subsequently photolysed to NO$_2$ (R22) or NO (R23). The photolysis rate for reaction R22 (flux at $\lambda{>}$345~nm)  is ${\sim}$8 times higher than R23 (flux at $\lambda{>}$412.5~nm) for the spectral flux distributions of the Sun and Proxima Centauri. The net result of this second catalytic cycle is the destruction of two ozone molecules. The termination step in the NO$_\mathrm{x}$ is the oxidation of NO$_2$ by OH:\\
\\
\ce{NO$_2$ + OH + M -> HNO$_3$ + M}. \hfill (R24)\\
\\
We know from the reaction rates of R10 and R14 (Figure~\ref{fig:rflux_b_addhox}) that OH (and thus HNO$_3$) production occurs predominantly on the dayside. NO$_2$ can further react with the NO$_3$ produced through reaction R21 to form N$_2$O$_5$:\\
\\
\ce{NO$_3$ + NO$_2$ + M -> N$_2$O$_5$ + M}. \hfill (R25)\\
\\
The species HNO$_3$ and N$_2$O$_5$ are more stable and therefore have a relatively long lifetime against chemical loss. Eventually, they are converted back into NO$_\mathrm{x}$, for example, via:\\
\\
\ce{HNO$_3$ + h$\nu$ -> NO$_2$ + OH} \hfill (R26)\\
\ce{N$_2$O$_5$ + h$\nu$ -> NO$_3$ + NO$_2$}. \hfill (R27)\\
\\
In the presence of water, N$_2$O$_5$ is also converted into HNO$_3$:\\
\\
\ce{N$_2$O$_5$ + H$_2$O -> 2HNO$_3$}. \hfill (R28)\\
\\
Furthermore, HNO$_3$ is subject to removal by deposition. On the nightside, the absence of stellar radiation further enhances the lifetime of NO$_3$, HNO$_3$ and N$_2$O$_5$. Therefore, these species serve as reservoirs for NO$_\mathrm{x}$. The reservoirs and NO$_\mathrm{x}$ together form the NO$_\mathrm{y}$ family. The tidally-locked configuration of Proxima Centauri b provides distinct radiation environments on the dayside and nightside, which leads to a dayside-nightside contrast in NO$_\mathrm{x}$ and its reservoirs. A schematic summary of the mechanism that is initiated by NO produced from lightning is shown in Figure~\ref{fig:nox_full_mech}, indicating chemical reactions that only occur on the dayside hemisphere in red.

\begin{figure}
	\includegraphics[width=\columnwidth]{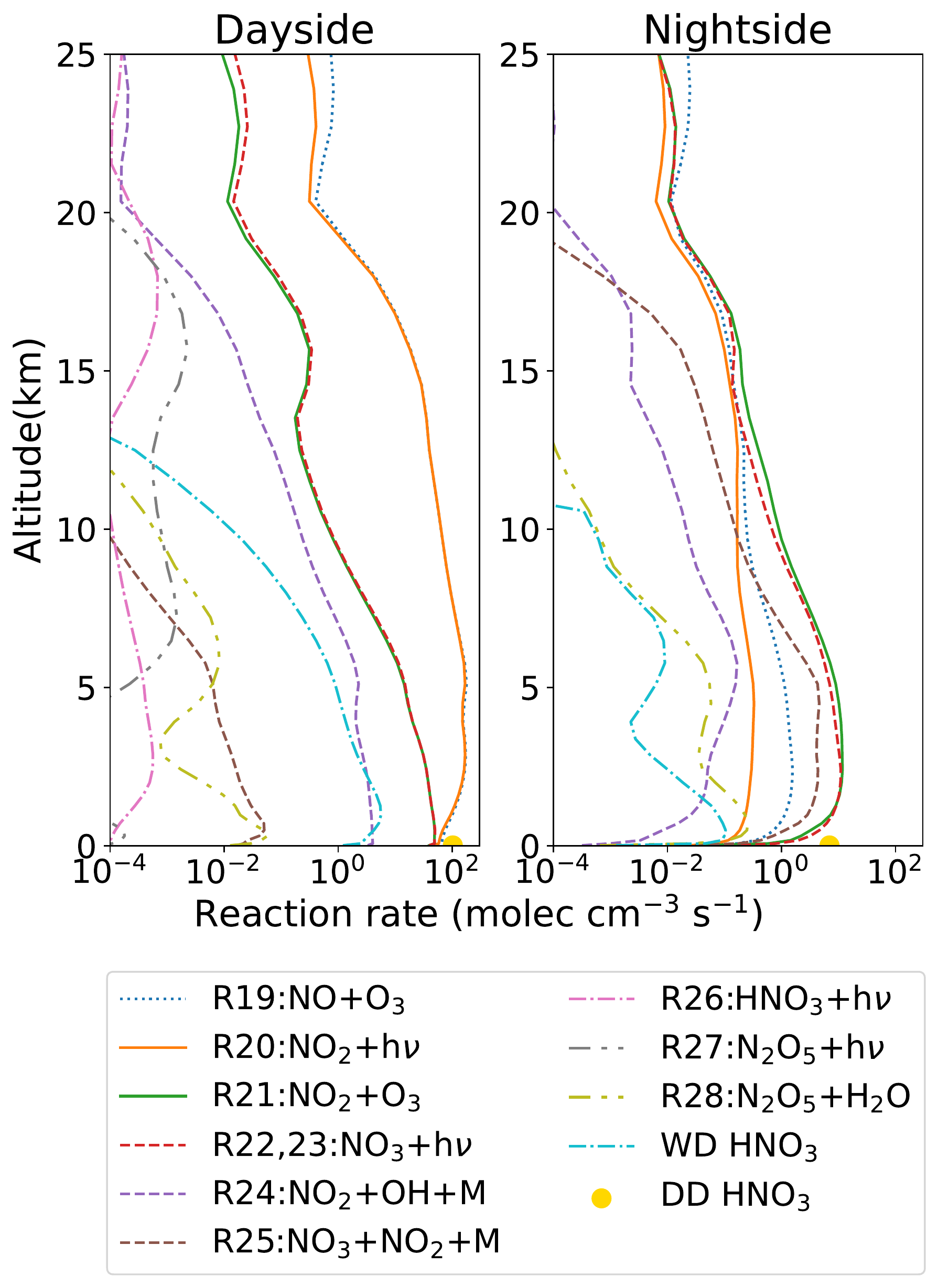}
    \caption{Hemispheric mean reaction rates (molecules~cm$^{-3}$~s$^{-1}$) for nitrogen chemistry, below altitudes of 40~km. Reaction numbers are explained in Section~\ref{subsec:noxchem} and Figure~\ref{fig:nox_full_mech}. The wet (WD) and dry deposition (DD) of HNO$_3$ are also shown. Wet deposition occurs up to altitudes of ${\sim}$12~km and dry deposition is limited to the surface.}
    \label{fig:pcb_noyflux}
\end{figure}

The dayside-nightside contrast in atmospheric chemistry involving nitrogen species is explored using the hemispheric-mean reaction rates in Figure~\ref{fig:pcb_noyflux}. Following the production of NO by lightning, NO$_2$ is predominantly formed on the dayside of the planet, with the reaction rate that describes the conversion from NO to NO$_2$ (R19) being about 200 times higher on the dayside than on the nightside. The photolysis of NO$_2$ (R20) also peaks on the dayside and is smaller by a factor of ${\sim}$2000 on the nightside. Hence, NO$_2$ is available for nightside NO$_\mathrm{y}$ chemistry, initiated by R21. We also find a strong decrease ($10^3$--$10^4$) in nightside rates that describe the photolysis of HNO$_3$ (R26) and N$_2$O$_5$ (R27). Notably, the reaction rate associated with the photolysis of NO$_3$ in R22 and R23 only triples on the dayside. This is because NO$_3$ mainly absorbs radiation at wavelengths between ${\sim}$550 and ${\sim}$700~nm, which is more likely to be scattered in the planet's atmosphere than shorter wavelength radiation. The reaction rate of R21 is twice as large on the dayside, and oxidation (by O($^3$P) and OH) further destroys dayside NO$_3$. The production of N$_2$O$_5$ through R25 is ${\sim}$700 times larger on the nightside. The reaction with water vapour through R28 can again lead to the production of HNO$_3$. HNO$_3$ is rapidly removed by dry deposition (DD) and wet deposition (WD), which is stronger on the dayside of the planet due to a higher precipitation rate. Figure~\ref{fig:pcb_noyflux} shows that wet deposition provides a loss term for HNO$_3$ up to 12~km and dry deposition is limited to the surface. These combined effects of chemistry and deposition on the NO$_\mathrm{x}$ reservoirs indicate an accumulation on the nightside.

\begin{figure*}
	\includegraphics[width=1.8\columnwidth]{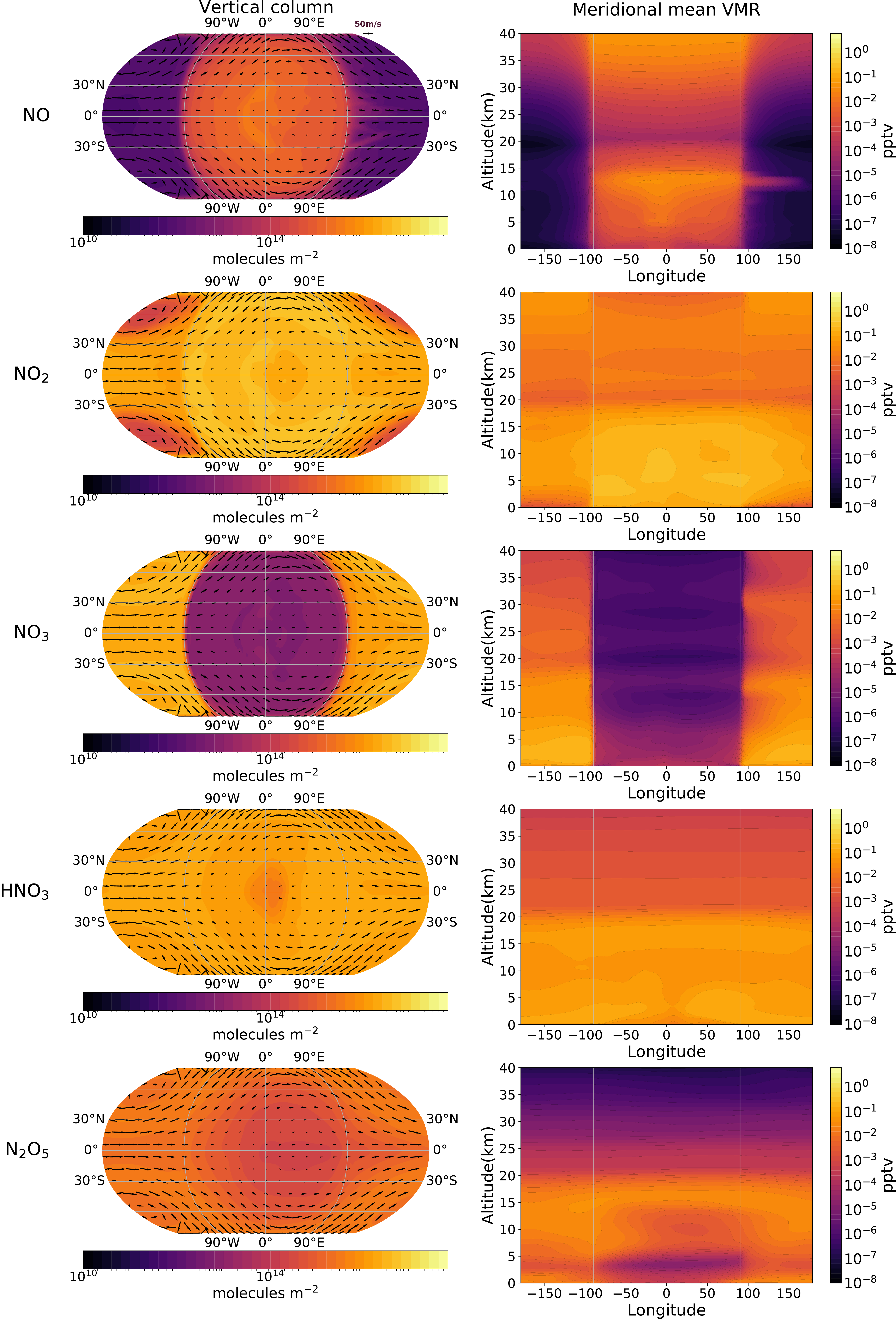}
    \caption{Spatial distributions of NO, NO$_2$, NO$_3$, HNO$_3$, and N$_2$O$_5$. Left-hand side panels show the vertical column densities calculated from a mass-weighted vertical integral. Furthermore, they show the wind vectors at altitudes of 6850~m, where the superrotating jet speed is the largest. Right-hand side panels show the meridional mean VMR (mean over latitude) as a function of longitude and altitude.}
    \label{fig:pcbnoxydistrib}
\end{figure*}

Figure~\ref{fig:pcbnoxydistrib} shows the resulting spatial distribution of these important NO$_\mathrm{x}$ and NO$_\mathrm{y}$ species. Left-hand side panels show the vertically integrated column densities (in molec~m$^{-2}$) of each species, with the wind vectors at an altitude of 6850~m, where the superrotating jet speed is the largest (up to 40~km~s$^{-1}$). The column densities are calculated as the pressure-weighted sum of the number densities in each vertical layer. Right-hand side panels show the meridional mean VMR for each of the species. The first row of Figure~\ref{fig:pcbnoxydistrib} shows that lightning-induced NO is produced on the dayside, most strongly in the crescent shape determined by the lightning flashes in convective clouds (Figure~\ref{fig:pcb_lfr}). To quantify hemispheric differences, we define a day-to-nightside contrast following \citet{chen_biosignature_2018} and \citet{koll_temperature_2016}. However, instead of using mean mixing ratios, we use hemispheric mean vertical column densities of species to calculate the relative difference between hemispheres:
\begin{equation}
    r_\mathrm{diff}=\frac{r_\mathrm{day}-r_\mathrm{night}}{ r_\mathrm{global}},
\end{equation}
where $r_\mathrm{day}$, $r_\mathrm{night}$ and $r_\mathrm{global}$ are the dayside, nightside and global mean vertical column densities (in molec~m$^{-2}$) of a chemical species, respectively.
There is enhanced NO on the dayside of the planet (Figure~\ref{fig:pcbnoxydistrib}) with an associated $r_\mathrm{diff}$ value of $198\%$. NO is most abundant between 5.5 and 15~km, with a maximum of 0.117 ppt at 12.5~km. This is a consequence of our parameterisation of NO emissions per lightning flash: emitted NO is distributed between 500~hPa and the cloud top, as described in Section~\ref{subsec:lightning}. Furthermore, the tidally-locked orbit, NO lifetime and wind structure allow for a maximum near the NO emission source. Differences in NO at the eastward terminator between 12--14~km are due to eastward advection and subsequent photolysis of NO$_2$ lower in the atmosphere (R20). The balance of chemical reactions R19-R23 in a tidally-locked configuration determines the vertical column densities of NO, NO$_2$ and NO$_3$ (first three rows of Figure~\ref{fig:pcbnoxydistrib}). The lightning-produced NO leads to NO$_2$ and subsequently NO$_3$ that spread around the dayside of the planet, following atmospheric transport. NO$_3$ is rapidly photolysed, releasing NO$_2$ back. Consequently, NO$_2$ ($r_\mathrm{diff}$=80\%) has the largest vertical column densities, peaking at $6.2{\times}10^{16}$ molecules~m$^{-2}$, westward of the substellar point. 
We know from reaction R24 that the production of HNO$_3$ occurs predominantly on the dayside. The decrease in the column density of HNO$_3$ in Figure~\ref{fig:pcbnoxydistrib}, towards the minimum value of $3.0{\times}10^{15}$ molecules~m$^{-2}$ close to the substellar point, shows that HNO$_3$ is susceptible to photolysis as well as to wet deposition following cloud formation and precipitation. 

In the absence of stellar radiation and lightning flashes, atmospheric chemistry is different on the nightside. Here, the chemistry depends exclusively on the advection of chemical species from the dayside of the planet. This can be seen from the NO$_2$ and HNO$_3$ columns aligning with the wind vectors in Figure~\ref{fig:pcbnoxydistrib}, as these species are advected from the dayside to the nightside. This dayside-to-nightside transport of chemical species was also discussed in previous work on terrestrial exoplanets \citep[][]{chen_biosignature_2018, yates_ozone_2020} and hot Jupiters \citep[e.g.][]{drummond_implications_2020}. NO column densities are much smaller on the nightside because the source is found on the dayside and because of the absence of photolysis of the other nitrogen-bearing species. Without stellar radiation and with relatively low nightside O($^3$P) abundances, the dominant loss pathways for nightside NO$_2$ result in the production of NO$_\mathrm{y}$ reservoir species. The oxidation by O$_3$ (R21) still produces a significant amount of NO$_3$. In the absence of stellar radiation, NO$_3$ and HNO$_3$ are no longer efficiently destroyed. This results in a much thicker nightside NO$_3$ column, evident from the $r_\mathrm{diff}$ of -200\% for NO$_3$, and a slightly thicker mean HNO$_3$ column on the nightside, with $r_\mathrm{diff}$=-8\%. HNO$_3$ has a smaller $r_\mathrm{diff}$ since it depends on dayside production and is subject to deposition. Nightside NO$_2$ and NO$_3$ are sufficiently abundant to initiate the production of N$_2$O$_5$ (R25). The fifth row in Figure~\ref{fig:pcbnoxydistrib} shows that, following the nightside production, N$_2$O$_5$ is advected towards the dayside. In doing so, N$_2$O$_5$ passes across the western terminator after which it is rapidly destroyed due to photolysis or the reaction with water vapour. The relative depletion of N$_2$O$_5$ in the equatorial regions between $90$ and $180^\circ$E is linked to the same reaction with water vapour (transported from the dayside). The tendency of N$_2$O$_5$ to accumulate on the nightside can also be seen from its $r_\mathrm{diff}$ of ${-}95\%$. Since the conversion from NO$_3$, HNO$_3$ or N$_2$O$_5$ back into NO$_\mathrm{x}$ involves either photolysis or the interaction with OH, the nitrogen reservoirs (NO$_3$, HNO$_3$ and N$_2$O$_5$) have an extremely long lifetime on the planet's nightside. It is this accumulation of the reservoir species that is seen by the contrasting dayside and nightside vertical columns in the last three rows of Figure~\ref{fig:pcbnoxydistrib}. It should be stated that while the day-to-night contrasts are big in relative terms, they are small in absolute terms. 

Combined, lightning-induced NO$_\mathrm{y}$ chemistry and the configuration of a tidally-locked M-dwarf planet lead to dayside-nightside distinctions. The chemical mechanism in Figure~\ref{fig:nox_full_mech} summarizes the main reactions responsible for these distinctions as well as where they occur on the planet, initiated by lightning flashes that produce NO. Lightning and chemical reactions that are limited to the dayside of the planet are shown in red and indicate why the accumulation of reservoir species on the nightside occurs. 

\section{Discussion}\label{sec:discussion}
Here we reflect on key uncertainties associated with our study, including 1) the importance of stellar radiation, particularly the UV wavelengths; 2) describing the distribution of LFRs, including the thermal decomposition of chemical species; and 3) the prospects of observing the changes in atmospheric chemistry predicted by our simulations using observations from the James Webb Space Telescope.

\subsection{The importance of photochemistry}
Photolytic reactions are the main driver of atmospheric chemistry on M-dwarf planets. The host stars irradiate a large UV contribution to the bolometric flux compared to other stars (Figure~\ref{fig:spectra_irrad}), and therefore many chemical reactions on the orbiting planets are driven by UV radiation.

We use the Fast-JX model to describe photolysis in the 177--850~nm region. The upper limit is based on photons being sufficiently energetic to break molecular bonds \citep[][]{bian_fast-j2_2002}. Fast-JX is optimized for the calculation of photolysis rates in the troposphere and stratosphere on Earth, motivating the lower cutoff at 177~nm. Radiation with lower wavelengths is attenuated above the mesosphere (except for O$_2$ Lyman $\alpha$ absorption at 121~nm). Furthermore, the solar flux falls off precipitously for wavelengths shorter than 177~nm (Figure~\ref{fig:spectra_irrad}). Our current assumption of an Earth-like atmosphere on Proxima Centauri b intrinsically assumes attenuation of incoming radiative fluxes at these lower wavelength regions. The MUSCLES spectrum that we use is based on observations, but uncertainties remain. First, the region between 175--200~nm is based on a quadratic fit to the surrounding continuum \citep[][]{loyd_muscles_2016}. Besides that, MUSCLES continuum fluxes are often based on the photon-limited noise floor of the data rather than actual measurements of the stellar output, because of the intrinsic faintness of UV radiation from M-dwarfs \citep[][]{loyd_muscles_2016}. Future measurements will help to improve the quality and range of the spectra being used, as well as help test our assumption about using an Earth-like atmosphere.

The relation between UV fluxes and photochemistry in 3-D simulations of exoplanets has been explored across a range of stellar energy distributions (SEDs) \citep[][]{chen_biosignature_2018, chen_habitability_2019} and for time-dependent stellar activity \citep[][Ridgway et al. submitted]{chen_persistence_2021}. Furthermore, 1-D photochemical models have been used to quantify the amount of ozone resulting from different UV distributions in SEDs \citep[e.g.][]{segura_biosignatures_2005, segura_effect_2010, domagal-goldman_abiotic_2014, omalley-james_uv_2017, teal_effects_2022} and alongside varying molecular oxygen abundances \citep[][]{kozakis_is_2022}. To illustrate the dependence of assuming different SEDs on ozone photochemistry we compare in Appendix~\ref{app:o3chem} results from the BT-Settl spectrum and the improved bin distribution, with the results we report here using the MUSCLES spectrum. We find that the stronger UV flux from the MUSCLES spectrum leads to ozone mixing ratios that are enhanced by a factor of 15.

As described in Section~\ref{subsec:o3chem}, the 3-D spatial distribution of ozone is broadly similar to \citet{yates_ozone_2020}. The similarities include a lower dayside mean ozone column ($302$ DU) as compared to the nightside ($483$ DU) and a buildup of ozone at the locations of the nightside Rossby gyres. This behaviour is also shown in Figure~\ref{fig:o3distrib_nox}. Therefore, in the case of ozone and agreeing with the findings from \citet{chen_biosignature_2018}, we can conclude that the magnitudes of volume mixing ratios are controlled by a balance of photochemical production and destruction. On the other hand, the lifetime of ozone is long enough for the spatial distribution to be controlled by transport processes.

In agreement with \citet{boutle_exploring_2017}, we find a substantial fraction of the planetary surface on the dayside to be potentially habitable (${\sim}$45\%), with a temperature exceeding 273.15~K. This fraction could change with the inclusion of ocean heat transport \citep[][]{hu_role_2014} or ice-albedo feedback, though the latter is shown to play only a marginal role for M-dwarf planets \citep[][]{shields_effect_2013}. Furthermore, a thick ozone layer increases the likelihood that any surface life is protected from harmful UV radiation \citep[e.g.][]{shields_habitability_2016}. To quantify this, we compare the UV flux ($\lambda{<}$320~nm) at 51~km (just above the ozone layer) with the flux at the surface. We find that only 15\% reaches the surface levels, after being attenuated by the ozone layer (and water clouds closer to the surface). As mentioned in Section~\ref{sec:plan_clim}, this amounts to about 0.2\% of Earth's surface UV radiation levels. Hence, in line with the results from \citet{omalley-james_uv_2017} for a 1-D model of an `active' M-dwarf spectrum, the thick ozone layer provides sufficient protection against harmful UV radiation. This ozone layer can also protect against enhanced UV irradiation during stellar flares, although strong flares and the absence of a magnetic field might also produce strong perturbations of the ozone layer \citep[][]{chen_persistence_2021}. The perturbing of atmospheric chemistry by the enhanced UV flux and stellar energetic particles from flares is further explored in a separate publication (Ridgway et al., submitted). Lightning may also have played a role in the origin of life, which was shown by the production of amino acids in a reducing atmosphere following electrical discharges in the Miller-Urey experiment \citep[][]{miller_production_1953}. However, the formation of organic molecules following lightning is more likely in a reducing atmosphere \citep[e.g.][]{chameides_rates_1981}, which would require starting conditions that differ from our assumed oxygen-rich composition.

The motivation for assuming Earth-like atmospheric chemistry in this study was our aim of studying ozone chemistry and the impact of lightning-induced species. The addition of atmospheric chemistry to a 3-D GCM significantly increases the computational expense, particularly if, as done here, it is considered self consistently with the radiative transfer and dynamics. Therefore, we have included only the most essential chemical reactions (based on our broader knowledge of the reactions occurring in an Earth-like atmosphere). Moving away from this photochemical regime involves a significant increase in the size of the associated chemical network that would consequently be difficult to run in a 3-D CCM. As such, for this initial study, we used Earth-like atmospheric chemistry. Including accurate and sufficiently small chemical networks that support other photochemical regimes is an important topic for future investigation.

\subsection{CCM comparisons}
The previous CCM studies of terrestrial exoplanets made use of the CAM-Chem 3-D model \citep[][]{chen_biosignature_2018} for an atmosphere with 26 levels extending up to 50~km and WACCM \citep[][]{chen_habitability_2019, chen_persistence_2021} for an atmosphere up to 145~km in 66 levels, thus including the thermosphere. Our UM-UKCA CCM falls in between this range with 60 levels describing an atmosphere up to 85~km. Other differences in the previous studies include landmasses in the form of Earth's continental distribution (though with the substellar point in the middle of the Pacific Ocean) that can impact the climate \citep[][]{lewis_influence_2018, rushby_effect_2020}. \citet{chen_biosignature_2018} use a stellar insolation of 1360~W~m$^{-2}$, compared to 881.7~W~m$^{-2}$ used here. The insolation is varied in \citet{chen_habitability_2019, chen_persistence_2021}, but not to values as low as ours. Our study uses a reduced description of atmospheric chemistry, focusing on individual processes, whereas theirs include 97 species connected by 196 reactions \citep[][]{chen_biosignature_2018} and 58 species connected by 217 reactions \citep[][]{chen_habitability_2019, chen_persistence_2021}. We turned off surface emissions and, as opposed to the earlier work, included wet deposition, potentially further impacting the atmospheric composition. Finally, the aforementioned differences in UV fluxes play a major role in the chemical abundances.

The vertical distribution of ozone in Figure~\ref{fig:pcb_o3vmrz} can be directly compared to the previous studies. As mentioned before, the observation of a thinner O$_3$ layer for a weaker stellar UV irradiation is in agreement with \citet{chen_habitability_2019}. Furthermore, the influence of the HO$_\mathrm{x}$ catalytic cycle 1 as a sink for O$_3$ on tidally-locked, Earth-like exoplanets was also found by \citet{yates_ozone_2020} and \citet{chen_biosignature_2018}, though not extensively elaborated on in the latter study. The finding of the O$_3$ sink due to H$_2$O photolysis and the resulting HO$_\mathrm{x}$ catalytic cycle 2 in the upper stratosphere, as described in Section~\ref{subsec:o3chem}, further expands our knowledge of the impact of HO$_\mathrm{x}$ catalytic cycles on tidally-locked, Earth-like exoplanets. \citet{chen_biosignature_2018} find an O$_3$ layer that peaks at ${\sim}$10~ppm above altitudes of 30~km (see their figure 3), which agrees roughly with our findings for the HO$_\mathrm{x}$ and NO$_\mathrm{x}$ cases in Figure~\ref{fig:pcb_o3vmrz}. One notable difference is the decrease that can be seen upwards of 45~km in our simulation, whereas their O$_3$ abundance seems to stay constant. This difference can be caused by stronger O$_3$ photolysis in these atmospheric layers and above, in our case. The inactive stellar spectra in \citet{chen_habitability_2019} result in more reduced ozone at ${\sim}$0.1~ppm (see their figure 5). A potential cause of this is the increase of the model-top to 145~km \citep[][]{chen_habitability_2019}. The ozone abundance for a planet orbiting a pre-flare M star in \citet{chen_persistence_2021} peaks at a few ppm again. 

As discussed in Section~\ref{sec:unmodel} we performed this study, focusing on lightning and the time-averaged irradiation, alongside a complementary study focusing on stellar flares (Ridgway et al., submitted). These two studies used the same underlying model, the UM, but differing photolysis and chemistry schemes which provided an excellent mechanism to increase confidence in the robustness of our model setups. 

Inter-model differences can have a wide range of causes. Firstly, the exact impact of photochemistry is influenced by the choice of stellar SEDs, the extent of wavelength ranges and the specific distribution of fluxes over wavelength bins. Next to that, the complexity of the chemical network (as well as reaction rate constants), initial chemical abundances, treatment of dry and/or wet deposition and the potential inclusion of surface emissions can alter the resulting atmospheric composition. Finally, the vertical extent of the atmosphere and the possible inclusion of landmasses affect dynamics and thus the chemistry directly and indirectly. The inter-model differences and variety of potential causes motivate the need for further model intercomparisons for CCMs, as mentioned by \citep[][]{cooke_variability_2022} and following those done for 3-D GCM simulations of TRAPPIST-1e assuming static atmospheric compositions \citep[the THAI project,][]{turbet_trappist-1_2022, sergeev_trappist-1_2022, fauchez_trappist-1_2022}.

\subsection{Parameterising lightning}
Atmospheric lightning is a complex physical phenomenon that is observed on Earth and other Solar System planets
\citep[e.g.][]{aplin_atmospheric_2006, hodosan_lightning_2016}, but is difficult to model explicitly, partly due to the uncertainties associated with the responsible processes, e.g. initiation of lightning. To keep the process tractable within a GCM, lightning parameterisations have been developed in terms of different convection parameters \citep[e.g.][]{allen_evaluation_2002, finney_using_2014, etten-bohm_evaluating_2021, stolz_evaluating_2021}, but the parameterisation in terms of cloud-top height remains the most used in CCMs for Earth \citep[][]{luhar_assessing_2021}. 

The occurrence of lightning on Earth is generally higher than on the giant planets or Venus \citep[][]{hodosan_lightning_2016}. Nevertheless, the processes of charging of cloud particles, charge separation due to gravitational settling and the buildup of electrostatic potential differences that lead to discharges are expected to remain the same, and the electric field breakdown does not strongly depend on the chemical composition of the atmosphere \citep[][]{helling_dust_2013, helling_exoplanet_2019}. Simulating an Earth-like atmosphere on a tidally-locked exoplanet further supports the assumption of a similar process for the emergence of lightning discharges. The lightning parameterisation we use in terms of cloud-top height (Equation~\ref{eq:lfr_o_l20}) is evaluated successfully for Earth and based on the fundamental laws of electricity \citep[][]{vonnegut_facts_1963, williams_large-scale_1985}, as described in Section~\ref{subsec:lightning}. Therefore, it is reasonable that this parameterisation will also deliver reasonable LFRs for an Earth-like exoplanet, although we acknowledge that the coefficients used in Equation~\ref{eq:lfr_o_l20} have been tuned to match observed LFRs on Earth \citep[][]{price_simple_1992, luhar_assessing_2021} and therefore could represent a potential error. Nevertheless, the distribution of LFRs on our tidally-locked exoplanet (Figure~\ref{fig:pcb_lfr}) makes physical sense and is consistent with the predictions of a thick convective cloud deck covering the dayside of the planet (see e.g. \citealt{yang_stabilizing_2013, boutle_exploring_2017, sergeev_atmospheric_2020} and Figure~\ref{fig:pcb_conv}). 

The effect of decreasing LFR with lower atmospheric temperature was described in Section~\ref{subsec:lightning_results}. To predict the effects of climate change on the initiation of lightning, several studies have investigated the effects of a higher atmospheric temperature, but no consensus has been reached based on the different parameterisations \citep[][]{clark_parameterization-based_2017}. On the one hand, a higher temperature leads to fewer ice clouds and therefore fewer mixed-phase collisions, limiting the number of lightning flashes and NO$_\mathrm{x}$ produced by them \citep[e.g.][]{finney_projected_2018}. On the other hand, the depth of convection and precipitation rate increase in a warming climate, potentially increasing the number of lightning flashes \citep[e.g.][]{banerjee_lightning_2014}. The response of lightning initiation to changes in the atmospheric pressure is complex due to the competing effects of clouds, pressure broadening, heat capacity, lapse rate, and relative humidity, as shown by \citet{zhang_how_2020} for a tidally-locked planet. To build upon these results, further simulations of the initiation of lightning are currently being conducted, in a range of atmospheric conditions and using an improved description of the electrification process, following the high-resolution simulations of tidally-locked exoplanets by \citet{sergeev_atmospheric_2020}.

Another factor of uncertainty is the thermal decomposition of chemical species following a lightning discharge. The species affected by thermal decomposition will depend on the ambient atmospheric composition \citep[][]{harman_abiotic_2018, helling_exoplanet_2019}. This is further illustrated by studies of lightning-induced chemistry on Venus and Mars \citep[][]{nna_mvondo_production_2001} and Early Earth \citep[][]{chameides_rates_1981, navarro-gonzalez_possible_2001, ardaseva_lightning_2017}. Since we simulate a background composition dominated by N$_2$ and O$_2$, thermal decomposition of these species and subsequent NO production, following the Zel'dovich mechanism \citep[][]{zeldovich_oxidation_1947}, is a reasonable expectation. The exact amount of NO produced per flash is uncertain on Earth \citep[e.g.][]{schumann_global_2007, miyazaki_global_2014, bucsela_midlatitude_2019, allen_observations_2021, allen_observations_2021-1}, and is related to unknowns in the distinction between CG and IC flashes and dependencies on e.g. peak current, channel length, rate of energy dissipation and the number of strokes per flash \citep[][]{murray_lightning_2016, luhar_assessing_2021}. We performed a sensitivity calculation in which we increased the NO production to 830 moles NO per flash, which did not result in significant changes to the spatial distribution of NO$_\mathrm{y}$ as shown in Figure~\ref{fig:pcbnoxydistrib}. \citet{harman_abiotic_2018} show that NO production rates from lightning on temperate terrestrial planets with N$_2$-O$_2$-CO$_2$ atmospheres will stay within an order of magnitude of the rate on Earth. Other sources of NO include stellar flares and coronal mass ejections through enhanced UV flux and energetic particles \citep[][Ridgway et al. submitted]{segura_effect_2010, chen_persistence_2021}, volcanic eruptions \citep[][]{von_glasow_effects_2009}, and cosmic rays \citep[][]{scalo_m_2007}, and their combined effects alongside lightning have to be studied in future work. 

Comparing our results to the 1-D models of \citet{ardaseva_lightning_2017} and \citet{harman_abiotic_2018}, we find a much smaller impact of lightning-induced NO$_\mathrm{x}$ on atmospheric chemistry, explained by three major differences. First, flash rates up to $5\times10^{4}$ flashes $\mathrm{km}^{-2}\mathrm{hr}^{-1}$ are used by \citet{ardaseva_lightning_2017} and both 1-D models assume global thunderstorms emitting NO throughout the atmosphere, resulting in an increased magnitude and spatial extent of flash rates. Second, we find NO from lightning confined to the lower 20~km of the atmosphere, whereas NO has a strong impact up to ${\sim}$80~km for \citet{ardaseva_lightning_2017} and \citet{harman_abiotic_2018}. Lastly, the effect of lightning-induced NO on the distribution of O$_2$ and ozone strongly depends on catalytic cycles following the photolysis of CO$_2$ \citep[][]{harman_abiotic_2018}, and future work will be directed to analysing these effects in a 3-D CCM. This will be part of a wider investigation of exoplanet atmospheric chemistry and physics such as lightning in the context of different atmospheric composition and mass, following GCM simulations such as \citet{turbet_habitability_2016, paradise_climate_2021}. Through the variation of atmospheric mass, we can take into account the potential effects of stellar wind-driven escape \citep[][]{airapetian_impact_2020}. 

\subsection{Observational prospects}
Lightning discharges are known to emit both optical \citep[e.g.][]{borucki_spectra_1985} and radio signals \citep[e.g.][]{zarka_study_2004}. Our limited knowledge of lightning across the Solar System indicates the difficulties of detecting this phenomenon on other planets. Therefore, direct detection of exoplanetary lightning will be difficult with current observing technology. However, as described in Section~\ref{subsec:noxchem}, lightning produces NO that will subsequently result in a cascade of perturbed chemical reactions. Among the lightning-induced chemical species, NO, NO$_2$, N$_2$O and HNO$_3$ are known to have absorption features (e.g. \citealt{tabataba-vakili_atmospheric_2016, kopparapu_nitrogen_2021}) in the spectral range that will be observed by JWST. The presence of high disequilibrium abundances of N$_2$O has also been proposed as a potential biosignature since N$_2$O in Earth's atmosphere is mainly produced by biotic sources \citep[e.g.][]{sagan_search_1993, schwieterman_exoplanet_2018}. The possibility of lightning on exoplanets broadens the observational context of trace gases, and ignoring this may result in a false positive in the search for biosignatures.

We use the NASA Planetary Spectrum Generator (PSG) \citep[][]{villanueva_planetary_2018} to test whether our combined UM-UKCA model can potentially result in atmospheric phenomena that would be observable via transmission spectroscopy. The PSG is a radiative transfer model suite for forward modelling and retrievals of planetary spectra. For the purpose of our study, we generate transmission spectra for our UM-UKCA description of Proxima Centauri b (assuming it transits), using pressure, temperature, altitude and abundances of N$_2$, H$_2$O, CO$_2$, O$_3$, NO, NO$_2$, N$_2$O, and HNO$_3$. Recognising that transmission spectra will observe the atmosphere in the terminator regions, we extract model values for 1) all latitudes at longitudes between 75--105$^\circ$E and 75--105$^\circ$W and 2) all longitudes at latitudes between 85--90$^\circ$N and 85--90$^\circ$S. After that, we calculate the zonal and meridional mean to find a mean vertical profile for the terminator regions. 

\begin{figure}
	\includegraphics[width=\columnwidth]{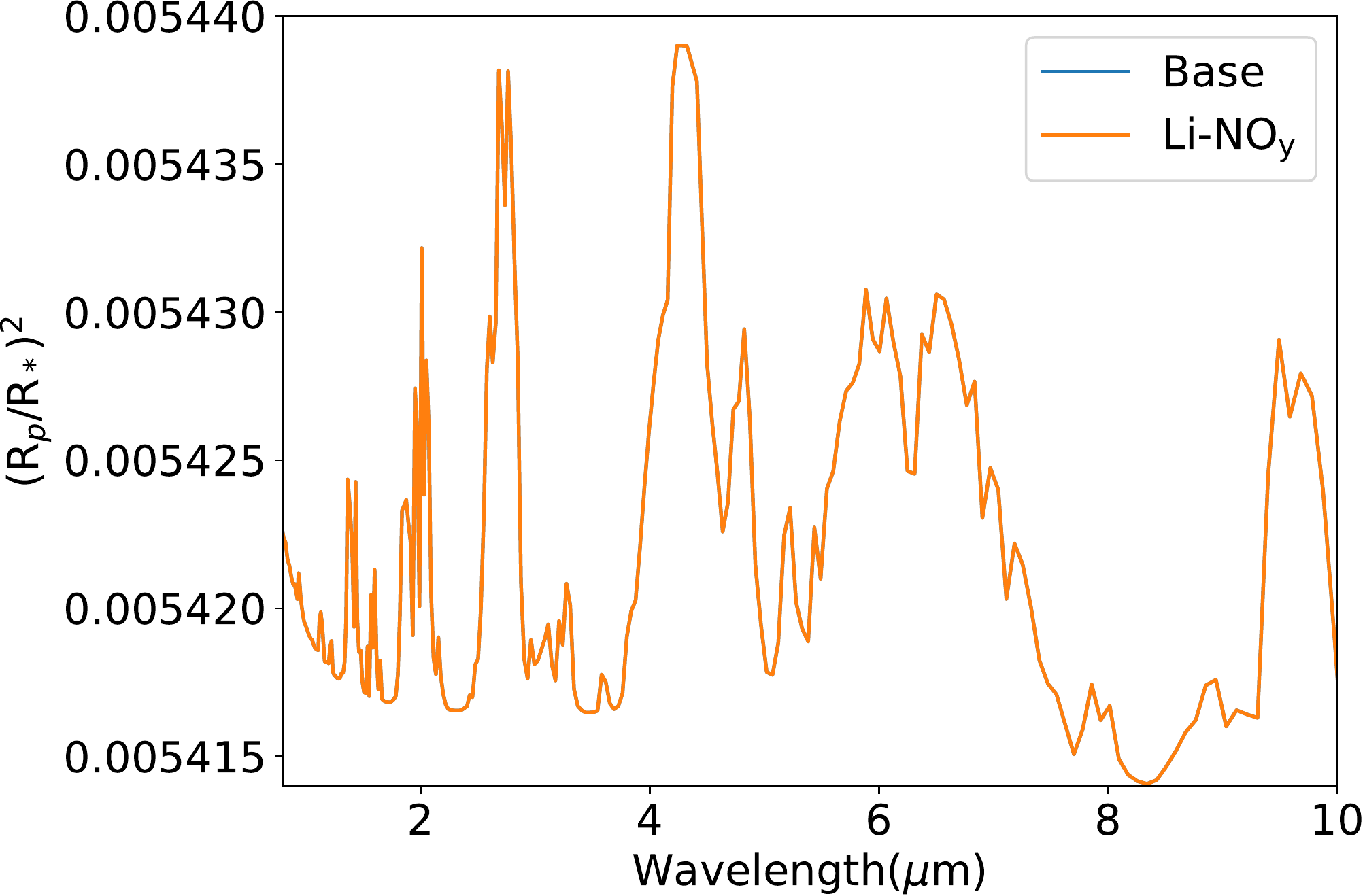}
    \caption{Transmission spectra for our setup of Proxima Centauri b, assuming it transits. Spectra are shown with and without lightning-induced NO$_\mathrm{y}$ present in the atmosphere.}
    \label{fig:linoxspec}
\end{figure}

As shown in Figure~\ref{fig:linoxspec}, we find that lightning-induced NO$_\mathrm{y}$ species are not sufficiently abundant to leave detectable signals in the transmission spectrum, which agrees with previous 1-D model results \citep[][]{ardaseva_lightning_2017}. Although Proxima Centauri b does not transit, the spectra are shown to be illustrative for other M-dwarf orbiting planets that do transit. We acknowledge that extending this study to exoplanets that support larger-scale and more vigorous convective updrafts and/or different chemical composition may result in large chemical perturbations that could be observable.

\section{Conclusions}\label{sec:conclusion}
We used a 3-D CCM (Met Office Unified Model coupled to the UK Chemistry and Aerosol framework) to study the emergence of lightning, driven by vigorous convection over and around the substellar point, and the subsequent impact on atmospheric chemistry on an Earth-like, tidally-locked exoplanet. Building on previous work \citep[][]{yates_ozone_2020} we have: 1) updated our treatment of photochemistry, by including the MUSCLES spectrum for Proxima Centauri in the Fast-JX photolysis code that provides more UV flux than our previous study \citep[][]{yates_ozone_2020} with implications for ozone chemistry, and by correcting the distribution of these fluxes over different wavelength bins; and 2) expanded the atmospheric chemistry to improve the description of ozone chemistry, including a more comprehensive description of the HO$_\mathrm{x}$ catalytic cycle and NO$_\mathrm{y}$ chemistry that is initiated by lightning as the source of NO.

We find that the incoming stellar flux supports an ozone layer across the planet, with a spatial distribution determined by atmospheric dynamics. The distribution is dominated by the substellar point and two local ozone hotspots on the nightside driven by two cyclonic Rossby gyres. The ozone column ranges from values found on Earth (${\sim}$300~DU) on the dayside to values that are five times higher over the nightside cold traps. For our Earth-like exoplanet, we parameterise lightning flashes as a function of cloud-top height, using parameterisations widely applied to modelling lightning on Earth. The flashes result in the production of NO from the thermal decomposition of N$_2$ and O$_2$ in the lightning channel. We find that:
\begin{itemize}
    \item Lightning flash rates peak at 0.16 flashes km$^{-2}$yr$^{-1}$, concentrated in a crescent-like shape westward of the substellar point, coinciding with regions of vigorous dayside convection.
    \item Lightning flashes result in a dayside atmosphere that is rich in NO$_\mathrm{x}({=}\mathrm{NO}+\mathrm{NO}_2$) due to the rapid interconversion between the species in the presence of stellar radiation. The lightning-induced NO$_\mathrm{x}$ is found below altitudes of 20~km and thus is vertically separated from the peak ozone layer at 40~km. The ozone profile is determined mainly by the Chapman mechanism and the HO$_\mathrm{x}$ catalytic cycle resulting from the oxidation and photolysis of water.
    \item Following strong eastward winds that result from thermal gradients between the day- and nightside of the planet, NO$_\mathrm{x}$ is advected to the nightside. Here, the absence of stellar radiation allows NO$_\mathrm{x}$ to be stored in reservoir species such as NO$_3$, HNO$_3$, and N$_2$O$_5$.
\end{itemize}
We find that part of the planetary surface is potentially habitable in terms of the surface temperature and absorption of harmful UV radiation by the ozone layer, in agreement with \citet{boutle_exploring_2017} and \citet{yates_ozone_2020}. However, the amount of ozone and thus surface habitability of a planet strongly depend on the amount of UV flux, illustrating the need for a systematic characterisation of host star spectra (such as the MUSCLES spectral survey). Given the large range of causes for inter-model variability, exoplanetary CCMs would greatly benefit from further model intercomparisons. Furthermore, the emergence and chemical effects of lightning can be further explored in convection resolving models \citep[][]{sergeev_atmospheric_2020} and in different atmospheric compositions \citep[e.g.][]{chameides_rates_1981, ardaseva_lightning_2017}, to further assess observability and potential relevance in the search for biosignatures. The development of accurate chemical networks describing non-Earth-like regimes is essential in understanding exoplanet atmospheric chemistry.

\section*{Acknowledgements}
We kindly thank Michael Prather for the provision of the Fast-JX code and Oliver Wild and Martyn Chipperfield for useful discussions and input on its operation. We kindly thank Christiane Helling for useful comments on the manuscript.

MB, PIP and LD are part of the CHAMELEON MC ITN EJD which received funding from the European Union’s Horizon 2020 research and innovation programme under the Marie Sklodowska-Curie grant agreement no. 860470. PIP acknowledges funding from the STFC consolidator grant \#ST/V000594/1. LD acknowledges support from the KU Leuven IDN grant IDN/19/028 and from the FWO research grant G086217N. RR is funded through a University of Exeter, College of Engineering, Mathematics and Physical Sciences PhD scholarship.  This work was partly supported by a Science and Technology Facilities Council Consolidated Grant [ST/R000395/1], the Leverhulme Trust through a research project grant [RPG-2020-82] and a UKRI Future Leaders Fellowship [grant number MR/T040866/1]. 

We gratefully acknowledge the use of the MONSooN2 system, a collaborative facility supplied under the Joint Weather and Climate Research Programme, a strategic partnership between the Met Office and the Natural Environment Research Council. Our research was performed as part of the project space ‘Using UKCA to investigate atmospheric composition on extra-solar planets (ExoChem)'. For the purpose of open access, the authors have applied a Creative Commons Attribution (CC BY) licence to any Author Accepted Manuscript version arising from this submission.
\section*{Data Availability}
All the CCM data was generated using the Met Office Unified Model and UK Chemistry and Aerosol model (\href{https://www.ukca.ac.uk/}{https://www.ukca.ac.uk/}), which are available for use under licence; see \href{http://www.metoffice.gov.uk/research/modelling-systems/unified-model}{http://www.metoffice.gov.uk/research/modelling-systems/unified-model}. The offline version of Fast-JX is publicly available at \href{https://www.ess.uci.edu/researchgrp/prather/scholar\_software/fast-jx}{https://www.ess.uci.edu/researchgrp/prather/scholar\_software/fast-jx}. The data underlying this article will be shared on reasonable request to the corresponding author.

We used the iris \citep[][]{met_office_iris_2022} and aeolus \citep[][]{sergeev_aeolus_2022} python packages for the post-processing of model output. Scripts to process and visualize the data are available on github: \href{https://github.com/marrickb/exo\_lightning\_code}{https://github.com/marrickb/exo\_lightning\_code}.


\bibliographystyle{mnras}
\bibliography{linox} 

\begin{thebibliography}{}
\makeatletter
\relax
\def\mn@urlcharsother{\let\do\@makeother \do\$\do\&\do\#\do\^\do\_\do\%\do\~}
\def\mn@doi{\begingroup\mn@urlcharsother \@ifnextchar [ {\mn@doi@}
  {\mn@doi@[]}}
\def\mn@doi@[#1]#2{\def\@tempa{#1}\ifx\@tempa\@empty \href
  {http://dx.doi.org/#2} {doi:#2}\else \href {http://dx.doi.org/#2} {#1}\fi
  \endgroup}
\def\mn@eprint#1#2{\mn@eprint@#1:#2::\@nil}
\def\mn@eprint@arXiv#1{\href {http://arxiv.org/abs/#1} {{\tt arXiv:#1}}}
\def\mn@eprint@dblp#1{\href {http://dblp.uni-trier.de/rec/bibtex/#1.xml}
  {dblp:#1}}
\def\mn@eprint@#1:#2:#3:#4\@nil{\def\@tempa {#1}\def\@tempb {#2}\def\@tempc
  {#3}\ifx \@tempc \@empty \let \@tempc \@tempb \let \@tempb \@tempa \fi \ifx
  \@tempb \@empty \def\@tempb {arXiv}\fi \@ifundefined
  {mn@eprint@\@tempb}{\@tempb:\@tempc}{\expandafter \expandafter \csname
  mn@eprint@\@tempb\endcsname \expandafter{\@tempc}}}

\bibitem[\protect\citeauthoryear{Abe, Abe-Ouchi, Sleep  \& Zahnle}{Abe
  et~al.}{2011}]{abe_habitable_2011}
Abe Y.,  Abe-Ouchi A.,  Sleep N.~H.,   Zahnle K.~J.,  2011, \mn@doi
  [Astrobiology] {10.1089/ast.2010.0545}, 11, 443

\bibitem[\protect\citeauthoryear{Airapetian, Glocer, Khazanov, Loyd, France,
  Sojka, Danchi  \& Liemohn}{Airapetian et~al.}{2017}]{airapetian_how_2017}
Airapetian V.~S.,  Glocer A.,  Khazanov G.~V.,  Loyd R. O.~P.,  France K.,
  Sojka J.,  Danchi W.~C.,   Liemohn M.~W.,  2017, \mn@doi [The Astrophysical
  Journal] {10.3847/2041-8213/836/1/L3}, 836, L3

\bibitem[\protect\citeauthoryear{Airapetian et~al.,}{Airapetian
  et~al.}{2020}]{airapetian_impact_2020}
Airapetian V.~S.,  et~al., 2020, \mn@doi [International Journal of
  Astrobiology] {10.1017/S1473550419000132}, 19, 136

\bibitem[\protect\citeauthoryear{Allen \& Pickering}{Allen \&
  Pickering}{2002}]{allen_evaluation_2002}
Allen D.~J.,  Pickering K.~E.,  2002, \mn@doi [Journal of Geophysical Research:
  Atmospheres] {10.1029/2002JD002066}, 107, ACH 15

\bibitem[\protect\citeauthoryear{Allen et~al.,}{Allen
  et~al.}{2021a}]{allen_observations_2021}
Allen D.~J.,  et~al., 2021a, \mn@doi [Journal of Geophysical Research:
  Atmospheres] {10.1029/2020JD033769}, 126, e2020JD033769

\bibitem[\protect\citeauthoryear{Allen, Pickering, Bucsela, Van~Geffen,
  Lapierre, Koshak  \& Eskes}{Allen et~al.}{2021b}]{allen_observations_2021-1}
Allen D.,  Pickering K.~E.,  Bucsela E.,  Van~Geffen J.,  Lapierre J.,  Koshak
  W.,   Eskes H.,  2021b, \mn@doi [Journal of Geophysical Research:
  Atmospheres] {10.1029/2020JD034174}, 126, e2020JD034174

\bibitem[\protect\citeauthoryear{Anglada-Escudé et~al.,}{Anglada-Escudé
  et~al.}{2016}]{anglada-escude_terrestrial_2016}
Anglada-Escudé G.,  et~al., 2016, \mn@doi [Nature] {10.1038/nature19106}, 536,
  437

\bibitem[\protect\citeauthoryear{Aplin}{Aplin}{2006}]{aplin_atmospheric_2006}
Aplin K.~L.,  2006, \mn@doi [Surveys in Geophysics]
  {10.1007/s10712-005-0642-9}, 27, 63

\bibitem[\protect\citeauthoryear{Archibald et~al.,}{Archibald
  et~al.}{2020}]{archibald_description_2020}
Archibald A.~T.,  et~al., 2020, \mn@doi [Geoscientific Model Development]
  {https://doi.org/10.5194/gmd-13-1223-2020}, 13, 1223

\bibitem[\protect\citeauthoryear{Ardaseva, Rimmer, Waldmann, Rocchetto,
  Yurchenko, Helling  \& Tennyson}{Ardaseva
  et~al.}{2017}]{ardaseva_lightning_2017}
Ardaseva A.,  Rimmer P.~B.,  Waldmann I.,  Rocchetto M.,  Yurchenko S.~N.,
  Helling C.,   Tennyson J.,  2017, \mn@doi [Monthly Notices of the Royal
  Astronomical Society] {10.1093/mnras/stx1012}, 470, 187

\bibitem[\protect\citeauthoryear{Bacmeister \& Stephens}{Bacmeister \&
  Stephens}{2011}]{bacmeister_spatial_2011}
Bacmeister J.~T.,  Stephens G.~L.,  2011, \mn@doi [Journal of Geophysical
  Research: Atmospheres] {10.1029/2010JD014444}, 116

\bibitem[\protect\citeauthoryear{Banerjee, Archibald, Maycock, Telford,
  Abraham, Yang, Braesicke  \& Pyle}{Banerjee
  et~al.}{2014}]{banerjee_lightning_2014}
Banerjee A.,  Archibald A.~T.,  Maycock A.~C.,  Telford P.,  Abraham N.~L.,
  Yang X.,  Braesicke P.,   Pyle J.~A.,  2014, \mn@doi [Atmospheric Chemistry
  and Physics] {10.5194/acp-14-9871-2014}, 14, 9871

\bibitem[\protect\citeauthoryear{Barnes}{Barnes}{2017}]{barnes_tidal_2017}
Barnes R.,  2017, \mn@doi [Celestial Mechanics and Dynamical Astronomy]
  {10.1007/s10569-017-9783-7}, 129, 509

\bibitem[\protect\citeauthoryear{Bian \& Prather}{Bian \&
  Prather}{2002}]{bian_fast-j2_2002}
Bian H.,  Prather M.~J.,  2002, \mn@doi [Journal of Atmospheric Chemistry]
  {10.1023/A:1014980619462}, 41, 281

\bibitem[\protect\citeauthoryear{Boccippio}{Boccippio}{2002}]{boccippio_lightning_2002}
Boccippio D.~J.,  2002, \mn@doi [Journal of the Atmospheric Sciences]
  {10.1175/1520-0469(2002)059<1086:LSRR>2.0.CO;2}, 59, 1086

\bibitem[\protect\citeauthoryear{Borucki, Mc~Kenzie, McKay, Duong  \&
  Boac}{Borucki et~al.}{1985}]{borucki_spectra_1985}
Borucki W.~J.,  Mc~Kenzie R.~L.,  McKay C.~P.,  Duong N.~D.,   Boac D.~S.,
  1985, \mn@doi [Icarus] {10.1016/0019-1035(85)90087-9}, 64, 221

\bibitem[\protect\citeauthoryear{Boutle, Mayne, Drummond, Manners, Goyal,
  Hugo~Lambert, Acreman  \& Earnshaw}{Boutle
  et~al.}{2017}]{boutle_exploring_2017}
Boutle I.~A.,  Mayne N.~J.,  Drummond B.,  Manners J.,  Goyal J.,  Hugo~Lambert
  F.,  Acreman D.~M.,   Earnshaw P.~D.,  2017, \mn@doi [Astronomy \&
  Astrophysics] {10.1051/0004-6361/201630020}, 601, A120

\bibitem[\protect\citeauthoryear{Brown, Beare, Edwards, Lock, Keogh, Milton  \&
  Walters}{Brown et~al.}{2008}]{brown_upgrades_2008}
Brown A.~R.,  Beare R.~J.,  Edwards J.~M.,  Lock A.~P.,  Keogh S.~J.,  Milton
  S.~F.,   Walters D.~N.,  2008, \mn@doi [Boundary-Layer Meteorology]
  {10.1007/s10546-008-9275-0}, 128, 117

\bibitem[\protect\citeauthoryear{Brune et~al.,}{Brune
  et~al.}{2021}]{brune_extreme_2021}
Brune W.~H.,  et~al., 2021, \mn@doi [Science] {10.1126/science.abg0492}, 372,
  711

\bibitem[\protect\citeauthoryear{Bucsela, Pickering, Allen, Holzworth  \&
  Krotkov}{Bucsela et~al.}{2019}]{bucsela_midlatitude_2019}
Bucsela E.~J.,  Pickering K.~E.,  Allen D.~J.,  Holzworth R.~H.,   Krotkov
  N.~A.,  2019, \mn@doi [Journal of Geophysical Research: Atmospheres]
  {10.1029/2019JD030561}, 124, 13475

\bibitem[\protect\citeauthoryear{Burrows, Richter, Dehn, Deters, Himmelmann,
  Voigt  \& Orphal}{Burrows et~al.}{1999}]{burrows_atmospheric_1999}
Burrows J.~P.,  Richter A.,  Dehn A.,  Deters B.,  Himmelmann S.,  Voigt S.,
  Orphal J.,  1999, \mn@doi [Journal of Quantitative Spectroscopy and Radiative
  Transfer] {10.1016/S0022-4073(98)00037-5}, 61, 509

\bibitem[\protect\citeauthoryear{Carone, Keppens  \& Decin}{Carone
  et~al.}{2014}]{carone_connecting_2014}
Carone L.,  Keppens R.,   Decin L.,  2014, \mn@doi [Monthly Notices of the
  Royal Astronomical Society] {10.1093/mnras/stu1793}, 445, 930

\bibitem[\protect\citeauthoryear{Carone, Keppens  \& Decin}{Carone
  et~al.}{2015}]{carone_connecting_2015}
Carone L.,  Keppens R.,   Decin L.,  2015, \mn@doi [Monthly Notices of the
  Royal Astronomical Society] {10.1093/mnras/stv1752}, 453, 2412

\bibitem[\protect\citeauthoryear{Carone, Keppens, Decin  \& Henning}{Carone
  et~al.}{2018}]{carone_stratosphere_2018}
Carone L.,  Keppens R.,  Decin L.,   Henning T.,  2018, \mn@doi [Monthly
  Notices of the Royal Astronomical Society] {10.1093/mnras/stx2732}, 473, 4672

\bibitem[\protect\citeauthoryear{Cecil, Buechler  \& Blakeslee}{Cecil
  et~al.}{2014}]{cecil_gridded_2014}
Cecil D.~J.,  Buechler D.~E.,   Blakeslee R.~J.,  2014, \mn@doi [Atmospheric
  Research] {10.1016/j.atmosres.2012.06.028}, 135-136, 404

\bibitem[\protect\citeauthoryear{Chameides \& Walker}{Chameides \&
  Walker}{1981}]{chameides_rates_1981}
Chameides W.~L.,  Walker J. C.~G.,  1981, \mn@doi [Origins of life]
  {10.1007/BF00931483}, 11, 291

\bibitem[\protect\citeauthoryear{Chapman}{Chapman}{1930}]{chapman_xxxv_1930}
Chapman S.,  1930, The London, Edinburgh, and Dublin Philosophical Magazine and
  Journal of Science, 10, 369

\bibitem[\protect\citeauthoryear{Chen, Wolf, Kopparapu, Domagal-Goldman  \&
  Horton}{Chen et~al.}{2018}]{chen_biosignature_2018}
Chen H.,  Wolf E.~T.,  Kopparapu R.,  Domagal-Goldman S.,   Horton D.~E.,
  2018, \mn@doi [The Astrophysical Journal] {10.3847/2041-8213/aaedb2}, 868, L6

\bibitem[\protect\citeauthoryear{Chen, Wolf, Zhan  \& Horton}{Chen
  et~al.}{2019}]{chen_habitability_2019}
Chen H.,  Wolf E.~T.,  Zhan Z.,   Horton D.~E.,  2019, \mn@doi [The
  Astrophysical Journal] {10.3847/1538-4357/ab4f7e}, 886, 16

\bibitem[\protect\citeauthoryear{Chen, Zhan, Youngblood, Wolf, Feinstein  \&
  Horton}{Chen et~al.}{2021}]{chen_persistence_2021}
Chen H.,  Zhan Z.,  Youngblood A.,  Wolf E.~T.,  Feinstein A.~D.,   Horton
  D.~E.,  2021, \mn@doi [Nature Astronomy] {10.1038/s41550-020-01264-1}, 5, 298

\bibitem[\protect\citeauthoryear{Chipperfield et~al.,}{Chipperfield
  et~al.}{2010}]{chipperfield_chapter_2010}
Chipperfield M.,  et~al., 2010, Chapter 6 - Stratospheric Chemistry in SPARC
  Report No. 5 on the Evaluation of Chemistry-Climate Models

\bibitem[\protect\citeauthoryear{Clark, Ward  \& Mahowald}{Clark
  et~al.}{2017}]{clark_parameterization-based_2017}
Clark S.~K.,  Ward D.~S.,   Mahowald N.~M.,  2017, \mn@doi [Geophysical
  Research Letters] {10.1002/2017GL073017}, 44, 2893

\bibitem[\protect\citeauthoryear{Cohen, Bollasina, Palmer, Sergeev, Boutle,
  Mayne  \& Manners}{Cohen et~al.}{2022}]{cohen_longitudinally_2022}
Cohen M.,  Bollasina M.~A.,  Palmer P.~I.,  Sergeev D.~E.,  Boutle I.~A.,
  Mayne N.~J.,   Manners J.,  2022, \mn@doi [The Astrophysical Journal]
  {10.3847/1538-4357/ac625d}, 930, 152

\bibitem[\protect\citeauthoryear{Cooke, Marsh, Walsh, Rugheimer  \&
  Villanueva}{Cooke et~al.}{2022a}]{cooke_variability_2022}
Cooke G.,  Marsh D.,  Walsh C.,  Rugheimer S.,   Villanueva G.,  2022a, \mn@doi
  [Monthly Notices of the Royal Astronomical Society]
  {10.48550/arXiv.2209.07566}

\bibitem[\protect\citeauthoryear{Cooke, Marsh, Walsh, Black  \& Lamarque}{Cooke
  et~al.}{2022b}]{cooke_revised_2022}
Cooke G.~J.,  Marsh D.~R.,  Walsh C.,  Black B.,   Lamarque J.-F.,  2022b,
  \mn@doi [Royal Society Open Science] {10.1098/rsos.211165}, 9, 211165

\bibitem[\protect\citeauthoryear{Crutzen}{Crutzen}{1970}]{crutzen_influence_1970}
Crutzen P.~J.,  1970, \mn@doi [Quarterly Journal of the Royal Meteorological
  Society] {https://doi.org/10.1002/qj.49709640815}, 96, 320

\bibitem[\protect\citeauthoryear{Del~Genio, Way, Amundsen, Aleinov, Kelley,
  Kiang  \& Clune}{Del~Genio et~al.}{2019}]{del_genio_habitable_2019}
Del~Genio A.~D.,  Way M.~J.,  Amundsen D.~S.,  Aleinov I.,  Kelley M.,  Kiang
  N.~Y.,   Clune T.~L.,  2019, \mn@doi [Astrobiology] {10.1089/ast.2017.1760},
  19, 99

\bibitem[\protect\citeauthoryear{Dessler, Palm  \& Spinhirne}{Dessler
  et~al.}{2006}]{dessler_tropical_2006}
Dessler A.~E.,  Palm S.~P.,   Spinhirne J.~D.,  2006, \mn@doi [Journal of
  Geophysical Research: Atmospheres] {10.1029/2005JD006705}, 111

\bibitem[\protect\citeauthoryear{Diamond-Lowe, Berta-Thompson, Charbonneau  \&
  Kempton}{Diamond-Lowe et~al.}{2018}]{diamond-lowe_ground-based_2018}
Diamond-Lowe H.,  Berta-Thompson Z.,  Charbonneau D.,   Kempton E. M.-R.,
  2018, \mn@doi [The Astronomical Journal] {10.3847/1538-3881/aac6dd}, 156, 42

\bibitem[\protect\citeauthoryear{Domagal-Goldman, Segura, Claire, Robinson  \&
  Meadows}{Domagal-Goldman et~al.}{2014}]{domagal-goldman_abiotic_2014}
Domagal-Goldman S.~D.,  Segura A.,  Claire M.~W.,  Robinson T.~D.,   Meadows
  V.~S.,  2014, \mn@doi [The Astrophysical Journal]
  {10.1088/0004-637X/792/2/90}, 792, 90

\bibitem[\protect\citeauthoryear{Dressing \& Charbonneau}{Dressing \&
  Charbonneau}{2015}]{dressing_occurrence_2015}
Dressing C.~D.,  Charbonneau D.,  2015, \mn@doi [The Astrophysical Journal]
  {10.1088/0004-637X/807/1/45}, 807, 45

\bibitem[\protect\citeauthoryear{Drummond, Tremblin, Baraffe, Amundsen, Mayne,
  Venot  \& Goyal}{Drummond et~al.}{2016}]{drummond_effects_2016}
Drummond B.,  Tremblin P.,  Baraffe I.,  Amundsen D.~S.,  Mayne N.~J.,  Venot
  O.,   Goyal J.,  2016, \mn@doi [Astronomy \& Astrophysics]
  {10.1051/0004-6361/201628799}, 594, A69

\bibitem[\protect\citeauthoryear{Drummond, Mayne, Baraffe, Tremblin, Manners,
  Amundsen, Goyal  \& Acreman}{Drummond et~al.}{2018}]{drummond_effect_2018}
Drummond B.,  Mayne N.~J.,  Baraffe I.,  Tremblin P.,  Manners J.,  Amundsen
  D.~S.,  Goyal J.,   Acreman D.,  2018, \mn@doi [Astronomy \& Astrophysics]
  {10.1051/0004-6361/201732010}, 612, A105

\bibitem[\protect\citeauthoryear{Drummond et~al.,}{Drummond
  et~al.}{2020}]{drummond_implications_2020}
Drummond B.,  et~al., 2020, \mn@doi [Astronomy \& Astrophysics]
  {10.1051/0004-6361/201937153}, 636, A68

\bibitem[\protect\citeauthoryear{Eager et~al.,}{Eager
  et~al.}{2020}]{eager_implications_2020}
Eager J.~K.,  et~al., 2020, \mn@doi [Astronomy \& Astrophysics]
  {10.1051/0004-6361/202038089}, 639, A99

\bibitem[\protect\citeauthoryear{Edson, Lee, Bannon, Kasting  \& Pollard}{Edson
  et~al.}{2011}]{edson_atmospheric_2011}
Edson A.,  Lee S.,  Bannon P.,  Kasting J.~F.,   Pollard D.,  2011, \mn@doi
  [Icarus] {10.1016/j.icarus.2010.11.023}, 212, 1

\bibitem[\protect\citeauthoryear{Edwards \& Slingo}{Edwards \&
  Slingo}{1996}]{edwards_studies_1996}
Edwards J.~M.,  Slingo A.,  1996, \mn@doi [Quarterly Journal of the Royal
  Meteorological Society] {https://doi.org/10.1002/qj.49712253107}, 122, 689

\bibitem[\protect\citeauthoryear{Etten-Bohm, Yang, Schumacher  \&
  Jun}{Etten-Bohm et~al.}{2021}]{etten-bohm_evaluating_2021}
Etten-Bohm M.,  Yang J.,  Schumacher C.,   Jun M.,  2021, \mn@doi [Journal of
  Geophysical Research: Atmospheres] {10.1029/2020JD033990}, 126, e2020JD033990

\bibitem[\protect\citeauthoryear{Fauchez et~al.,}{Fauchez
  et~al.}{2022}]{fauchez_trappist-1_2022}
Fauchez T.~J.,  et~al., 2022, \mn@doi [The Planetary Science Journal]
  {10.3847/PSJ/ac6cf1}, 3, 213

\bibitem[\protect\citeauthoryear{Finney, Doherty, Wild, Huntrieser, Pumphrey
  \& Blyth}{Finney et~al.}{2014}]{finney_using_2014}
Finney D.~L.,  Doherty R.~M.,  Wild O.,  Huntrieser H.,  Pumphrey H.~C.,
  Blyth A.~M.,  2014, \mn@doi [Atmospheric Chemistry and Physics]
  {10.5194/acp-14-12665-2014}, 14, 12665

\bibitem[\protect\citeauthoryear{Finney, Doherty, Wild, Stevenson, MacKenzie
  \& Blyth}{Finney et~al.}{2018}]{finney_projected_2018}
Finney D.~L.,  Doherty R.~M.,  Wild O.,  Stevenson D.~S.,  MacKenzie I.~A.,
  Blyth A.~M.,  2018, \mn@doi [Nature Climate Change]
  {10.1038/s41558-018-0072-6}, 8, 210

\bibitem[\protect\citeauthoryear{France et~al.,}{France
  et~al.}{2016}]{france_muscles_2016}
France K.,  et~al., 2016, \mn@doi [The Astrophysical Journal]
  {10.3847/0004-637X/820/2/89}, 820, 89

\bibitem[\protect\citeauthoryear{Garcia-Sage, Glocer, Drake, Gronoff  \&
  Cohen}{Garcia-Sage et~al.}{2017}]{garcia-sage_magnetic_2017}
Garcia-Sage K.,  Glocer A.,  Drake J.~J.,  Gronoff G.,   Cohen O.,  2017,
  \mn@doi [The Astrophysical Journal] {10.3847/2041-8213/aa7eca}, 844, L13

\bibitem[\protect\citeauthoryear{Garraffo, Drake  \& Cohen}{Garraffo
  et~al.}{2016}]{garraffo_space_2016}
Garraffo C.,  Drake J.~J.,   Cohen O.,  2016, \mn@doi [The Astrophysical
  Journal] {10.3847/2041-8205/833/1/L4}, 833, L4

\bibitem[\protect\citeauthoryear{Giannakopoulos, Chipperfield, Law  \&
  Pyle}{Giannakopoulos et~al.}{1999}]{giannakopoulos_validation_1999}
Giannakopoulos C.,  Chipperfield M.~P.,  Law K.~S.,   Pyle J.~A.,  1999,
  \mn@doi [Journal of Geophysical Research: Atmospheres]
  {10.1029/1999JD900392}, 104, 23761

\bibitem[\protect\citeauthoryear{Gregory \& Rowntree}{Gregory \&
  Rowntree}{1990}]{gregory_mass_1990}
Gregory D.,  Rowntree P.~R.,  1990, \mn@doi [Monthly Weather Review]
  {10.1175/1520-0493(1990)118<1483:AMFCSW>2.0.CO;2}, 118, 1483

\bibitem[\protect\citeauthoryear{Grenfell, Lehmann, Mieth, Langematz  \&
  Steil}{Grenfell et~al.}{2006}]{grenfell_chemical_2006}
Grenfell J.~L.,  Lehmann R.,  Mieth P.,  Langematz U.,   Steil B.,  2006,
  \mn@doi [Journal of Geophysical Research: Atmospheres]
  {10.1029/2004JD005713}, 111

\bibitem[\protect\citeauthoryear{Hammond \& Lewis}{Hammond \&
  Lewis}{2021}]{hammond_rotational_2021}
Hammond M.,  Lewis N.~T.,  2021, \mn@doi [Proceedings of the National Academy
  of Sciences] {10.1073/pnas.2022705118}, 118, e2022705118

\bibitem[\protect\citeauthoryear{Hammond, Tsai  \& Pierrehumbert}{Hammond
  et~al.}{2020}]{hammond_equatorial_2020}
Hammond M.,  Tsai S.-M.,   Pierrehumbert R.~T.,  2020, \mn@doi [The
  Astrophysical Journal] {10.3847/1538-4357/abb08b}, 901, 78

\bibitem[\protect\citeauthoryear{Han, Luo, Wu, Zhang  \& Dong}{Han
  et~al.}{2021}]{han_cloud_2021}
Han Y.,  Luo H.,  Wu Y.,  Zhang Y.,   Dong W.,  2021, \mn@doi [Communications
  Earth \& Environment] {10.1038/s43247-021-00233-4}, 2, 1

\bibitem[\protect\citeauthoryear{Harman, Schwieterman, Schottelkotte  \&
  Kasting}{Harman et~al.}{2015}]{harman_abiotic_2015}
Harman C.~E.,  Schwieterman E.~W.,  Schottelkotte J.~C.,   Kasting J.~F.,
  2015, \mn@doi [The Astrophysical Journal] {10.1088/0004-637X/812/2/137}, 812,
  137

\bibitem[\protect\citeauthoryear{Harman, Felton, Hu, Domagal-Goldman, Segura,
  Tian  \& Kasting}{Harman et~al.}{2018}]{harman_abiotic_2018}
Harman C.~E.,  Felton R.,  Hu R.,  Domagal-Goldman S.~D.,  Segura A.,  Tian F.,
    Kasting J.~F.,  2018, \mn@doi [The Astrophysical Journal]
  {10.3847/1538-4357/aadd9b}, 866, 56

\bibitem[\protect\citeauthoryear{Helling}{Helling}{2019}]{helling_exoplanet_2019}
Helling C.,  2019, \mn@doi [Annual Review of Earth and Planetary Sciences]
  {10.1146/annurev-earth-053018-060401}, 47, 583

\bibitem[\protect\citeauthoryear{Helling, Jardine, Diver  \& Witte}{Helling
  et~al.}{2013}]{helling_dust_2013}
Helling C.,  Jardine M.,  Diver D.,   Witte S.,  2013, \mn@doi [Planetary and
  Space Science] {10.1016/j.pss.2012.07.003}, 77, 152

\bibitem[\protect\citeauthoryear{Hodosán, Helling, Asensio-Torres, Vorgul  \&
  Rimmer}{Hodosán et~al.}{2016}]{hodosan_lightning_2016}
Hodosán G.,  Helling C.,  Asensio-Torres R.,  Vorgul I.,   Rimmer P.~B.,
  2016, \mn@doi [Monthly Notices of the Royal Astronomical Society]
  {10.1093/mnras/stw1571}, 461, 3927

\bibitem[\protect\citeauthoryear{Hsu, Ford, Ragozzine  \& Ashby}{Hsu
  et~al.}{2019}]{hsu_occurrence_2019}
Hsu D.~C.,  Ford E.~B.,  Ragozzine D.,   Ashby K.,  2019, \mn@doi [The
  Astronomical Journal] {10.3847/1538-3881/ab31ab}, 158, 109

\bibitem[\protect\citeauthoryear{Hu \& Yang}{Hu \& Yang}{2014}]{hu_role_2014}
Hu Y.,  Yang J.,  2014, \mn@doi [Proceedings of the National Academy of
  Sciences] {10.1073/pnas.1315215111}, 111, 629

\bibitem[\protect\citeauthoryear{Joshi}{Joshi}{2003}]{joshi_climate_2003}
Joshi M.,  2003, \mn@doi [Astrobiology] {10.1089/153110703769016488}, 3, 415

\bibitem[\protect\citeauthoryear{Joshi \& Haberle}{Joshi \&
  Haberle}{2012}]{joshi_suppression_2012}
Joshi M.~M.,  Haberle R.~M.,  2012, \mn@doi [Astrobiology]
  {10.1089/ast.2011.0668}, 12, 3

\bibitem[\protect\citeauthoryear{Joshi, Haberle  \& Reynolds}{Joshi
  et~al.}{1997}]{joshi_simulations_1997}
Joshi M.~M.,  Haberle R.~M.,   Reynolds R.~T.,  1997, \mn@doi [Icarus]
  {10.1006/icar.1997.5793}, 129, 450

\bibitem[\protect\citeauthoryear{Kasting \& Catling}{Kasting \&
  Catling}{2003}]{kasting_evolution_2003}
Kasting J.~F.,  Catling D.,  2003, \mn@doi [Annual Review of Astronomy and
  Astrophysics] {10.1146/annurev.astro.41.071601.170049}, 41, 429

\bibitem[\protect\citeauthoryear{Kasting, Whitmire  \& Reynolds}{Kasting
  et~al.}{1993}]{kasting_habitable_1993}
Kasting J.~F.,  Whitmire D.~P.,   Reynolds R.~T.,  1993, \mn@doi [Icarus]
  {10.1006/icar.1993.1010}, 101, 108

\bibitem[\protect\citeauthoryear{Koll \& Abbot}{Koll \&
  Abbot}{2016}]{koll_temperature_2016}
Koll D. D.~B.,  Abbot D.~S.,  2016, \mn@doi [The Astrophysical Journal]
  {10.3847/0004-637X/825/2/99}, 825, 99

\bibitem[\protect\citeauthoryear{Kopparapu et~al.,}{Kopparapu
  et~al.}{2013}]{kopparapu_habitable_2013}
Kopparapu R.~K.,  et~al., 2013, \mn@doi [The Astrophysical Journal]
  {10.1088/0004-637X/765/2/131}, 765, 131

\bibitem[\protect\citeauthoryear{Kopparapu, Wolf, Haqq-Misra, Yang, Kasting,
  Meadows, Terrien  \& Mahadevan}{Kopparapu
  et~al.}{2016}]{kopparapu_inner_2016}
Kopparapu R.~k.,  Wolf E.~T.,  Haqq-Misra J.,  Yang J.,  Kasting J.~F.,
  Meadows V.,  Terrien R.,   Mahadevan S.,  2016, \mn@doi [The Astrophysical
  Journal] {10.3847/0004-637X/819/1/84}, 819, 84

\bibitem[\protect\citeauthoryear{Kopparapu, Arney, Haqq-Misra, Lustig-Yaeger
  \& Villanueva}{Kopparapu et~al.}{2021}]{kopparapu_nitrogen_2021}
Kopparapu R.,  Arney G.,  Haqq-Misra J.,  Lustig-Yaeger J.,   Villanueva G.,
  2021, \mn@doi [The Astrophysical Journal] {10.3847/1538-4357/abd7f7}, 908,
  164

\bibitem[\protect\citeauthoryear{Kozakis, Mendonça  \& Buchhave}{Kozakis
  et~al.}{2022}]{kozakis_is_2022}
Kozakis T.,  Mendonça J.~M.,   Buchhave L.~A.,  2022, \mn@doi [Astronomy \&
  Astrophysics] {10.1051/0004-6361/202244164}

\bibitem[\protect\citeauthoryear{Kreidberg et~al.,}{Kreidberg
  et~al.}{2014}]{kreidberg_precise_2014}
Kreidberg L.,  et~al., 2014, \mn@doi [The Astrophysical Journal Letters]
  {10.1088/2041-8205/793/2/L27}, 793, L27

\bibitem[\protect\citeauthoryear{Lewis, Lambert, Boutle, Mayne, Manners  \&
  Acreman}{Lewis et~al.}{2018}]{lewis_influence_2018}
Lewis N.~T.,  Lambert F.~H.,  Boutle I.~A.,  Mayne N.~J.,  Manners J.,
  Acreman D.~M.,  2018, \mn@doi [The Astrophysical Journal]
  {10.3847/1538-4357/aaad0a}, 854, 171

\bibitem[\protect\citeauthoryear{Lincowski, Lustig-Yaeger  \&
  Meadows}{Lincowski et~al.}{2019}]{lincowski_observing_2019}
Lincowski A.~P.,  Lustig-Yaeger J.,   Meadows V.~S.,  2019, \mn@doi [The
  Astronomical Journal] {10.3847/1538-3881/ab2385}, 158, 26

\bibitem[\protect\citeauthoryear{Liu, Cecil, Zipser, Kronfeld  \&
  Robertson}{Liu et~al.}{2012}]{liu_relationships_2012}
Liu C.,  Cecil D.~J.,  Zipser E.~J.,  Kronfeld K.,   Robertson R.,  2012,
  \mn@doi [Journal of Geophysical Research: Atmospheres]
  {10.1029/2011JD017123}, 117

\bibitem[\protect\citeauthoryear{Lock, Brown, Bush, Martin  \& Smith}{Lock
  et~al.}{2000}]{lock_new_2000}
Lock A.~P.,  Brown A.~R.,  Bush M.~R.,  Martin G.~M.,   Smith R. N.~B.,  2000,
  \mn@doi [Monthly Weather Review]
  {10.1175/1520-0493(2000)128<3187:ANBLMS>2.0.CO;2}, 128, 3187

\bibitem[\protect\citeauthoryear{Loyd et~al.,}{Loyd
  et~al.}{2016}]{loyd_muscles_2016}
Loyd R. O.~P.,  et~al., 2016, \mn@doi [The Astrophysical Journal]
  {10.3847/0004-637X/824/2/102}, 824, 102

\bibitem[\protect\citeauthoryear{Luhar, Galbally, Woodhouse  \& Abraham}{Luhar
  et~al.}{2021}]{luhar_assessing_2021}
Luhar A.~K.,  Galbally I.~E.,  Woodhouse M.~T.,   Abraham N.~L.,  2021, \mn@doi
  [Atmospheric Chemistry and Physics] {10.5194/acp-21-7053-2021}, 21, 7053

\bibitem[\protect\citeauthoryear{Lustig-Yaeger, Meadows  \&
  Lincowski}{Lustig-Yaeger et~al.}{2019}]{lustig-yaeger_detectability_2019}
Lustig-Yaeger J.,  Meadows V.~S.,   Lincowski A.~P.,  2019, \mn@doi [The
  Astronomical Journal] {10.3847/1538-3881/ab21e0}, 158, 27

\bibitem[\protect\citeauthoryear{Manners, Edwards, Hill  \& Thelen}{Manners
  et~al.}{2021}]{manners_socrates_2021}
Manners J.,  Edwards J.~M.,  Hill P.,   Thelen J.-C.,  2021, {SOCRATES}
  ({Suite} {Of} {Community} {RAdiative} {Transfer} codes based on {Edwards} and
  {Slingo}) {Technical} {Guide}, \url
  {https://code.metoffice.gov.uk/trac/socrates}

\bibitem[\protect\citeauthoryear{Mao, Zhao, Keller, Wang, McFarland, Jenkins
  \& Brune}{Mao et~al.}{2021}]{mao_global_2021}
Mao J.,  Zhao T.,  Keller C.~A.,  Wang X.,  McFarland P.~J.,  Jenkins J.~M.,
  Brune W.~H.,  2021, \mn@doi [Geophysical Research Letters]
  {10.1029/2021GL095740}, 48, e2021GL095740

\bibitem[\protect\citeauthoryear{Mayne, Baraffe, Acreman, Smith, Wood,
  Amundsen, Thuburn  \& Jackson}{Mayne et~al.}{2014a}]{mayne_using_2014}
Mayne N.~J.,  Baraffe I.,  Acreman D.~M.,  Smith C.,  Wood N.,  Amundsen D.~S.,
   Thuburn J.,   Jackson D.~R.,  2014a, \mn@doi [Geoscientific Model
  Development] {10.5194/gmd-7-3059-2014}, 7, 3059

\bibitem[\protect\citeauthoryear{Mayne et~al.,}{Mayne
  et~al.}{2014b}]{mayne_unified_2014}
Mayne N.~J.,  et~al., 2014b, \mn@doi [Astronomy \& Astrophysics]
  {10.1051/0004-6361/201322174}, 561, A1

\bibitem[\protect\citeauthoryear{Mayne et~al.,}{Mayne
  et~al.}{2017}]{mayne_results_2017}
Mayne N.~J.,  et~al., 2017, \mn@doi [Astronomy \& Astrophysics]
  {10.1051/0004-6361/201730465}, 604, A79

\bibitem[\protect\citeauthoryear{Merlis \& Schneider}{Merlis \&
  Schneider}{2010}]{merlis_atmospheric_2010}
Merlis T.~M.,  Schneider T.,  2010, \mn@doi [Journal of Advances in Modeling
  Earth Systems] {10.3894/JAMES.2010.2.13}, 2

\bibitem[\protect\citeauthoryear{{Met Office}}{{Met
  Office}}{2022}]{met_office_iris_2022}
{Met Office} 2022, Iris: {A} {Python} library for analysing and visualising
  meteorological and oceanographic data sets, \url {http://scitools.org.uk/}

\bibitem[\protect\citeauthoryear{Miller}{Miller}{1953}]{miller_production_1953}
Miller S.~L.,  1953, Science, New Series, 117, 528

\bibitem[\protect\citeauthoryear{Miyazaki, Eskes, Sudo  \& Zhang}{Miyazaki
  et~al.}{2014}]{miyazaki_global_2014}
Miyazaki K.,  Eskes H.~J.,  Sudo K.,   Zhang C.,  2014, \mn@doi [Atmospheric
  Chemistry and Physics] {10.5194/acp-14-3277-2014}, 14, 3277

\bibitem[\protect\citeauthoryear{Morgenstern, Braesicke, O'Connor, Bushell,
  Johnson, Osprey  \& Pyle}{Morgenstern
  et~al.}{2009}]{morgenstern_evaluation_2009}
Morgenstern O.,  Braesicke P.,  O'Connor F.~M.,  Bushell A.~C.,  Johnson C.~E.,
   Osprey S.~M.,   Pyle J.~A.,  2009, \mn@doi [Geoscientific Model Development]
  {https://doi.org/10.5194/gmd-2-43-2009}, 2, 43

\bibitem[\protect\citeauthoryear{Murray}{Murray}{2016}]{murray_lightning_2016}
Murray L.~T.,  2016, \mn@doi [Current Pollution Reports]
  {10.1007/s40726-016-0031-7}, 2, 115

\bibitem[\protect\citeauthoryear{Navarro-González, McKay  \&
  Mvondo}{Navarro-González et~al.}{2001}]{navarro-gonzalez_possible_2001}
Navarro-González R.,  McKay C.~P.,   Mvondo D.~N.,  2001, \mn@doi [Nature]
  {10.1038/35083537}, 412, 61

\bibitem[\protect\citeauthoryear{Neu, Prather  \& Penner}{Neu
  et~al.}{2007}]{neu_global_2007}
Neu J.~L.,  Prather M.~J.,   Penner J.~E.,  2007, \mn@doi [Journal of
  Geophysical Research: Atmospheres] {https://doi.org/10.1029/2006JD008007},
  112

\bibitem[\protect\citeauthoryear{Nna~Mvondo, Navarro-González, McKay, Coll  \&
  Raulin}{Nna~Mvondo et~al.}{2001}]{nna_mvondo_production_2001}
Nna~Mvondo D.,  Navarro-González R.,  McKay C.~P.,  Coll P.,   Raulin F.,
  2001, \mn@doi [Advances in Space Research] {10.1016/S0273-1177(01)00050-3},
  27, 217

\bibitem[\protect\citeauthoryear{O'Connor et~al.,}{O'Connor
  et~al.}{2014}]{oconnor_evaluation_2014}
O'Connor F.~M.,  et~al., 2014, \mn@doi [Geoscientific Model Development]
  {10.5194/gmd-7-41-2014}, 7, 41

\bibitem[\protect\citeauthoryear{O’Malley-James \&
  Kaltenegger}{O’Malley-James \& Kaltenegger}{2017}]{omalley-james_uv_2017}
O’Malley-James J.~T.,  Kaltenegger L.,  2017, \mn@doi [Monthly Notices of the
  Royal Astronomical Society: Letters] {10.1093/mnrasl/slx047}, 469, L26

\bibitem[\protect\citeauthoryear{Paradise, Fan, Menou  \& Lee}{Paradise
  et~al.}{2021}]{paradise_climate_2021}
Paradise A.,  Fan B.~L.,  Menou K.,   Lee C.,  2021, \mn@doi [Icarus]
  {10.1016/j.icarus.2020.114301}, 358, 114301

\bibitem[\protect\citeauthoryear{Pierrehumbert}{Pierrehumbert}{2010}]{pierrehumbert_palette_2010}
Pierrehumbert R.~T.,  2010, \mn@doi [The Astrophysical Journal]
  {10.1088/2041-8205/726/1/L8}, 726, L8

\bibitem[\protect\citeauthoryear{Pont, Sing, Gibson, Aigrain, Henry  \&
  Husnoo}{Pont et~al.}{2013}]{pont_prevalence_2013}
Pont F.,  Sing D.~K.,  Gibson N.~P.,  Aigrain S.,  Henry G.,   Husnoo N.,
  2013, \mn@doi [Monthly Notices of the Royal Astronomical Society]
  {10.1093/mnras/stt651}, 432, 2917

\bibitem[\protect\citeauthoryear{Price \& Rind}{Price \&
  Rind}{1992}]{price_simple_1992}
Price C.,  Rind D.,  1992, \mn@doi [Journal of Geophysical Research:
  Atmospheres] {https://doi.org/10.1029/92JD00719}, 97, 9919

\bibitem[\protect\citeauthoryear{Price \& Rind}{Price \&
  Rind}{1993}]{price_what_1993}
Price C.,  Rind D.,  1993, \mn@doi [Geophysical Research Letters]
  {10.1029/93GL00226}, 20, 463

\bibitem[\protect\citeauthoryear{Price \& Rind}{Price \&
  Rind}{1994}]{price_modeling_1994}
Price C.,  Rind D.,  1994, \mn@doi [Monthly Weather Review]
  {10.1175/1520-0493(1994)122<1930:MGLDIA>2.0.CO;2}, 122, 1930

\bibitem[\protect\citeauthoryear{Proedrou \& Hocke}{Proedrou \&
  Hocke}{2016}]{proedrou_characterising_2016}
Proedrou E.,  Hocke K.,  2016, \mn@doi [Earth, Planets and Space]
  {10.1186/s40623-016-0461-x}, 68, 96

\bibitem[\protect\citeauthoryear{Rakov \& Uman}{Rakov \&
  Uman}{2003}]{rakov_lightning_2003}
Rakov V.~A.,  Uman M.~A.,  2003, Lightning: {Physics} and {Effects}.
Cambridge University Press

\bibitem[\protect\citeauthoryear{Ribas et~al.,}{Ribas
  et~al.}{2016}]{ribas_habitability_2016}
Ribas I.,  et~al., 2016, \mn@doi [Astronomy \& Astrophysics]
  {10.1051/0004-6361/201629576}, 596, A111

\bibitem[\protect\citeauthoryear{Ribas, Gregg, Boyajian  \& Bolmont}{Ribas
  et~al.}{2017}]{ribas_full_2017}
Ribas I.,  Gregg M.~D.,  Boyajian T.~S.,   Bolmont E.,  2017, \mn@doi
  [Astronomy \& Astrophysics] {10.1051/0004-6361/201730582}, 603, A58

\bibitem[\protect\citeauthoryear{Rimmer \& Helling}{Rimmer \&
  Helling}{2016}]{rimmer_chemical_2016}
Rimmer P.~B.,  Helling C.,  2016, \mn@doi [The Astrophysical Journal Supplement
  Series] {10.3847/0067-0049/224/1/9}, 224, 9

\bibitem[\protect\citeauthoryear{Rushby, Shields, Wolf, Laguë  \&
  Burgasser}{Rushby et~al.}{2020}]{rushby_effect_2020}
Rushby A.~J.,  Shields A.~L.,  Wolf E.~T.,  Laguë M.,   Burgasser A.,  2020,
  \mn@doi [The Astrophysical Journal] {10.3847/1538-4357/abbe04}, 904, 124

\bibitem[\protect\citeauthoryear{Sagan, Thompson, Carlson, Gurnett  \&
  Hord}{Sagan et~al.}{1993}]{sagan_search_1993}
Sagan C.,  Thompson W.~R.,  Carlson R.,  Gurnett D.,   Hord C.,  1993, \mn@doi
  [Nature] {10.1038/365715a0}, 365, 715

\bibitem[\protect\citeauthoryear{Scalo et~al.,}{Scalo
  et~al.}{2007}]{scalo_m_2007}
Scalo J.,  et~al., 2007, \mn@doi [Astrobiology] {10.1089/ast.2006.0125}, 7, 85

\bibitem[\protect\citeauthoryear{Schumann \& Huntrieser}{Schumann \&
  Huntrieser}{2007}]{schumann_global_2007}
Schumann U.,  Huntrieser H.,  2007, Atmos. Chem. Phys., p.~85

\bibitem[\protect\citeauthoryear{Schwieterman et~al.,}{Schwieterman
  et~al.}{2018}]{schwieterman_exoplanet_2018}
Schwieterman E.~W.,  et~al., 2018, \mn@doi [Astrobiology]
  {10.1089/ast.2017.1729}, 18, 663

\bibitem[\protect\citeauthoryear{Segura, Krelove, Kasting, Sommerlatt, Meadows,
  Crisp, Cohen  \& Mlawer}{Segura et~al.}{2003}]{segura_ozone_2003}
Segura A.,  Krelove K.,  Kasting J.~F.,  Sommerlatt D.,  Meadows V.,  Crisp D.,
   Cohen M.,   Mlawer E.,  2003, \mn@doi [Astrobiology]
  {10.1089/153110703322736024}, 3, 689

\bibitem[\protect\citeauthoryear{Segura, Kasting, Meadows, Cohen, Scalo, Crisp,
  Butler  \& Tinetti}{Segura et~al.}{2005}]{segura_biosignatures_2005}
Segura A.,  Kasting J.~F.,  Meadows V.,  Cohen M.,  Scalo J.,  Crisp D.,
  Butler R.~A.,   Tinetti G.,  2005, \mn@doi [Astrobiology]
  {10.1089/ast.2005.5.706}, 5, 706

\bibitem[\protect\citeauthoryear{Segura, Walkowicz, Meadows, Kasting  \&
  Hawley}{Segura et~al.}{2010}]{segura_effect_2010}
Segura A.,  Walkowicz L.~M.,  Meadows V.,  Kasting J.,   Hawley S.,  2010,
  \mn@doi [Astrobiology] {10.1089/ast.2009.0376}, 10, 751

\bibitem[\protect\citeauthoryear{Seinfeld \& Pandis}{Seinfeld \&
  Pandis}{2016}]{seinfeld_atmospheric_2016}
Seinfeld J.~H.,  Pandis S.~N.,  2016, Atmospheric {Chemistry} and {Physics}:
  {From} {Air} {Pollution} to {Climate} {Change}.
John Wiley \& Sons

\bibitem[\protect\citeauthoryear{Sergeev \& Zamyatina}{Sergeev \&
  Zamyatina}{2022}]{sergeev_aeolus_2022}
Sergeev D.~E.,  Zamyatina M.,  2022, aeolus, \url
  {https://doi.org/10.5281/zenodo.5145604}

\bibitem[\protect\citeauthoryear{Sergeev, Lambert, Mayne, Boutle, Manners  \&
  Kohary}{Sergeev et~al.}{2020}]{sergeev_atmospheric_2020}
Sergeev D.~E.,  Lambert F.~H.,  Mayne N.~J.,  Boutle I.~A.,  Manners J.,
  Kohary K.,  2020, \mn@doi [The Astrophysical Journal]
  {10.3847/1538-4357/ab8882}, 894, 84

\bibitem[\protect\citeauthoryear{Sergeev et~al.,}{Sergeev
  et~al.}{2022}]{sergeev_trappist-1_2022}
Sergeev D.~E.,  et~al., 2022, \mn@doi [The Planetary Science Journal]
  {10.3847/PSJ/ac6cf2}, 3, 212

\bibitem[\protect\citeauthoryear{Shields, Meadows, Bitz, Pierrehumbert, Joshi
  \& Robinson}{Shields et~al.}{2013}]{shields_effect_2013}
Shields A.~L.,  Meadows V.~S.,  Bitz C.~M.,  Pierrehumbert R.~T.,  Joshi M.~M.,
    Robinson T.~D.,  2013, \mn@doi [Astrobiology] {10.1089/ast.2012.0961}, 13,
  715

\bibitem[\protect\citeauthoryear{Shields, Ballard  \& Johnson}{Shields
  et~al.}{2016}]{shields_habitability_2016}
Shields A.~L.,  Ballard S.,   Johnson J.~A.,  2016, \mn@doi [Physics Reports]
  {10.1016/j.physrep.2016.10.003}, 663, 1

\bibitem[\protect\citeauthoryear{Showman \& Guillot}{Showman \&
  Guillot}{2002}]{showman_atmospheric_2002}
Showman A.~P.,  Guillot T.,  2002, \mn@doi [Astronomy \& Astrophysics]
  {10.1051/0004-6361:20020101}, 385, 166

\bibitem[\protect\citeauthoryear{Showman \& Polvani}{Showman \&
  Polvani}{2011}]{showman_equatorial_2011}
Showman A.~P.,  Polvani L.~M.,  2011, \mn@doi [The Astrophysical Journal]
  {10.1088/0004-637X/738/1/71}, 738, 71

\bibitem[\protect\citeauthoryear{Stolz, Bilsback, Pierce  \& Rutledge}{Stolz
  et~al.}{2021}]{stolz_evaluating_2021}
Stolz D.~C.,  Bilsback K.~R.,  Pierce J.~R.,   Rutledge S.~A.,  2021, \mn@doi
  [Journal of Geophysical Research: Atmospheres] {10.1029/2020JD033695}, 126,
  e2020JD033695

\bibitem[\protect\citeauthoryear{Tabataba-Vakili, Grenfell, Grießmeier  \&
  Rauer}{Tabataba-Vakili et~al.}{2016}]{tabataba-vakili_atmospheric_2016}
Tabataba-Vakili F.,  Grenfell J.~L.,  Grießmeier J.-M.,   Rauer H.,  2016,
  \mn@doi [Astronomy \& Astrophysics] {10.1051/0004-6361/201425602}, 585, A96

\bibitem[\protect\citeauthoryear{Teal, Kempton, Bastelberger, Youngblood  \&
  Arney}{Teal et~al.}{2022}]{teal_effects_2022}
Teal D.~J.,  Kempton E. M.-R.,  Bastelberger S.,  Youngblood A.,   Arney G.,
  2022, \mn@doi [The Astrophysical Journal] {10.3847/1538-4357/ac4d99}, 927, 90

\bibitem[\protect\citeauthoryear{Telford et~al.,}{Telford
  et~al.}{2013}]{telford_implementation_2013}
Telford P.~J.,  et~al., 2013, \mn@doi [Geoscientific Model Development]
  {10.5194/gmd-6-161-2013}, 6, 161

\bibitem[\protect\citeauthoryear{Tian}{Tian}{2015}]{tian_history_2015}
Tian F.,  2015, \mn@doi [Earth and Planetary Science Letters]
  {10.1016/j.epsl.2015.09.051}, 432, 126

\bibitem[\protect\citeauthoryear{Tian, France, Linsky, Mauas  \& Vieytes}{Tian
  et~al.}{2014}]{tian_high_2014}
Tian F.,  France K.,  Linsky J.~L.,  Mauas P. J.~D.,   Vieytes M.~C.,  2014,
  \mn@doi [Earth and Planetary Science Letters] {10.1016/j.epsl.2013.10.024},
  385, 22

\bibitem[\protect\citeauthoryear{Turbet, Leconte, Selsis, Bolmont, Forget,
  Ribas, Raymond  \& Anglada-Escudé}{Turbet
  et~al.}{2016}]{turbet_habitability_2016}
Turbet M.,  Leconte J.,  Selsis F.,  Bolmont E.,  Forget F.,  Ribas I.,
  Raymond S.~N.,   Anglada-Escudé G.,  2016, \mn@doi [Astronomy \&
  Astrophysics] {10.1051/0004-6361/201629577}, 596, A112

\bibitem[\protect\citeauthoryear{Turbet et~al.,}{Turbet
  et~al.}{2018}]{turbet_modeling_2018}
Turbet M.,  et~al., 2018, \mn@doi [Astronomy \& Astrophysics]
  {10.1051/0004-6361/201731620}, 612, A86

\bibitem[\protect\citeauthoryear{Turbet, Bolmont, Bourrier, Demory, Leconte,
  Owen  \& Wolf}{Turbet et~al.}{2020}]{turbet_review_2020}
Turbet M.,  Bolmont E.,  Bourrier V.,  Demory B.-O.,  Leconte J.,  Owen J.,
  Wolf E.~T.,  2020, \mn@doi [Space Science Reviews]
  {10.1007/s11214-020-00719-1}, 216, 100

\bibitem[\protect\citeauthoryear{Turbet et~al.,}{Turbet
  et~al.}{2022}]{turbet_trappist-1_2022}
Turbet M.,  et~al., 2022, \mn@doi [The Planetary Science Journal]
  {10.3847/PSJ/ac6cf0}, 3, 211

\bibitem[\protect\citeauthoryear{Villanueva, Smith, Protopapa, Faggi  \&
  Mandell}{Villanueva et~al.}{2018}]{villanueva_planetary_2018}
Villanueva G.~L.,  Smith M.~D.,  Protopapa S.,  Faggi S.,   Mandell A.~M.,
  2018, \mn@doi [Journal of Quantitative Spectroscopy and Radiative Transfer]
  {10.1016/j.jqsrt.2018.05.023}, 217, 86

\bibitem[\protect\citeauthoryear{Von~Glasow, Bobrowski  \& Kern}{Von~Glasow
  et~al.}{2009}]{von_glasow_effects_2009}
Von~Glasow R.,  Bobrowski N.,   Kern C.,  2009, \mn@doi [Chemical Geology]
  {10.1016/j.chemgeo.2008.08.020}, 263, 131

\bibitem[\protect\citeauthoryear{Vonnegut}{Vonnegut}{1963}]{vonnegut_facts_1963}
Vonnegut B.,  1963, in Atlas D.,  et~al., eds, Meteorological {Monographs},
  Severe {Local} {Storms}.
American Meteorological Society, Boston, MA, pp 224--241

\bibitem[\protect\citeauthoryear{Walters et~al.,}{Walters
  et~al.}{2019}]{walters_met_2019}
Walters D.,  et~al., 2019, \mn@doi [Geoscientific Model Development]
  {https://doi.org/10.5194/gmd-12-1909-2019}, 12, 1909

\bibitem[\protect\citeauthoryear{Wild \& Prather}{Wild \&
  Prather}{2000}]{wild_excitation_2000}
Wild O.,  Prather M.~J.,  2000, \mn@doi [Journal of Geophysical Research:
  Atmospheres] {https://doi.org/10.1029/2000JD900399}, 105, 24647

\bibitem[\protect\citeauthoryear{Williams}{Williams}{1985}]{williams_large-scale_1985}
Williams E.~R.,  1985, \mn@doi [Journal of Geophysical Research: Atmospheres]
  {https://doi.org/10.1029/JD090iD04p06013}, 90, 6013

\bibitem[\protect\citeauthoryear{Williams, Chan  \& Boccippio}{Williams
  et~al.}{2004}]{williams_islands_2004}
Williams E.,  Chan T.,   Boccippio D.,  2004, \mn@doi [Journal of Geophysical
  Research: Atmospheres] {10.1029/2003JD003833}, 109

\bibitem[\protect\citeauthoryear{Wilson, Bushell, Kerr‐Munslow, Price  \&
  Morcrette}{Wilson et~al.}{2008}]{wilson_pc2_2008}
Wilson D.~R.,  Bushell A.~C.,  Kerr‐Munslow A.~M.,  Price J.~D.,   Morcrette
  C.~J.,  2008, \mn@doi [Quarterly Journal of the Royal Meteorological Society]
  {https://doi.org/10.1002/qj.333}, 134, 2093

\bibitem[\protect\citeauthoryear{Wolf, Shields, Kopparapu, Haqq-Misra  \&
  Toon}{Wolf et~al.}{2017}]{wolf_constraints_2017}
Wolf E.~T.,  Shields A.~L.,  Kopparapu R.~K.,  Haqq-Misra J.,   Toon O.~B.,
  2017, \mn@doi [The Astrophysical Journal] {10.3847/1538-4357/aa5ffc}, 837,
  107

\bibitem[\protect\citeauthoryear{Wood et~al.,}{Wood
  et~al.}{2014}]{wood_inherently_2014}
Wood N.,  et~al., 2014, \mn@doi [Quarterly Journal of the Royal Meteorological
  Society] {https://doi.org/10.1002/qj.2235}, 140, 1505

\bibitem[\protect\citeauthoryear{Wordsworth, Schaefer  \& Fischer}{Wordsworth
  et~al.}{2018}]{wordsworth_redox_2018}
Wordsworth R.~D.,  Schaefer L.~K.,   Fischer R.~A.,  2018, \mn@doi [The
  Astronomical Journal] {10.3847/1538-3881/aab608}, 155, 195

\bibitem[\protect\citeauthoryear{Yang \& Abbot}{Yang \&
  Abbot}{2014}]{yang_low-order_2014}
Yang J.,  Abbot D.~S.,  2014, \mn@doi [The Astrophysical Journal]
  {10.1088/0004-637X/784/2/155}, 784, 155

\bibitem[\protect\citeauthoryear{Yang, Cowan  \& Abbot}{Yang
  et~al.}{2013}]{yang_stabilizing_2013}
Yang J.,  Cowan N.~B.,   Abbot D.~S.,  2013, \mn@doi [The Astrophysical
  Journal] {10.1088/2041-8205/771/2/L45}, 771, L45

\bibitem[\protect\citeauthoryear{Yates, Palmer, Manners, Boutle, Kohary, Mayne
  \& Abraham}{Yates et~al.}{2020}]{yates_ozone_2020}
Yates J.~S.,  Palmer P.~I.,  Manners J.,  Boutle I.,  Kohary K.,  Mayne N.,
  Abraham L.,  2020, \mn@doi [Monthly Notices of the Royal Astronomical
  Society] {10.1093/mnras/stz3520}, 492, 1691

\bibitem[\protect\citeauthoryear{Youngblood et~al.,}{Youngblood
  et~al.}{2016}]{youngblood_muscles_2016}
Youngblood A.,  et~al., 2016, \mn@doi [The Astrophysical Journal]
  {10.3847/0004-637X/824/2/101}, 824, 101

\bibitem[\protect\citeauthoryear{Zarka, Farrell, Kaiser, Blanc  \& Kurth}{Zarka
  et~al.}{2004}]{zarka_study_2004}
Zarka P.,  Farrell W.~M.,  Kaiser M.~L.,  Blanc E.,   Kurth W.~S.,  2004,
  \mn@doi [Planetary and Space Science] {10.1016/j.pss.2004.09.011}, 52, 1435

\bibitem[\protect\citeauthoryear{Zeldovich, Frank-Kamenetskii  \&
  Sadovnikov}{Zeldovich et~al.}{1947}]{zeldovich_oxidation_1947}
Zeldovich Y.~B.,  Frank-Kamenetskii D.,   Sadovnikov P.,  1947, Oxidation of
  nitrogen in combustion.
Publishing House of the Acad of Sciences of USSR

\bibitem[\protect\citeauthoryear{Zhang \& Yang}{Zhang \&
  Yang}{2020}]{zhang_how_2020}
Zhang Y.,  Yang J.,  2020, \mn@doi [The Astrophysical Journal]
  {10.3847/2041-8213/abb87f}, 901, L36

\makeatother
\end{thebibliography}



\clearpage
\appendix
\section{Sensitivity of ozone chemistry to different SEDs}\label{app:o3chem}
\begin{figure}
	\includegraphics[width=\columnwidth]{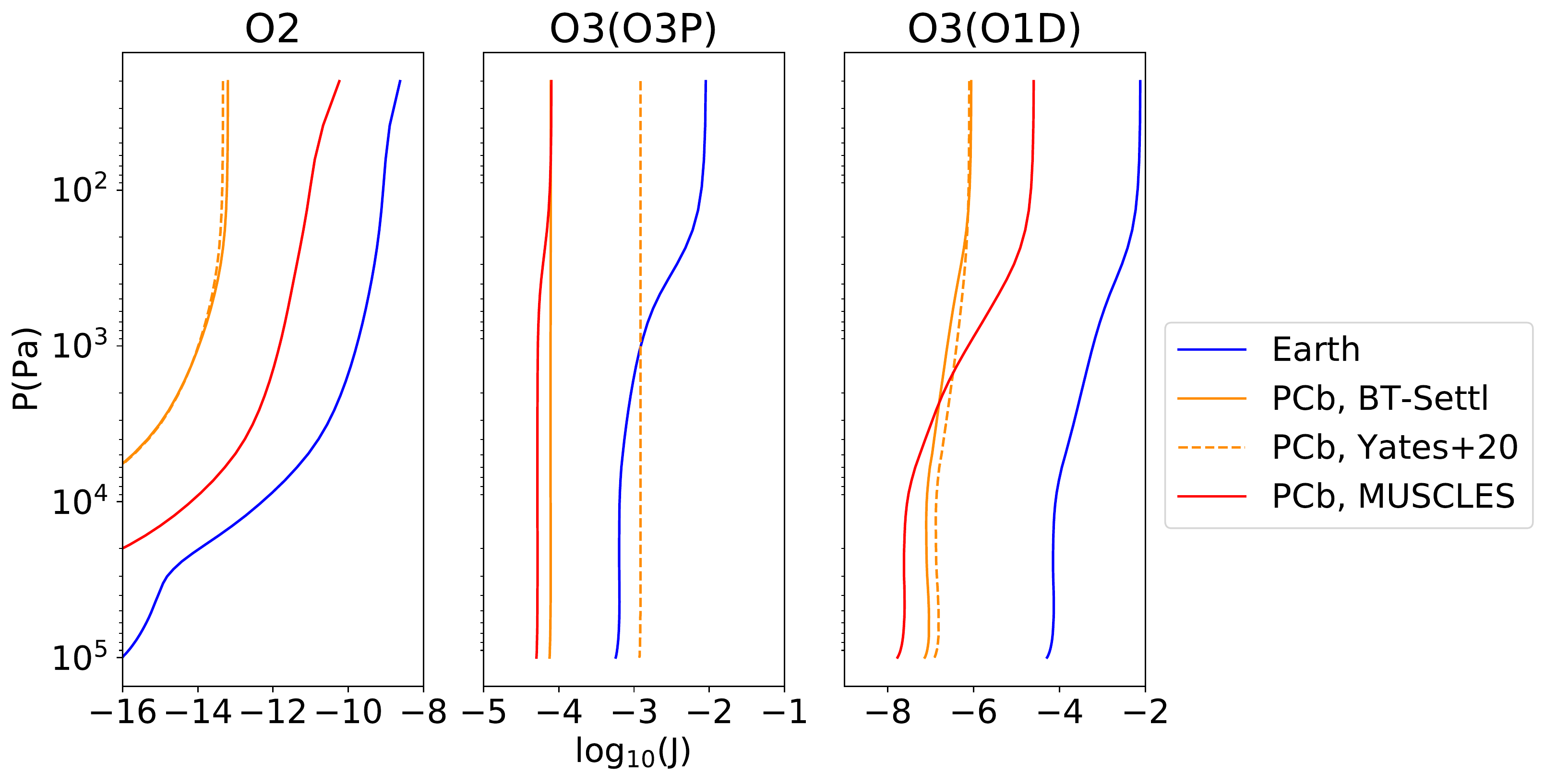}
    \caption{Photolysis rates (J, in s$^{-1}$) as computed with Fast-JX offline runs. Every run corresponds to 1-D column in an atmosphere assuming Earth-like conditions (e.g. O$_3$ layer, P-T structure, etc.). The only change between the runs is the stellar flux distribution.}
    \label{fig:jrates}
\end{figure}

The application of different stellar SEDs in our study (see Figure~\ref{fig:spectra_irrad}) and the comparison with \citet{yates_ozone_2020} allow for an assessment of the sensitivity of ozone chemistry to stellar SED. As can be seen in Figure~\ref{fig:spectra_irrad}, the BT-Settl spectrum represents a quiescent M-dwarf with minor UV output. The MUSCLES spectrum provides a better representation of the UV output from M-dwarfs, though it does not account for time-dependent stellar activity. To quantify the impact of UV fluxes on the ozone distribution, Figure~\ref{fig:jrates} shows the photolysis (or J-) rates for reactions R1 (O$_2$ photolysis), R3 (O$_3$ photolysis into O($^3$P)) and R4 (O$_3$ photolysis into O($^1$D)), respectively. The rates follow from:
\begin{equation}
    J = \int^{\lambda_{\mathrm{max}}}_{\lambda_{\mathrm{min}}}\sigma(\lambda)\phi(\lambda)F(\lambda)d\lambda,
\end{equation}
where $\sigma$ and $\phi$ are the cross-sections (in cm$^2$) and quantum yields (unitless) for constituents (from the JPL\footnote{\href{https://jpldataeval.jpl.nasa.gov/}{https://jpldataeval.jpl.nasa.gov/}} and IUPAC\footnote{\href{https://iupac-aeris.ipsl.fr/}{https://iupac-aeris.ipsl.fr/}} databases) and $F(\lambda)$ denotes the actinic flux travelling through the atmosphere in photons~cm$^{-2}$~s$^{-1}$. The photolysis rates have been calculated using the offline version of Fast-JX \citep[][]{wild_excitation_2000, bian_fast-j2_2002}, under the assumption of an Earth-like atmosphere (O$_3$ layer, P-T structure, cloud distributions, etc.). The Earth-like parameters were kept unchanged for the sake of comparison as well as computational efficiency, while only changing the stellar flux distributions according to the spectra shown in Figure~\ref{fig:spectra_irrad}. The photolysis rate is calculated in each of the 18 wavelength bins, with a $\lambda_{\mathrm{min}}$ and $\lambda_{\mathrm{max}}$ corresponding to each of the bins, as described in \citet{bian_fast-j2_2002}.

\begin{figure}
	\includegraphics[width=\columnwidth]{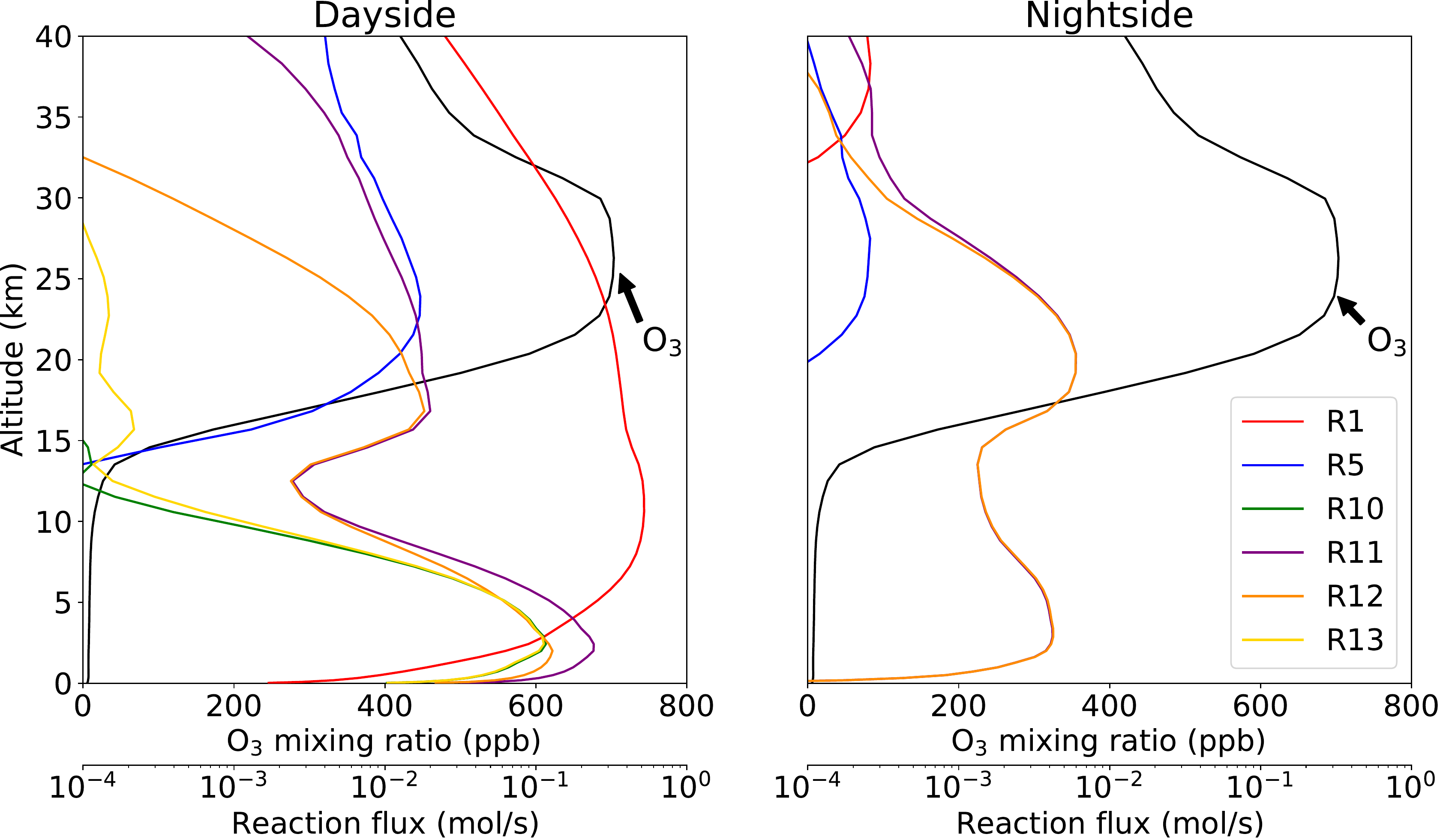}
    \caption{Reaction rates and ozone profile using the spectral flux distribution as presented in Table 2 of \citet{yates_ozone_2020}. Can be compared to the lower row of their figure 2. Note that the numbering of chemical reactions is different.}
    \label{fig:rflux_repy}
\end{figure}

\begin{figure}
	\includegraphics[width=\columnwidth]{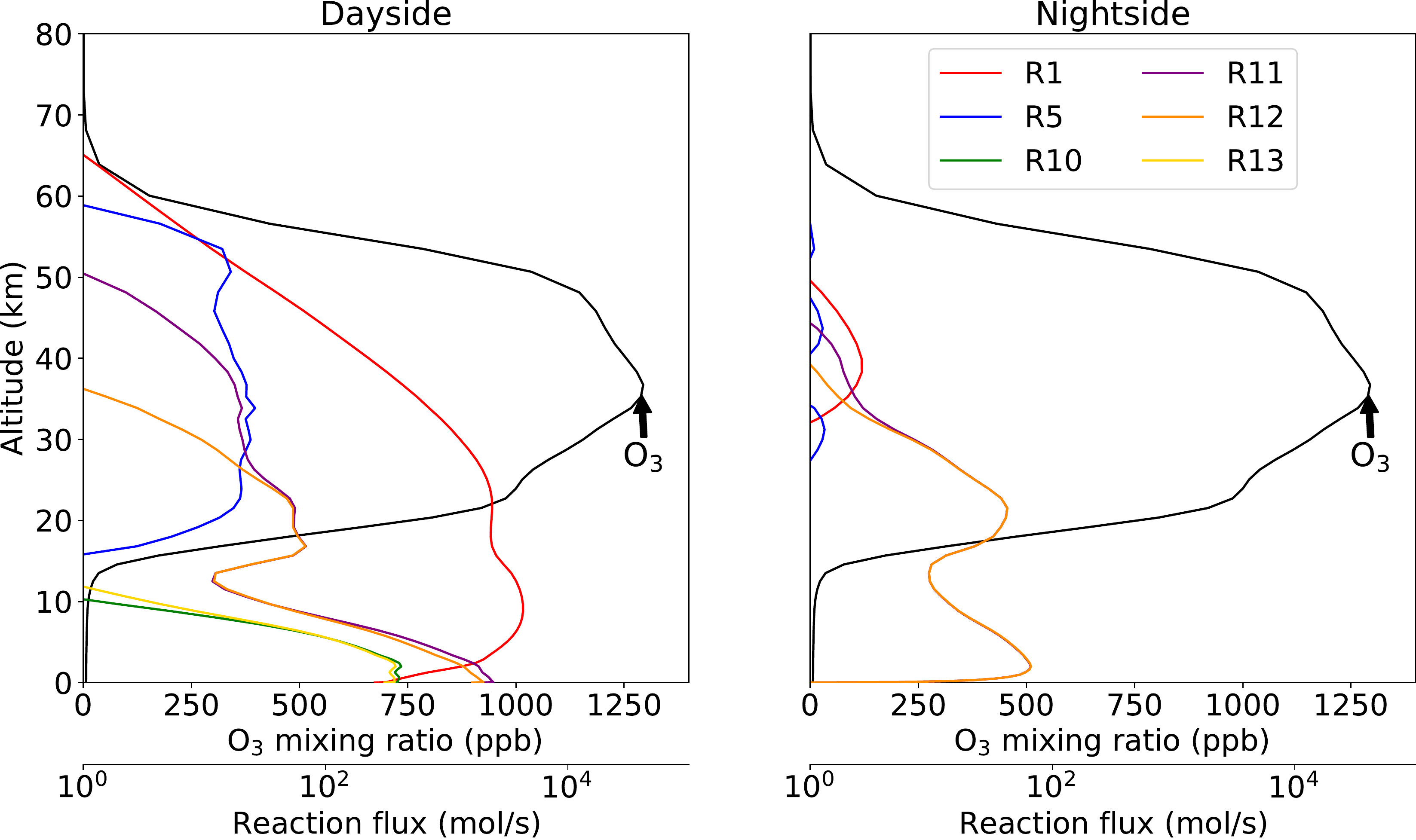}
    \caption{Reaction rates and ozone profile using the improved spectral flux distribution \citep[][]{bian_fast-j2_2002} for the BT-Settl spectrum. Results shown are from using the Chapman mechanism and basic HO$_\mathrm{x}$ chemistry. Can be compared to Figure~\ref{fig:rflux_repy} and the bottom panels of figure 2 in \citet{yates_ozone_2020}.}
    \label{fig:rflux_b_hox}
\end{figure}

First, the blue lines indicate the J-rates for the flux that Earth receives, and have been validated in intercomparison studies \citep[e.g.][]{chipperfield_chapter_2010}. We can compare the two orange lines in Figure~\ref{fig:jrates}, both corresponding to BT-Settl spectra for Proxima Centauri b. The dotted line shows the photolysis rates for the rebinning of fluxes by \citet{yates_ozone_2020}, while the solid line shows our improved rebinning for BT-Settl. The J-rates for O$_2$ and O$_3$(O$^1$D) are similar between the two cases, but O$_3$(O$^3$P) photolysis was a factor 10 stronger in the models from \citet{yates_ozone_2020}. As described in Section~\ref{subsec:fastjx}, this is caused by their extension of the upper limit of bin 18. The improved BT-Settl spectrum results in an ozone layer peaking at ${\sim}$1360~ppb (see Figure~\ref{fig:rflux_b_hox}), two times as high as the value found by \citet{yates_ozone_2020} (see Figure~\ref{fig:rflux_repy}). 

Basing our approach on observations, substituting the MUSCLES spectrum provides us with more flux in the UV regions important for O$_2$ photolysis. Consequently, the red line in Figure~\ref{fig:jrates} shows that J(O$_2$) is ${\sim}$100--1000 times higher than for the BT-Settl spectrum, still falling short by a factor of ${\sim}$100 as compared to a solar spectrum at Earth's orbit. O$_2$ photolysis for the MUSCLES spectrum is mainly driven by the stellar flux in the Schumann-Runge bands through bins 3 and 5 and the flux in bin 10 at wavelengths of 221.5--233~nm. For both the BT-Settl and a Solar spectrum, fluxes at wavelengths of 215.5--240~nm are predominantly driving O$_2$ photolysis. This illustrates the impact of the relatively flat MUSCLES spectrum as seen in Figure~\ref{fig:spectra_irrad}. Nevertheless, the ozone layer of Proxima Centauri b peaks at a higher VMR (${\sim}$21 compared to ${\sim}$12~ppm on Earth: \citet{seinfeld_atmospheric_2016}), since O$_3$ photolysis rates are significantly lower. J(O$_3$) is lower than the BT-Settl values for P${<}10^2$~Pa, which is caused by the respective flux distributions in bins 12--15. For MUSCLES, the flux in bin 12 (291--298.3~nm) is higher, whereas the fluxes in BT-Settl bins 13--15 (298.3--320.3~nm) are higher. The impact on the O$_3$ layer is in broad agreement with the results from \citet{chen_habitability_2019}, who also find a thinner O$_3$ layer for a quiescent M-dwarf as compared to UV irradiation from an active M-dwarf and enhanced Solar spectrum.


\section{Supplementary information}
\begin{table}
\centering
\begin{tabular}{lll} \hline
\textbf{Species} & \textbf{Initial MMR} & \textbf{Notes} \\ \hline
N$_2$       & $0.78084$     &  Uniform \\
O$_2$       & $0.2314$     &  Uniform, rad active \\
CO$_2$      & $5.941\times10^{-4}$     & Uniform, rad active \\
H$_2$O      & N/A$^a$   & Rad active \\
N$_2$O      & ($1\times10^{-15}$)     & Lightning-induced, rad active  \\
CH$_4$      & $0.0$    & \\
CFCs        & $0.0$     & \\
Cl-bearers$^b$ & 0.0 & \\
Br-bearers$^c$ & 0.0 & \\ 
H$_2$       & $1\times10^{-9}$  & \\
O$_3$      & $1\times10^{-9}$    &  Rad active \\
O($^3$P)      & $1\times10^{-9}$    &    \\
O($^1$D)      & $1\times10^{-9}$    &  \\
H$_2$O$_2$      & $1\times10^{-9}$  &    \\
H      & $1\times10^{-9}$    &    \\
OH      & $1\times10^{-9}$   &    \\
HO$_2$     & $1\times10^{-9}$  &    \\ 
NO         & ($1\times10^{-15}$)    & Lightning-induced   \\
NO$_2$      & ($1\times10^{-15}$)    & Lightning-induced   \\
NO$_3$      & ($1\times10^{-15}$)   & Lightning-induced  \\ 
N$_2$O$_5$      & ($1\times10^{-15}$)    & Lightning-induced   \\
HONO      & ($1\times10^{-15}$)    & Lightning-induced   \\
HNO$_3$      & ($1\times10^{-15}$)    & Lightning-induced   \\
HO$_2$NO$_2$      & ($1\times10^{-15}$)     & Lightning-induced  \\
N           & ($1\times10^{-15}$)    & Lightning-induced   \\
CO      & $0.0$   &    \\
HCHO      & $0.0$   &    \\
MeOO      & $0.0$     &    \\
MeOOH      & $0.0$    &    \\
H$_2$S      & $0.0$     &    \\
SO$_3$      & $0.0$  &    \\
DMS      & $0.0$    &    \\
SO$_2$      & $0.0$     &    \\
H$_2$SO$_4$      & $0.0$    &    \\
MSA      & $0.0$   &    \\
CS$_2$      & $0.0$   &    \\
NH$_3$      & $0.0$     &    \\
\hline
\end{tabular}
\caption[Initial abundances]{Abundances used for the chemistry initialisation. `Rad active' indicates that a species is used in the UM radiative transfer. The ozone abundance is fed back from UKCA to the UM for this radiative transfer. $^a$ H$_2$O follows from evaporation from the slab ocean; $^b$Cl-bearers include Cl, ClO, Cl$_2$O$_2$, OClO, BrCl, HOCl, ClONO$_2$, CFCl$_3$, CF$_2$Cl$_2$, HCl; $^c$Br-bearers include Br, BrO, BrCl, BrONO$_2$, HBr, HOBr, MeBr}
\label{tab:initabundances}
\end{table}

\begin{table}
\centering
\scriptsize
\begin{tabular}{lp{5cm}l} \hline
 & \textbf{Reaction} & \textbf{Scheme} \\ \hline
1      &  \ce{O($^1$D) + O$_2$ -> O($^3$P) + O$_2$}   & Chapman \\
2      &  \ce{O($^1$D) + N$_2$ -> O($^3$P) + N$_2$}   & Chapman \\
3      &  \ce{O($^1$D) + CO$_2$ -> O($^3$P) + CO$_2$}  & Chapman \\
4     & \ce{O($^1$D) + O$_3$ -> O$_2$ + O($^3$P) + O($^3$P)}  & Chapman \\
5     &  \ce{O($^1$D) + O$_3$ -> O$_2$ + O$_2$}  & Chapman \\
6     &  \ce{O($^3$P) + O$_3$ -> O$_2$ + O$_2$}    & Chapman \\
7      & \ce{O($^3$P) + O$_2$ + M -> O$_3$ + M}       & Chapman \\
8      & \ce{O$_2$ + h$\nu$ -> O($^3$P) + O($^3$P)} & Chapman \\
9      & \ce{O$_2$ + h$\nu$ -> O($^3$P) + O($^1$D)} & Chapman \\
10     &  \ce{O$_3$ + h$\nu$ -> O$_2$ + O($^3$P)}  & Chapman \\
11      & \ce{O$_3$ + h$\nu$ -> O$_2$ + O($^1$D)}   & Chapman \\ \hline
12     &  \ce{H$_2$O + O($^1$D) -> 2OH}  & HO$_\mathrm{x}$ \\
13     &  \ce{HO$_2$ + O($^3$P) -> OH + O$_2$}  & HO$_\mathrm{x}$ \\
14     &  \ce{HO$_2$ + O$_3$ -> OH + 2O$_2$}  & HO$_\mathrm{x}$ \\
15     &  \ce{OH + HO$_2$ -> H$_2$O + O$_2$}  & HO$_\mathrm{x}$ \\
16     &  \ce{OH + O$_3$ -> HO$_2$ + O$_2$}  & HO$_\mathrm{x}$ \\ 
17     &  \ce{H + HO$_2$ -> H$_2$ + O$_2$}  & HO$_\mathrm{x}$ \\ 
18     &  \ce{H + HO$_2$ -> O($^3$P) + H$_2$O}  & HO$_\mathrm{x}$ \\
19     &  \ce{H + HO$_2$ -> OH + OH}  & HO$_\mathrm{x}$ \\ 
20     &  \ce{H + O$_3$ -> OH + O$_2$}  & HO$_\mathrm{x}$ \\ 
21     &  \ce{HO$_2$ + HO$_2$ -> H$_2$O$_2$ + O$_2$}  & HO$_\mathrm{x}$ \\ 
22     &  \ce{O($^1$D) + H$_2$ -> OH + H}  & HO$_\mathrm{x}$ \\ 
23     &  \ce{O($^3$P) + H$_2$ -> OH + H}  & HO$_\mathrm{x}$ \\ 
24     &  \ce{O($^3$P) + H$_2$O$_2$ -> OH + HO$_2$}  & HO$_\mathrm{x}$ \\ 
25     &  \ce{O($^3$P) + OH -> O$_2$ + H}  & HO$_\mathrm{x}$ \\ 
26     &  \ce{OH + H$_2$ -> H$_2$O + HO$_2$}  & HO$_\mathrm{x}$ \\ 
27     &  \ce{OH + H$_2$O$_2$ -> HO$_2$ + H$_2$O}  & HO$_\mathrm{x}$ \\ 
28     &  \ce{OH + OH -> H$_2$O + O($^3$P)}  & HO$_\mathrm{x}$ \\ 
29     &  \ce{H + O$_2$ + M -> HO$_2$ + M}  & HO$_\mathrm{x}$ \\ 
30     &  \ce{HO$_2$ + HO$_2$ + M -> H$_2$O$_2$ + O$_2$ + M}  & HO$_\mathrm{x}$ \\ 
31     &  \ce{OH + OH + M -> H$_2$O$_2$ + M}  & HO$_\mathrm{x}$ \\ 
32     &  \ce{H$_2$O + h$\nu$ -> OH + H}  & HO$_\mathrm{x}$ \\ 
33     &  \ce{H$_2$O$_2$ + h$\nu$ -> OH + OH}  & HO$_\mathrm{x}$ \\ \hline
34     &  \ce{H + NO$_2$ -> OH + NO}  & NO$_\mathrm{x}$ \\ 
35     &  \ce{HNO$_3$ + OH -> NO$_3$ + H$_2$O}  & NO$_\mathrm{x}$ \\ 
36     &  \ce{HO$_2$ + NO -> OH + NO$_2$}  & NO$_\mathrm{x}$ \\ 
37     &  \ce{HO$_2$ + NO$_3$ -> OH + NO$_2$ + O$_2$}  & NO$_\mathrm{x}$ \\ 
38     &  \ce{N + NO -> N$_2$ + O($^3$P)}  & NO$_\mathrm{x}$ \\ 
39     &  \ce{N + NO$_2$ -> N$_2$O + O($^3$P)}  & NO$_\mathrm{x}$ \\ 
40     &  \ce{N + O$_2$ -> NO + O($^3$P)}  & NO$_\mathrm{x}$ \\ 
41     &  \ce{N$_2$O$_5$ + H$_2$O -> HONO$_2$ + HONO$_2$}  & NO$_\mathrm{x}$ \\ 
42     &  \ce{NO + NO$_3$ -> NO$_2$ + NO$_2$}  & NO$_\mathrm{x}$ \\ 
43     &  \ce{NO + O$_3$ -> NO$_2$ + O$_2$ }  & NO$_\mathrm{x}$ \\ 
44     &  \ce{NO$_2$ + NO$_3$ -> NO + NO$_2$ + O$_2$}  & NO$_\mathrm{x}$ \\ 
45     &  \ce{NO$_2$ + O($^3$P) -> NO + O$_2$}  & NO$_\mathrm{x}$ \\ 
46     &  \ce{NO$_2$ + O$_3$ -> NO$_3$ + O$_2$}   & NO$_\mathrm{x}$ \\ 
47     &  \ce{O($^1$D) + N$_2$O -> N$_2$ + O$_2$}  & NO$_\mathrm{x}$ \\ 
48     &  \ce{O($^1$D) + N$_2$O -> NO + NO}  & NO$_\mathrm{x}$ \\ 
49     &  \ce{O($^3$P) + NO$_3$ -> O$_2$ + NO$_2$}  & NO$_\mathrm{x}$ \\ 
50     &  \ce{OH + HO$_2$NO$_2$ -> H$_2$O + NO$_2$ + O$_2$}  & NO$_\mathrm{x}$ \\ 
51     &  \ce{OH + HONO -> H$_2$O + NO$_2$}  & NO$_\mathrm{x}$ \\ 
52     &  \ce{OH + NO$_3$ -> HO$_2$ + NO$_2$}  & NO$_\mathrm{x}$ \\ 
53     &  \ce{HO$_2$ + NO$_2$ + M -> HO$_2$NO$_2$ + M}  & NO$_\mathrm{x}$ \\ 
54     &  \ce{HO$_2$NO$_2$ + M -> HO$_2$ + NO$_2$ + M}  & NO$_\mathrm{x}$ \\ 
55     &  \ce{NO + NO + O$_2$ -> NO$_2$ + NO$_2$}  & NO$_\mathrm{x}$ \\ 
56     &  \ce{NO$_2$ + OH + M -> HNO$_3$ + M}  & NO$_\mathrm{x}$ \\ 
57     &  \ce{NO$_3$ + NO$_2$ + M -> N$_2$O$_5$ + M}  & NO$_\mathrm{x}$ \\ 
58     &  \ce{N$_2$O$_5$ + M -> NO$_2$ + NO$_3$ + M}  & NO$_\mathrm{x}$ \\ 
59     &  \ce{O($^1$D) + N$_2$ + M -> N$_2$O + M}  & NO$_\mathrm{x}$ \\ 
60     &  \ce{O($^3$P) + NO + M -> NO$_2$ + M}  & NO$_\mathrm{x}$ \\ 
61     &  \ce{O($^3$P) + NO$_2$ + M -> NO$_3$ + M}  & NO$_\mathrm{x}$ \\ 
62     &  \ce{OH + NO + M -> HONO + M}  & NO$_\mathrm{x}$ \\ 
63     &  \ce{HNO$_3$ + h$\nu$ -> NO$_2$ + OH}  & NO$_\mathrm{x}$ \\ 
64     &  \ce{HONO + h$\nu$ -> OH + NO}  & NO$_\mathrm{x}$ \\ 
65     &  \ce{HO$_2$NO$_2$ + h$\nu$ -> HO$_2$ + NO$_2$ } & NO$_\mathrm{x}$ \\ 
66     &  \ce{NO + h$\nu$ -> N + O($^3$P)}  & NO$_\mathrm{x}$ \\ 
67     &  \ce{NO$_2$ + h$\nu$ -> NO + O($^3$P)}  & NO$_\mathrm{x}$ \\ 
68     &  \ce{NO$_3$ + h$\nu$ -> NO + O$_2$}  & NO$_\mathrm{x}$ \\ 
69     &  \ce{NO$_3$ + h$\nu$ -> NO$_2$ + O($^3$P)}  & NO$_\mathrm{x}$ \\ 
70     &  \ce{N$_2$O + h$\nu$ -> N$_2$ + O($^1$D)}  & NO$_\mathrm{x}$ \\ 
71     &  \ce{N$_2$O$_5$ + h$\nu$ -> NO$_3$ + NO$_2$}  & NO$_\mathrm{x}$ \\ 
\\ \hline
\end{tabular}
\caption[Reaction lists]{Reactions included in the different chemical networks. The associated rate constants can be found in \citet{archibald_description_2020}.}
\label{tab:reactions}
\end{table}


\bsp	
\label{lastpage}
\end{document}